\documentclass[11pt,preprint,preprintnumbers,showpacs,groupedaddress,
  superscriptaddress, floatfix,longabstract]{revtex4-1}

\usepackage{lmodern}
\usepackage{amsmath}
\usepackage{amssymb}
\usepackage{amsbsy}
\usepackage{bbm}
\usepackage{multirow}
\usepackage{setspace}
\usepackage{graphicx}
\usepackage{epsfig}
\usepackage{dcolumn}
\usepackage{bm}
\usepackage[FIGBOTCAP]{subfigure}
\usepackage{slashed}
\usepackage{longtable}
\usepackage[colorlinks,
            linkbordercolor={0 0 0},
            pdfborder={0 0 0},
            linkcolor=blue,
            citecolor=blue,
            urlcolor=blue,
            breaklinks=true]{hyperref}

\allowdisplaybreaks

\def\rmsmall #1{\mbox{\scriptsize #1}}

\def\eqarray#1{\begin{eqnarray} #1 \end{eqnarray}}

\newlength{\x}
\newlength{\y}
\newlength{\z}
\newlength{\leftright}
\urldef\milcurl\url{http://www.physics.utah.edu/~detar/milc/}

\begin{document}

\title{A New Order Parameter for the Higgs Transition in \\
  \boldmath{$SU(2)$}-Higgs Theory}

\author{Mridupawan Deka} %
\email{mpdeka@theor.jinr.ru}%
\affiliation{Bogoliubov Laboratory of Theoretical Physics, JINR, 141980 Dubna, Russia}%
\affiliation{Institute of Mathematical Sciences, Chennai 600113, India}%
\author{Sanatan Digal}%
\email{digal@imsc.res.in}
\affiliation{Institute of Mathematical Sciences, Chennai 600113, India}%

{\linespread{0.95}
\begin{abstract}
\medskip
%
%\begin{center}Abstract\end{center}

We investigate the Higgs transition within the four dimensional $SU(2)-$ 
gauge-Higgs model in search for an order parameter as a function of the 
Higgs field hopping parameter,\ $\kappa$,\ using Lattice technique.\ We 
measure the Higgs condensate after applying Landau Gauge Fixing and 
study the corresponding susceptibility,\ magnetization and fourth order 
Binder cumulant using four different spatial volumes with $N_\tau\! =\! 2$.\ 
The computation is carried out with gauge coupling,\ $\beta_g\! =\! 8$,\
for a range of scalar self-coupling,\ $\lambda\! =  \! \{0.00010, 0.00350\}$,
with emphasis near the critical end-point. Finite size scaling analysis of
the gauge fixed condensate and its cumulants agree with the standard $3$d
Ising values $\nu=0.62997$, $\beta/\nu=0.518$, $\gamma/\nu=1.964$ at
$\lambda = 0.00150$. These results are in agreement with previous studies
suggesting $3$d Ising universality class. The numerical results also indicate
that, at the transition point, the gauge fixed condensate vanishes in the
infinite volume limit.

\vspace{1pc}
\end{abstract}
}

\pacs{11.10.Wx,11.15.Ha,11.15.-q}

\maketitle
\thispagestyle{empty}

\pagestyle{plain}
\setcounter{page}{1}
\pagenumbering{arabic}
\singlespacing
\parskip 1pt
\parindent 0pt

%%%%%%%%%%%%%%%%%%%%%%%%%%%%%%%%%%%%%%%%%%%%
%%%%%%%%%%%%%%%%%%%%%%%%%%%%%%%%%%%%%%%%%%%%

\section{Introduction}
\label{sec:intro}
%%%%%%%%%%%%%%%%%%%%%%%%%%%%%%%%%%%%%%%%%%%%
%%%%%%%%%%%%%%%%%%%%%%%%%%%%%%%%%%%%%%%%%%%%
The understanding of the phase structure of $SU(2)-$Higgs theory has
crucial cosmological and experimental consequences~\cite{Kuzmin:1985mm,Matveev:1988pj,
  Rubakov:1996vz, Fodor:1998if}. Over the years, several non-perturbative lattice
studies have been devoted in this regard~\cite{Damgaard:1986qe,Evertz:1986af,Bunk:1992xt,
  Farakos:1994xh, Kajantie:1995kf,Rummukainen:1998as,Bunk:1992kf,Bunk:1994xs,
  Kajantie:1996mn, Gurtler:1997hr,Gurtler:1997ki, Buchmuller:1994qy,Karsch:1996aw,
  Karsch:1996yh,Fodor:1994dm,Fodor:1994sj, Csikor:1995jj,Aoki:1996cu, Aoki:1999fi,
  Bonati:2009pf, DOnofrio:2014rug,DOnofrio:2015gop,DOnofrio:2012yxq,Laine:2012jy,
  Gould:2022ran} (For reviews, please
see~\cite{Kajantie:1994mt, Jansen:1995yg, Rummukainen:1996sx}).
These studies find that the nature of Higgs transition varies with Higgs quartic
coupling ($\lambda$). With increase in $\lambda$, the strength of this
transition weakens and eventually  becomes a crossover for large $\lambda$. The
$1$st order transition and crossover regions are separated by a $2$nd order critical
end-point at $\lambda_c$~\cite{Aoki:1999fi}. There are similar examples in QCD,
spin-models, condensed matter systems etc. where lines/surfaces of
second order transition that separate $1$st order and crossover regions in the
phase diagram~\cite{Pisarski:1983ms, Karsch:2000xv,Alonso:1993tv}. Many attempts
have been made to study the universality class of such critical end-points in
different theories~\cite{Pisarski:1983ms,Alonso:1993tv, Rummukainen:1998as,
  Karsch:2000xv}. These end-points are expected to belong to the universality class
of the Ising model~\cite{Alonso:1993tv, Rummukainen:1998as, Janke:1996qb,Karsch:2000xv,
  Wilding:1994zkq,Wilding:1994ud,Wilding:1996, Rehr:1973zz}. In the electroweak
theory, the critical end-point in dimensionally reduced $SU(2)-$Higgs theory is studied
in three dimensions~\cite{Rummukainen:1998as}. The analysis of the
critical behavior is done by identifying energy and magnetization like
observables~\cite{Alonso:1993tv}. The distribution of
these observables, higher cumulants and the extracted critical exponents, clearly 
show that the end-point is in the $3$d Ising universality class.  

\smallskip

In ref.~\cite{Rummukainen:1998as}, a linear combinations of various gauge invariant
observables plays the role of magnetization. Another orthogonal linear combination
plays the role of energy like observable. These linear combinations change with the
bare parameters of the theory. Though gauge invariant observables can be used to study
nature of transitions, it is desirable to have an order parameter for the Higgs
transition. In the absence of gauge symmetry, the Higgs condensate (volume average of 
the Higgs field) plays the role of an order parameter. On other hand, gauge symmetry
renders the condensate unphysical. However, it can be made well behaved by a suitable
choice of a gauge. It is expected that the choice of a gauge will not affect physical
quantities such as critical exponents. Note that the necessity of a suitable order parameter
exists in other gauge theories, such as in high density QCD to describe the normal to
color superconducting transition.

\smallskip

In the present work, we propose that the Higgs condensate in the Landau gauge,
  denoted by $\Phi^g$, as an order parameter for the Higgs transition in $SU(2)-$Higgs
theory.\ In $3+1$ dimensions, $\Phi^g$ and its cumulants are computed using Monte
Carlo simulation of the partition function by varying the Higgs hoping parameter
($\kappa$) and the quartic coupling. It is observed that the partition
function average of $\Phi^g$ behaves similar to that of magnetization in spin models.
For small $\lambda$, the Higgs transition is of $1$st order, and $\Phi^g$ is found to
vary discontinuously across the transition point. With the increase in $\lambda$, the
Higgs transition weakens which is also seen in $\Phi^g$ behavior. The jump in $\Phi^g$
across the transition point decreases with $\lambda$. At a critical point
$\lambda_c$, $\Phi^g$ varies continuously while its various cumulants 
show singular behavior. The
observed scaling behavior is consistent
with $3$d Ising universality class and deviates from the universality class of the $3$d
$O(4)$ spin models~\cite{Kanaya:1994qe,Toussaint:1996qr}. The Finite Size Scaling (FSS)
of $\Phi^g$ suggests that it vanishes at the critical point in the infinite volume
limit. It also vanishes in the high temperature or Higgs symmetric phase at least for
$\lambda < \lambda_c$. We would like to mention here that $\Phi^g$, as an order
parameter, reasonably reproduces the previous results. A linear combination along the
lines of previous studies may further fine tune the results.

\smallskip

The draft is organized as follows. In Sec.~\ref{sec:lattice.form}, we provide the lattice
definition of the $SU(2)$-Higgs theory. In Sec.~\ref{sec:numerical_param}, we discuss the
numerical procedures for our work. This includes the Landau Gauge Fixing in Lattice.
In Sec.~\ref{sec:fss}, we analyze our data using FSS, while we present the
magnetization study at various $\lambda$'s in Sec.~\ref{sec:mag_at_var_lambda}. Finally,
we present the discussion and conclusion of our study in Sec.~\ref{sec:discussion_conclusion}.

%%%%%%%%%%%%%%%%%%%%%%%%%%%%%%%%%%%%%%%%%%%%
%%%%%%%%%%%%%%%%%%%%%%%%%%%%%%%%%%%%%%%%%%%%

\section{Lattice Formalism}
\label{sec:lattice.form}
%%%%%%%%%%%%%%%%%%%%%%%%%%%%%%%%%%%%%%%%%%%%
%%%%%%%%%%%%%%%%%%%%%%%%%%%%%%%%%%%%%%%%%%%%

The continuum action of $SU(2)$ gauge-Higgs theory is discretized in Lattice as,
\eqarray{
  \displaystyle S 
  &=& \sum_x \left[ \sum_{\mu > \nu} \frac{\beta}{2}\, \mbox{Tr}\, U_{x,\mu\nu} 
    + \sum_\mu \kappa\, \mbox{Tr}\, \left(\phi^\dag_x\, U_{x,\mu}\, \phi_{x + {\hat\mu}}\right)
    - \frac{1}{2}\, \mbox{Tr} \left(\phi^\dag_x \phi_x\right) 
    - \lambda \left(\frac{1}{2}\, \mbox{Tr} \left(\phi^\dag_x \phi_x\right) - 1\right)^2
    \right]\, , \nonumber\\
  \label{eq:lattice_action}
}
where $U_{x,\mu} \in SU(2)$ is a link variable,\ and $U_{x,\mu\nu}$ is a product of four link 
variables which form a plaquette,\ and $\phi_x$ is $2\otimes 2$ matrix in isospin 
space describing the Higgs scalar field.\ The bare parameters $\beta_g \equiv4/g^2$ is the 
gauge coupling, $\lambda$ is the scalar quartic coupling,\ and $\kappa$ is the scalar
hopping parameter which is related to the bare mass square by the relation,\  
$\mu^2_0 = (1 - 2\lambda)\, \kappa^{-1} - 8$. 

\smallskip

In order to search for an order parameter,\ we apply Landau gauge  fixing,\
$\partial_\mu \, A_\mu \! =\! 0$,\ and construct the condensate, $\Phi^g$. We then study
the behavior of the magnetization, susceptibility and fourth order Binder cumulant of
$\Phi^g$ (defined below) at various values of $\kappa$ and $\lambda$ while
$\beta_g\! = \! 8$ is kept fixed.
\eqarray{
  \label{eqn:magtn}
  \mbox{Magnetization} \hspace{5mm} &:&  \hspace{5mm} \Phi^g
  \, \equiv \, \langle \tilde\Phi^g \rangle
  \, = \, \left\langle \frac{1}{N_s^3 N_t}
  \sqrt{\sum_{i = 1}^4 \left(\displaystyle\sum_x \phi^g_{i,x}\right)^2} \right\rangle, \\
  \label{eqn:suscept}
  \mbox{Susceptibility,}\,\, \chi^c \hspace{5mm} &:&  \hspace{5mm}
  V\, \left(\langle {\tilde\Phi^g}{}^2\rangle - \langle \tilde\Phi^g \rangle^2\right) ,\\
  \label{eqn:tot.suscept}
  \mbox{Total Susceptibility,}\,\,\chi \hspace{5mm} &:&  \hspace{5mm}
  V\, \langle {\tilde\Phi^g}{}^2 \rangle , \\
  \label{eqn:binder.cum}
  \mbox{Binder Cumulant,}\, \, B_{4} \hspace{5mm} &:&  \hspace{5mm}
  1 - \frac{\langle {\tilde\Phi^g}{}^4 \rangle}{3\,\langle {\tilde\Phi^g}{}^2 \rangle^2} .
}
where $\langle\cdots\rangle$ stands for gauge average, and $V = N_s^3$ is the spatial
volume of the lattice.\ We use the superscript,\ $c$,\ in Eq.~(\ref{eqn:suscept}) to
specify it as the Connected Susceptibility.

\smallskip

We devote a significant part of our study near the end-point for $1$st order transition,
namely at $\lambda = 0.00116, 0.00132$ and $0.00150$.\ Note that the first value of $\lambda$
is found to be the critical end-point in Ref.~\cite{Aoki:1999fi}. The two higher $\lambda$
values lie around $\sim +2\sigma$ of the first one.\ The reason to include these higher values
of $\lambda$ is discussed in the Sec.~\ref{subsec:lambda_1160}.\ We list the run parameters in
Table~\ref{tab:run_para_1}.

%%%%%%%%%%%%%%%%%%%%%%%%%%%%%%%%%%%%%%%%%%%%
\section{Numerical Procedure and Parameters}
\label{sec:numerical_param}
%%%%%%%%%%%%%%%%%%%%%%%%%%%%%%%%%%%%%%%%%%%%

For Monte Carlo simulations, we use the pure gauge part of the publicly available
{\tt MILC} code~\cite{milc} and modify it to accommodate the Higgs fields.\ To update
the gauge fields,\ we first use the standard heat bath algorithm~\cite{Creutz:1980zw,
  Cabibbo:1982zn},\ and then update Higgs fields using pseudo heat bath
algorithm~\cite{Bunk:1994xs}.\ We then again update the gauge fields using $4$ 
overrelaxation steps~\cite{Whitmer:1984he} after which Higgs fields are updated again
using pseudo heat bath algorithm.\ To reduce auto-correlation between successive
configurations, we carry out $50$ cycles of this updating procedure between two
subsequent measurements. We test our codes to reproduce histogram profiles similar to
Ref.~\cite{Aoki:1996cu}. For our simulations, we use four spatial volumes, $N_s \! =\! 20,
24, 32, 40$ while the temporal lattice extent is kept at $N_\tau = 2$~\cite{Aoki:1999fi}.

\smallskip

To implement Landau gauge fixing,\ we maximize the following quantity
\eqarray{
  H &=& \displaystyle \sum_x \sum^4_{\mu = 1} \mbox{Re}\, \left[\mbox{Tr}\, U_{x,\mu}\right]\, .
  \label{eq:gauge-fix_1}
}
In order to maximize $H$,\ we separately apply two different methods as described 
in~\cite{Aoki:2003ip}.\ One of the methods is the standard $SU(2)$ subgroup method,\ and 
the other one is Overrelaxed Steepest Descent method.\ This was done to check whether both 
the methods arrive at the same global maxima.\ In addition,\ we use single,\ double and 
quadruple precisions in our simulations so that we are able to set our convergence
conditions accordingly.\ We find that both the methods for maximization of $H$ as well as
the different precisions produce similar results.\ Eventually, as described
in~\cite{Aoki:2003ip},\ we combine both $SU(2)$ subgroup and Overrelaxed Steepest Descent
methods to speed up gauge fixing.\ First,\ we apply $SU(2)$ subgroup method to bring the
configuration close to the maximum,\ and then apply Overrelaxed Steepest Descent method
for the final convergence.\ We set our overrelaxation parameter,\ $\omega = 1.98$,\ and
we use double precision for all our simulations.\ For convergence of $SU(2)$ subgroup
method,\ we measure
\eqarray{
  h &\sim& \frac{H}{N^3_s\, N_\tau} ,
  \label{eq:gauge-fix_2}
}
We set the convergence condition at the $i$th iteration as,
\eqarray{
  \left|h_i - h_{i-1}\right| &<& 10^{-12} \, .
  \label{eq:gauge-fix_3}
}
For the Steepest Descent method,\ we measure
\eqarray{
  \Delta &=& \displaystyle \frac{1}{N^3_s\, N_\tau} \sum_x \frac{1}{2}\sum^4_{\mu =1}
  \frac{1}{4}\, \mbox{Tr}\, \left[\Delta_{x, \mu} \Delta^\dag_{x, \mu}\right] ,
  \label{eq:gauge-fix_4}
}
where
\eqarray{
  \Delta_{x, \mu} &=& U_{x - \hat\mu, \mu} - U_{x,\mu} - \mbox{h.c.} - \mbox{trace}\, .
  \label{eq:gauge-fix_5}
}
We set the convergence condition for $\Delta_j$ at the $j$th iteration as,
\eqarray{
  \left|\Delta_j\right| &<& 10^{-15} . 
  \label{eq:gauge-fix_6}
}
Simulations are performed for a wide range of $\kappa$ and $\lambda$ values. We focus
our studies at $\lambda \! = 0.00116, 0.00132$ and 
$0.00150$. For each of them, we generate $2.5$ million trajectories out of which gauge
fixing is performed on every $50$-th trajectory at each $\kappa$. This gives us a total
$50\, 000$ gauge-fixed configurations for each $\kappa$. The  parameters are listed 
in Table~\ref{tab:run_para_1}. For each of other values of $\lambda$, we generate
$500\, 000$ trajectories instead out of which every $50$-th trajectory is selected for
gauge fixing giving $10\, 000$ gauge-fixed configurations for each $\kappa$. The reduced
statistics is due to the limitations in computational resources. The parameters are listed
in Table~\ref{tab:run_para_2}. All the error analyses in this study are carried out using
Jackknife method with bin size determined such a way that the total number of bins is
$100$.
\begin{table}[htbp]
  \centering
  \caption{Simulation parameters near $\lambda_c$.}
  \setlength{\tabcolsep}{9pt}
  \small
  \begin{tabular}{ccccccc}
    \hline\hline
    \multirow{4}{*}{\boldmath$\beta_g$}
    &\multirow{4}{*}{\boldmath$\lambda$}
    & \multirow{4}{*}{\boldmath$N_s$}
    & \multirow{4}{*}{\boldmath$N_\tau$}
    & \multirow{4}{*}{\boldmath$\kappa$}
    & \multirow{4}{*}{\bf Confs.}
    & {\bf Trajectories} \\
    &&&&&& {\bf between} \\
    &&&&&& {\bf consecutive} \\
    &&&&&& {\bf confs.} \\
    \hline
    \multirow{3}{*}{8.0}
    & 0.00116
    & 20, 24, 32, 40  &  2 & \{0.129390, 0.129550\} & 50 000 & 50 \\
    %\hline
    & 0.00132
    & 20, 24, 32, 40  &  2 & \{0.129420, 0.129820\} & 50 000 & 50 \\
    %\hline
    & 0.00150
    & 20, 24, 32, 40  &  2 & \{0.129500, 0.130000\} & 50 000 & 50 \\
    \hline\hline
  \end{tabular}
  \normalsize
  \label{tab:run_para_1}
\end{table}
%
%\newpage

%%%%%%%%%%%%%%%%%%%%%%%%%%%%%%%%%%%%%%%%%%%%
%\section{Determination of \lowercase{{\boldmath{$\kappa_{c}$}}} from 
%\lowercase{\boldmath{${\kappa_\chi}_{\rmsmall{max}}$}}}
\section{\boldmath$M$-like and \boldmath$E$-like observables}
\label{sec:MEobservables}
%%%%%%%%%%%%%%%%%%%%%%%%%%%%%%%%%%%%%%%%%%%%

The $SU(2)-$Higgs theory is similar to the case of the liquid-gas system, i.e none of the
gauge invariant observables can be directly identified with the $M$-like and $E$-like
observables. In the previous studies of the critical endpoint, observables such as 
$S_K = \displaystyle\sum_{x,\mu}\mbox{Tr} \left(\phi^\dag_x\, U_{x,\mu}\, \phi_{x + {\hat\mu}}\right)$, 
$S_\phi = \displaystyle\sum_x \left(\frac{1}{2}\,\mbox{Tr}\
\left(\phi^\dag_x \phi_x\right) - 1\right)^2$ etc. have been used to define the $M$-like and
$E$-like observables. It has been pointed out that $S_K$, $S_\phi$ etc. are not necessarily
orthogonal in the vicinity of the critical point, i.e.
$\langle\Delta S_K \Delta S_\phi\rangle \neq 0$ where
$\Delta S_K =  S_K - \langle S_K\rangle$ and
$\Delta S_\phi = S_\phi - \langle S_\phi\rangle$~\cite{Rummukainen:1998as,Alonso:1993tv,
  Karsch:2000xv, Wilding:1994zkq, Wilding:1994ud, Wilding:1996}. However, linear and orthogonal
combinations of them have been found to behave as $M$-like and $E$-like observables. In the
present work, we include the Landau gauge fixed Higgs condensate, $\Phi^g$, along with other
gauge invariant observables to define $M$-like and $E$-like observables. The studies of
correlations between $\Phi^g$ and other gauge invariant observables seem to suggest that
$\langle\Delta\Phi^g\Delta S_K \rangle\approx 0$
and $\langle\Delta\Phi^g\Delta S_\phi\rangle \approx 0$, are satisfied within the
uncertainties.

\smallskip

\begin{figure}[htbp]
  \centering
  \subfigure[]{%
    %    \subfigtopskip 5pt
    \centering
    \rotatebox{270}{\includegraphics[width=0.24\hsize]
      {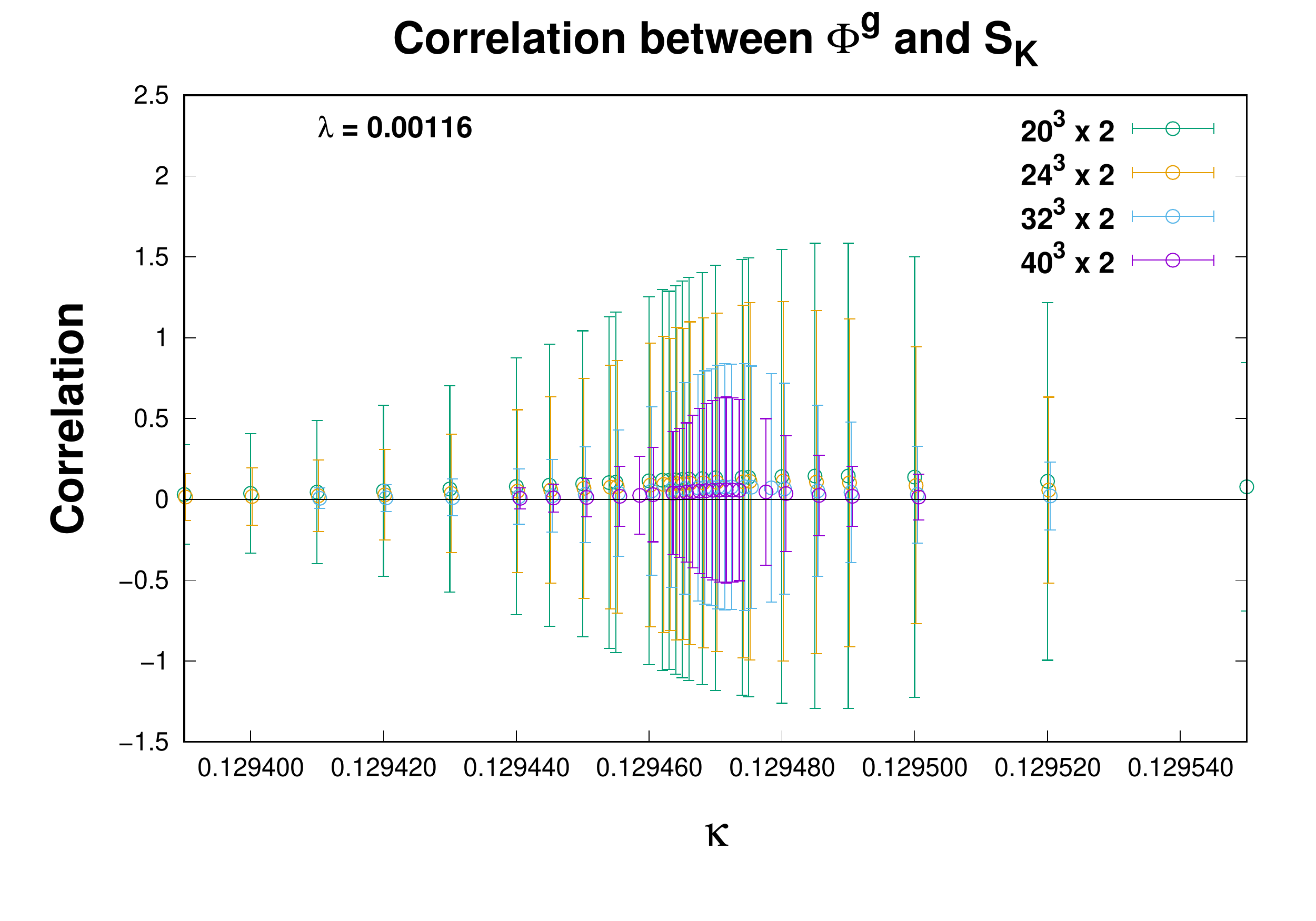}}
    \label{fig:orig_magtn_1160}
  }
  \subfigure[]{%
    \centering
    \rotatebox{270}{\includegraphics[width=0.24\hsize]
      {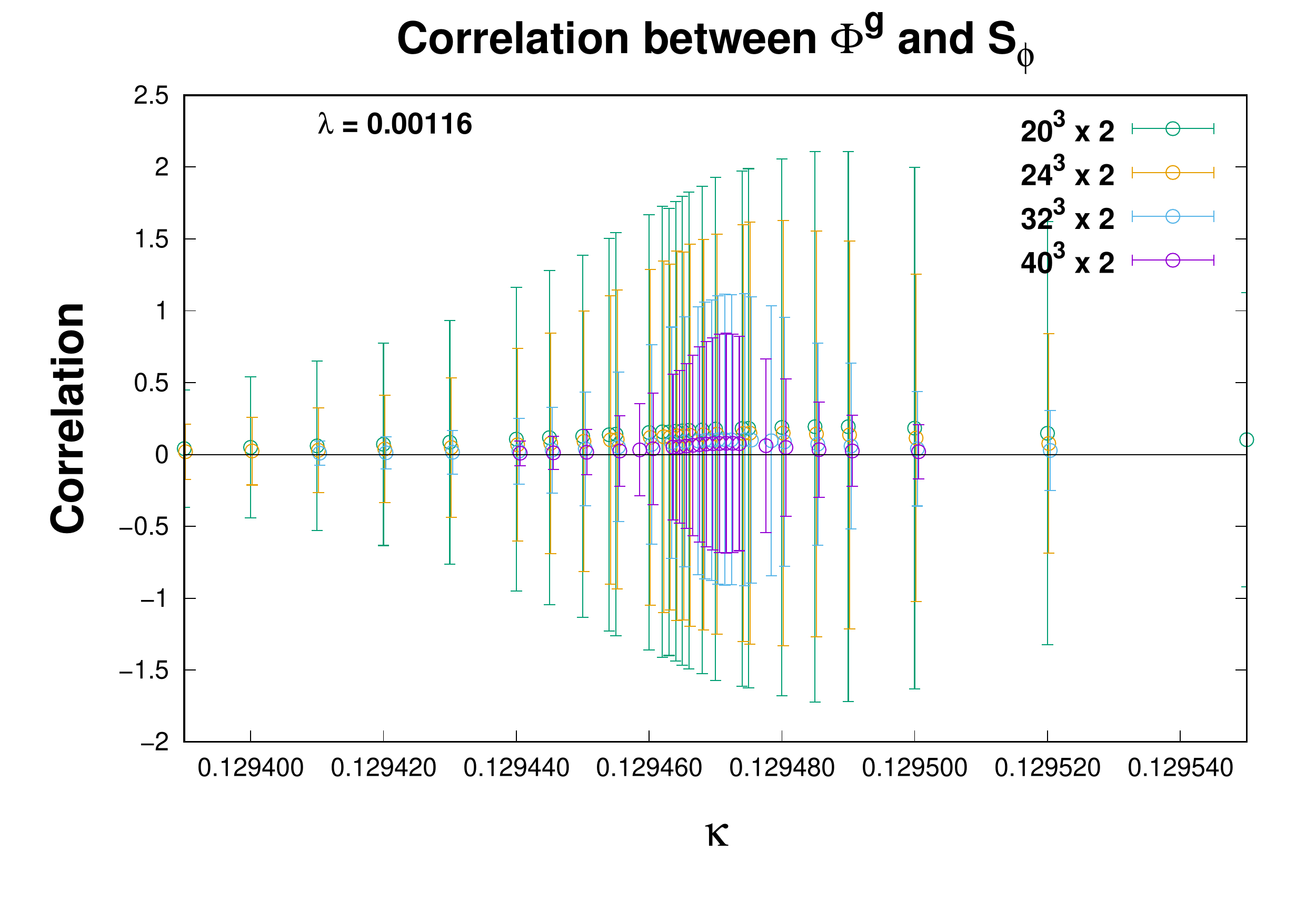}}
    \label{fig:orig_suscept_1160}
  }
  \caption{Correlations between
    (a) $\Phi^g$ and $S_K$
    (b) $\Phi^g$ and $S_\phi$ % $S_{\left(\phi^2 - 1\right)^2}$
    at $\lambda = 0.00116$.}
  \label{fig:corr_1160}
\end{figure}

\begin{figure}[htbp]
  \centering
  \subfigure[]{%
    %    \subfigtopskip 5pt
    \centering
    \rotatebox{270}{\includegraphics[width=0.24\hsize]
      {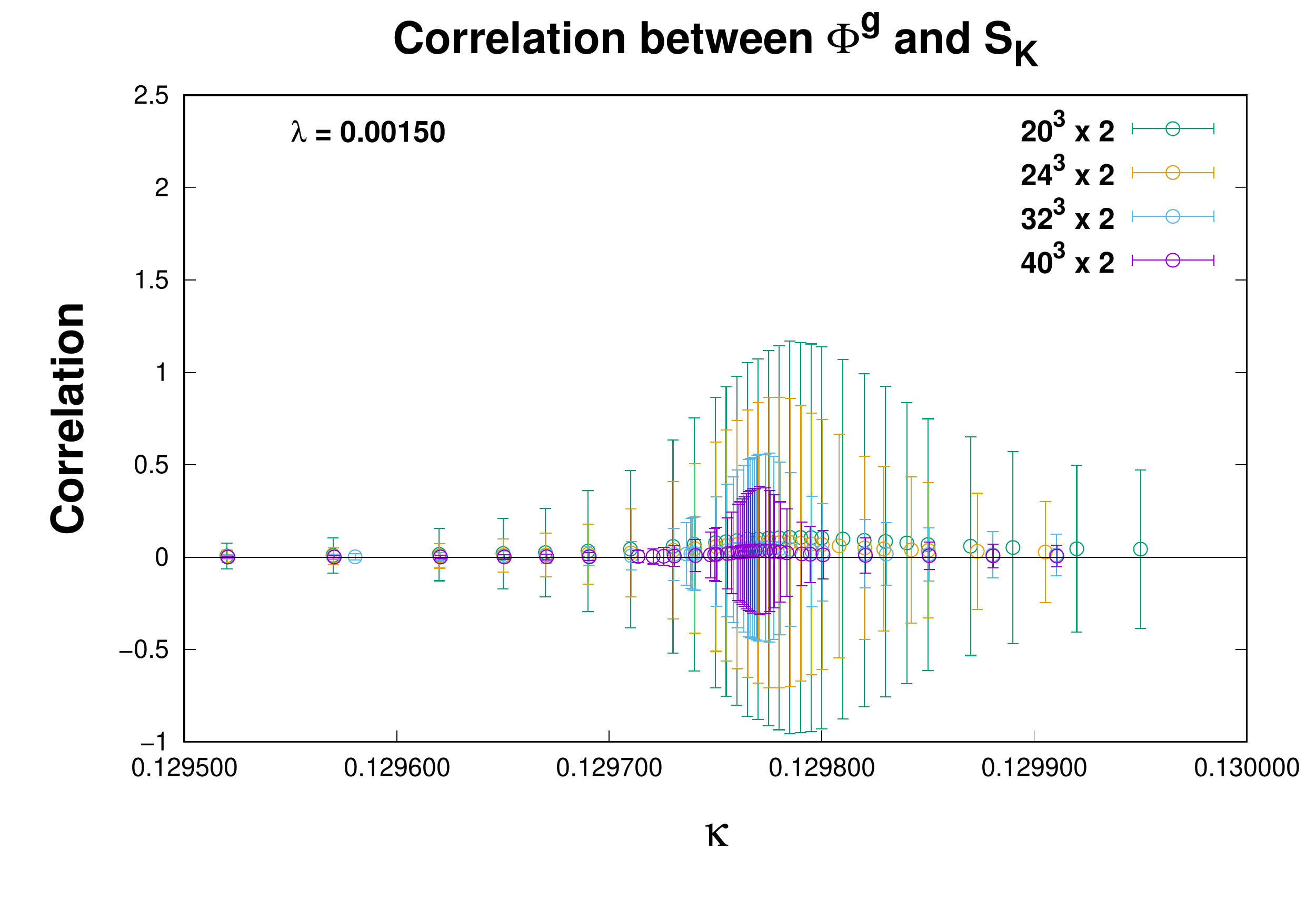}}
    \label{fig:orig_magtn_1500}
  }
  \subfigure[]{%
    \centering
    \rotatebox{270}{\includegraphics[width=0.24\hsize]
      {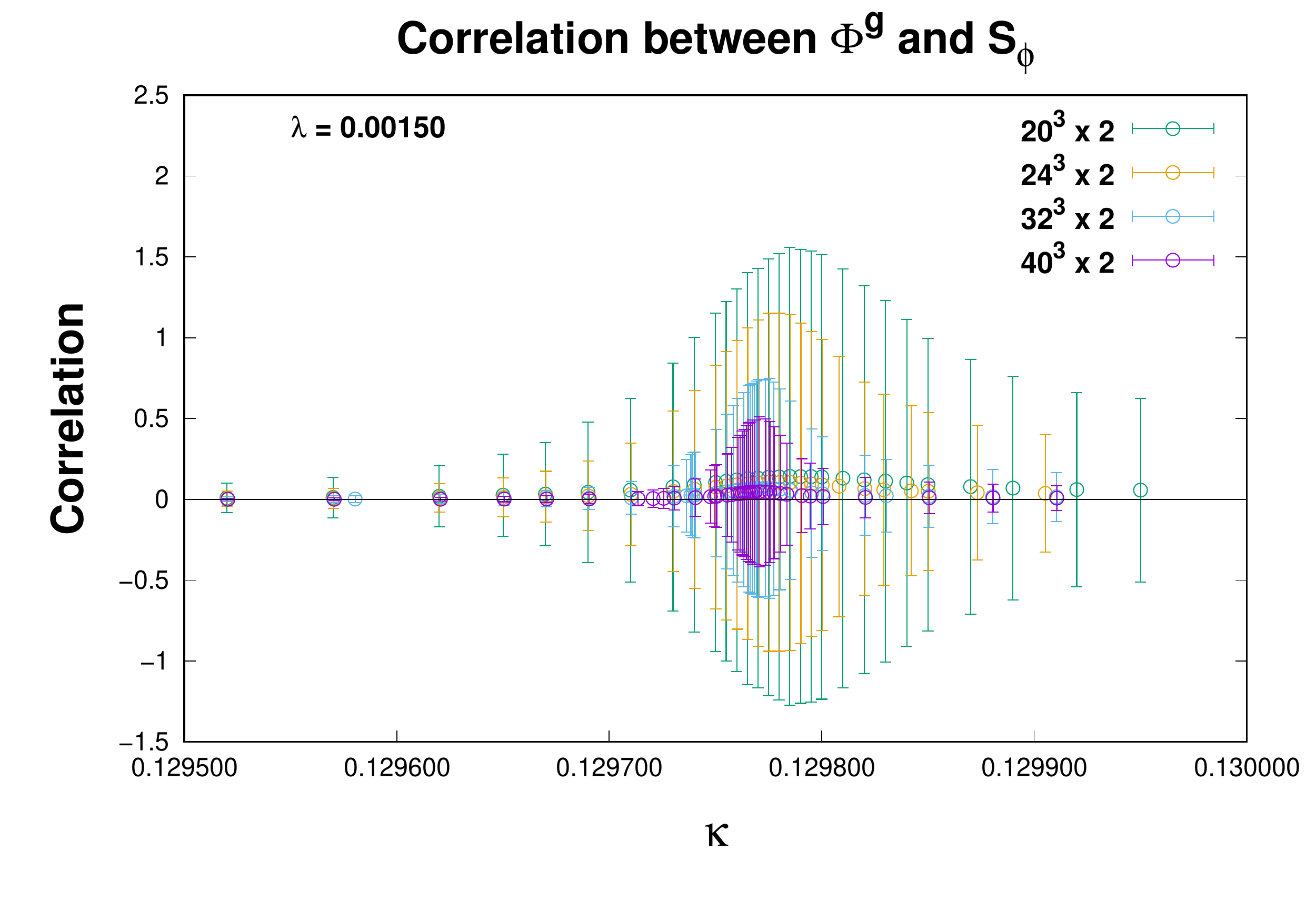}}
    \label{fig:orig_suscept_1500}
  }
  \caption{Correlations between
    (a) $\Phi^g$ and $S_K$
    (b) $\Phi^g$ and $S_\phi$ %$S_{\left(\phi^2 - 1\right)^2}$
    at $\lambda = 0.00150$.}
  \label{fig:corr_1500}
\end{figure}

We compute the correlation relations, $\langle\Delta\Phi^g\Delta S_K \rangle$ and
$\langle\Delta\Phi^g\Delta S_\phi\rangle$, over the whole range of respective
$\kappa$ values at $\lambda = 0.00116$ and $0.00150$. They are plotted in
Figs.~\ref{fig:corr_1160}~and~\ref{fig:corr_1500}, respectively. We clearly see that
there is no correlation between the $\Phi^g$ and other observables. We also observe that
the  sizes of uncertainties decrease as the volume increases. This suggests that
$\Phi^g$ is a $M$-like observable.

\smallskip

\begin{figure}[htbp]
  \centering
  \subfigure[]{%
    %    \subfigtopskip 5pt
    \centering
    \rotatebox{270}{\includegraphics[width=0.24\hsize]
      {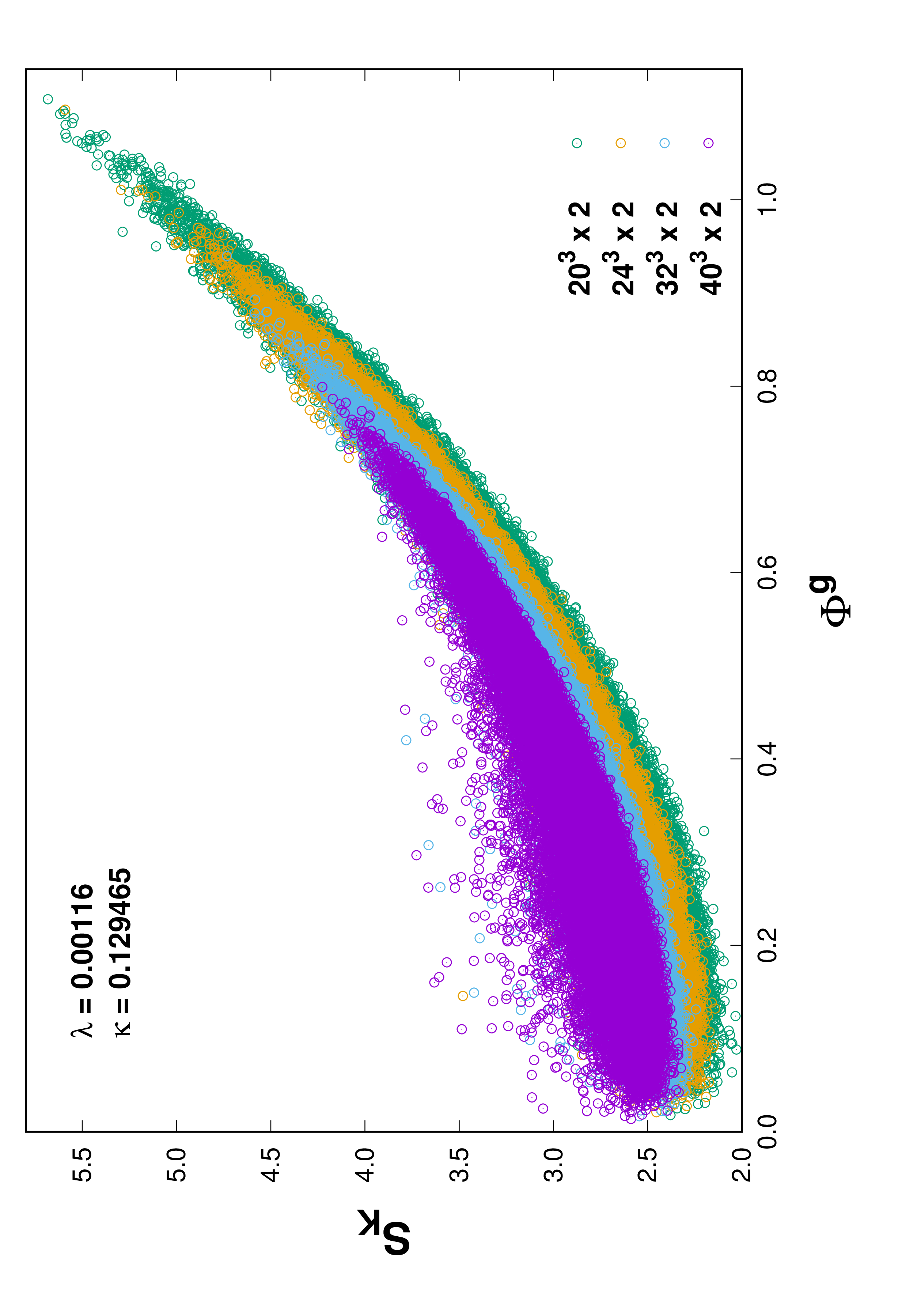}}
    \label{fig:phig_vs_pup_k_129465_1160}
  }
  \subfigure[]{%
    %    \subfigtopskip 5pt
    \centering
    \rotatebox{270}{\includegraphics[width=0.24\hsize]
      {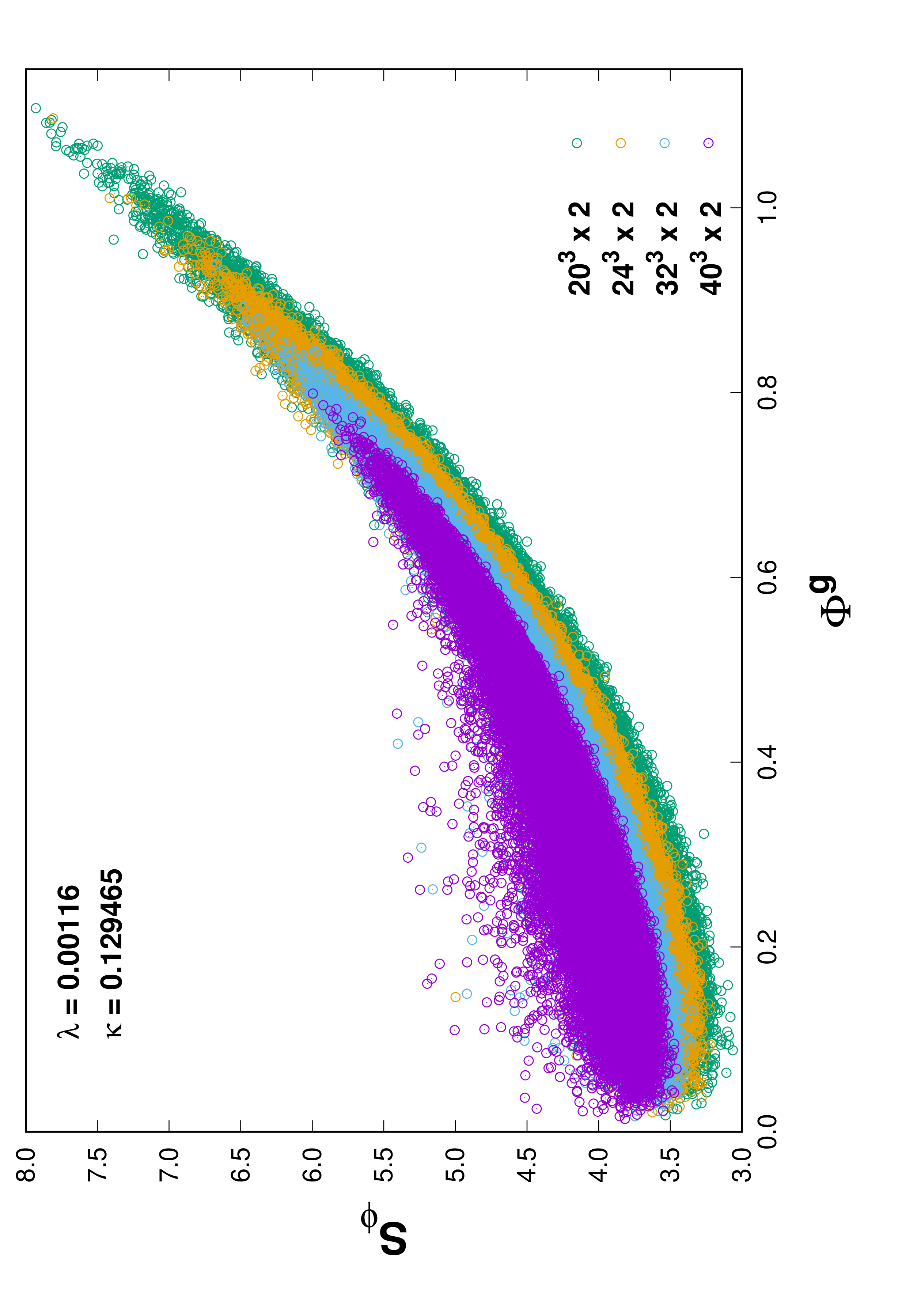}}
    \label{fig:phig_vs_ph4_k_129465_1160}
  }
  \caption{
    $\Phi^g$ vs.
    (a) $S_K$
    and
    (b) $S_\phi$ % $S_{(\phi^2 - 1)^2}$
    at $\lambda = 0.001160$ and $\kappa = 0.129465$.}
  \label{fig:phig_vs_pup_ph4_1160}
\end{figure}

\begin{figure}[htbp]
  \centering 
  \subfigure[]{%
    \centering
    \rotatebox{270}{\includegraphics[width=0.24\hsize]
      {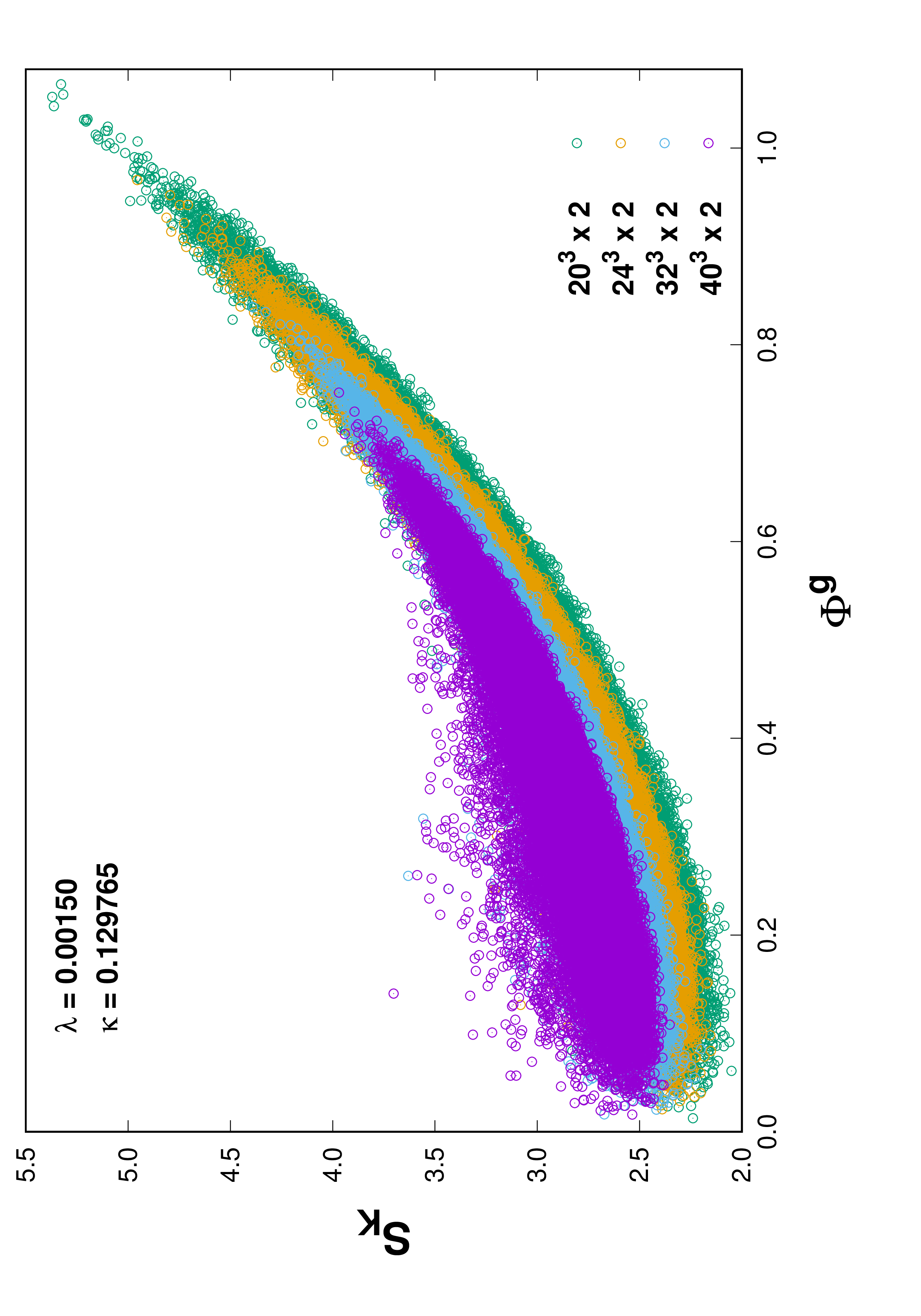}}
    \label{fig:phig_vs_pup_k_129765_1500}
  }
  \subfigure[]{%
    %    \subfigtopskip 5pt
    \centering
    \rotatebox{270}{\includegraphics[width=0.24\hsize]
      {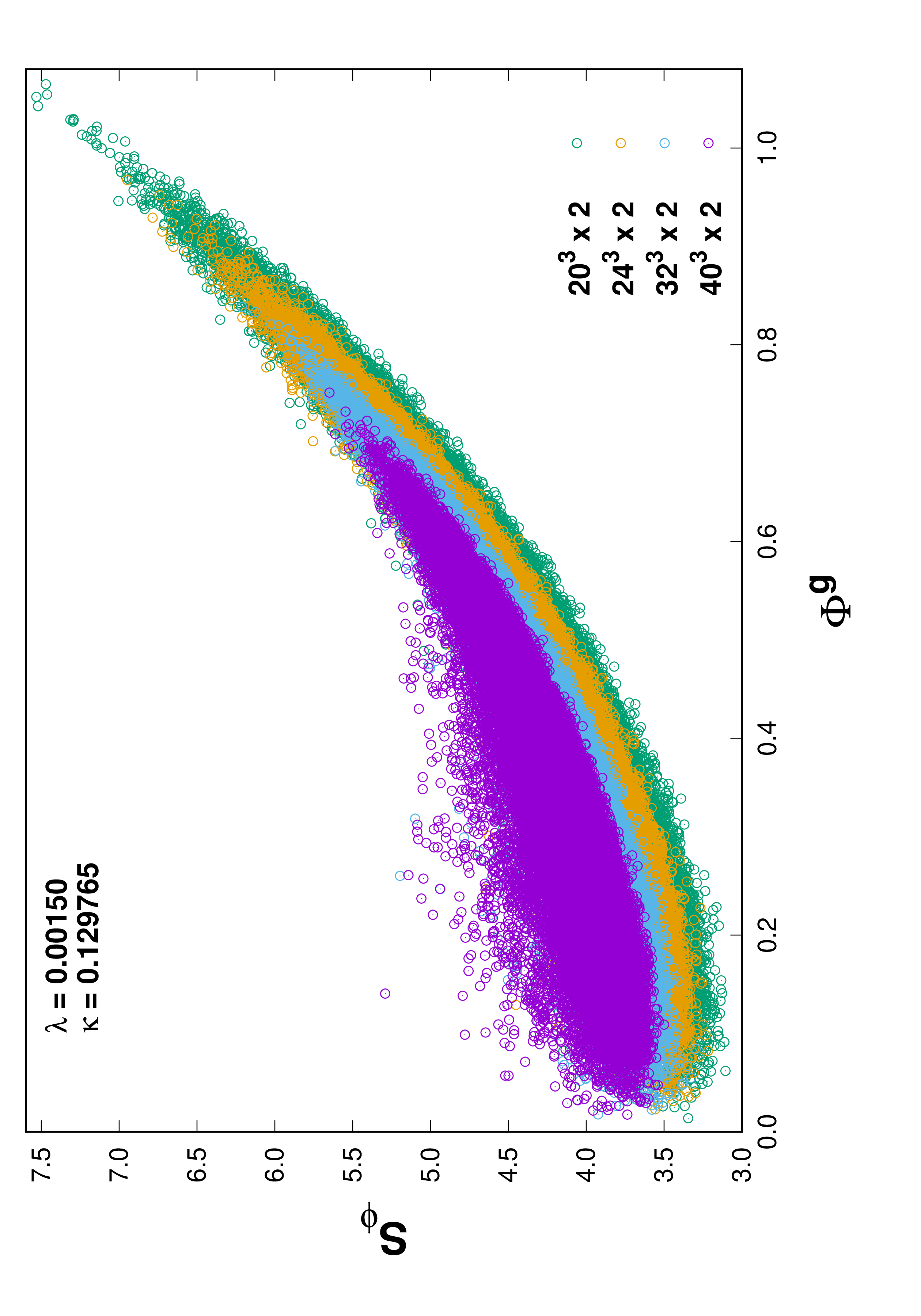}}
    \label{fig:phig_vs_ph4_k_129765_1500}
  }
  \caption{
    $\Phi^g$ vs.
    (a) $S_K$
    and
    (b) $S_\phi$ % $S_{(\phi^2 - 1)^2}$ 
    at $\lambda = 0.001500$ and $\kappa = 0.129765$.}
  \label{fig:phig_vs_pup_ph4_1500}
\end{figure}

We also study the behaviour of the $\Phi^g$ distributions vs. $S_K$ as well as
$S_\phi$ distributions. In Fig.~\ref{fig:phig_vs_pup_ph4_1160}, we plot $\Phi^g$ vs. $S_K$
and $S_\phi$ for $\lambda = 0.00116$ and $\kappa = 0.129465$. In
Fig.~\ref{fig:phig_vs_pup_ph4_1500}, the same quantities are plotted for
$\lambda = 0.001500$ and $\kappa = 0.129765$. From these figures, we see that these
distributions behave similarly to those of energy vs. magnetization distributions of
the $3$d Ising Model. Note that $\Phi^g$ has four components. The individual components
have no physical significance, as there are no Goldstone modes in this case. The
magnitude of $\Phi^g$ is physical and as a consequence, we can only reproduce the part of
the Ising model figure for which $M > 0$. In
Figs.~\ref{fig:phig_vs_pup_ph4_1160}~and~\ref{fig:phig_vs_pup_ph4_1500}, we observe that
the distributions are not linear, unlike the distribution of $S_k$ vs. $S_\phi$. In this
situation, no unique linear combination is possible.

%%%%%%%%%%%%%%%%%%%%%%%%%%%%%%%%%%%%%%%%%%%%
%\section{Determination of \lowercase{{\boldmath{$\kappa_{c}$}}} from 
%\lowercase{\boldmath{${\kappa_\chi}_{\rmsmall{max}}$}}}
\section{Finite Size Scaling Analysis}
\label{sec:fss}
%%%%%%%%%%%%%%%%%%%%%%%%%%%%%%%%%%%%%%%%%%%%
%
%
\begin{figure}[htbp]
  \centering
  \subfigure[]{%
    %    \subfigtopskip 5pt
    \centering
    \rotatebox{270}{\includegraphics[width=0.26\hsize]
      {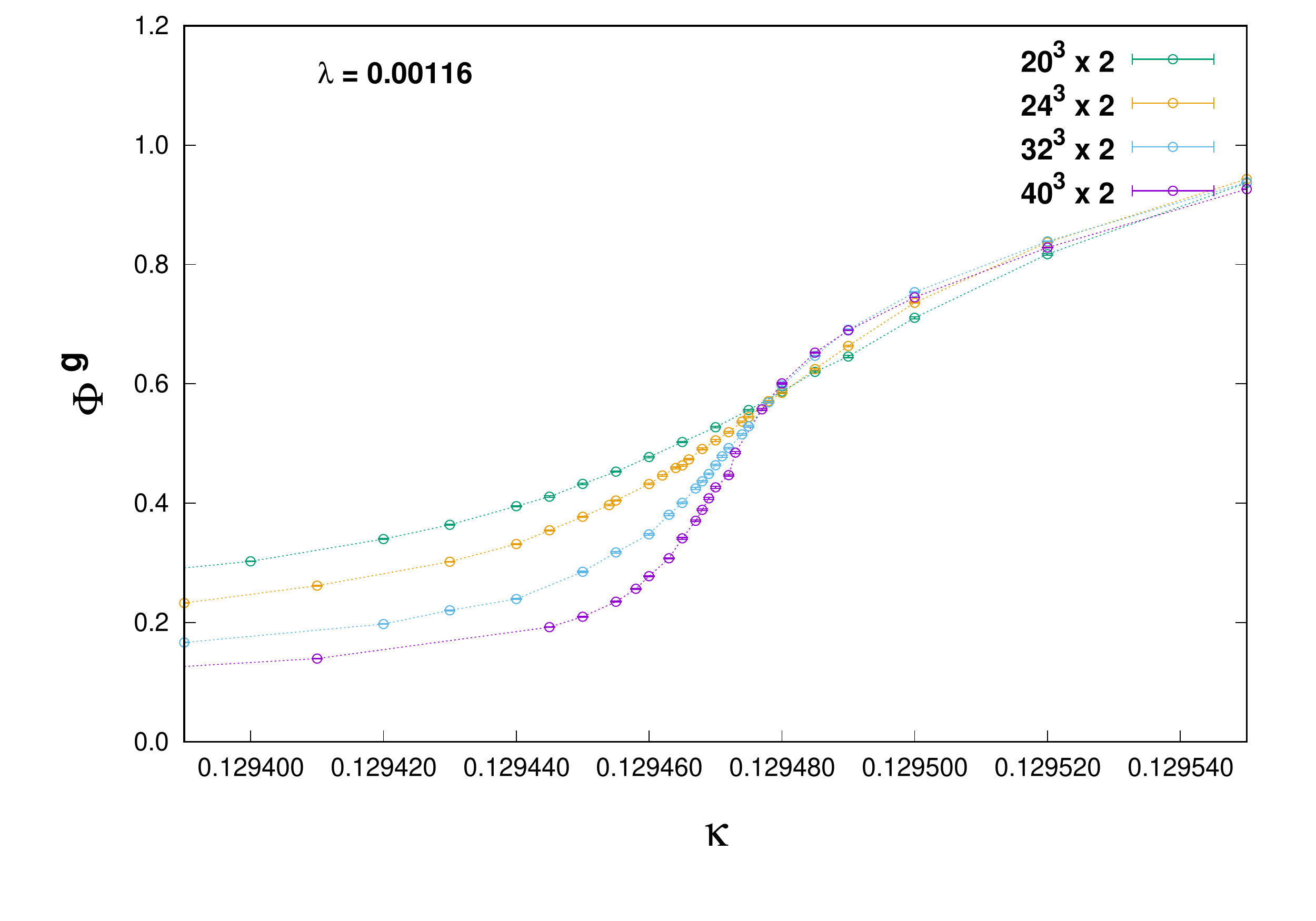}}
    \label{fig:orig_magtn_1160}
  }
  \subfigure[]{%
    \centering
    \rotatebox{270}{\includegraphics[width=0.26\hsize]
      {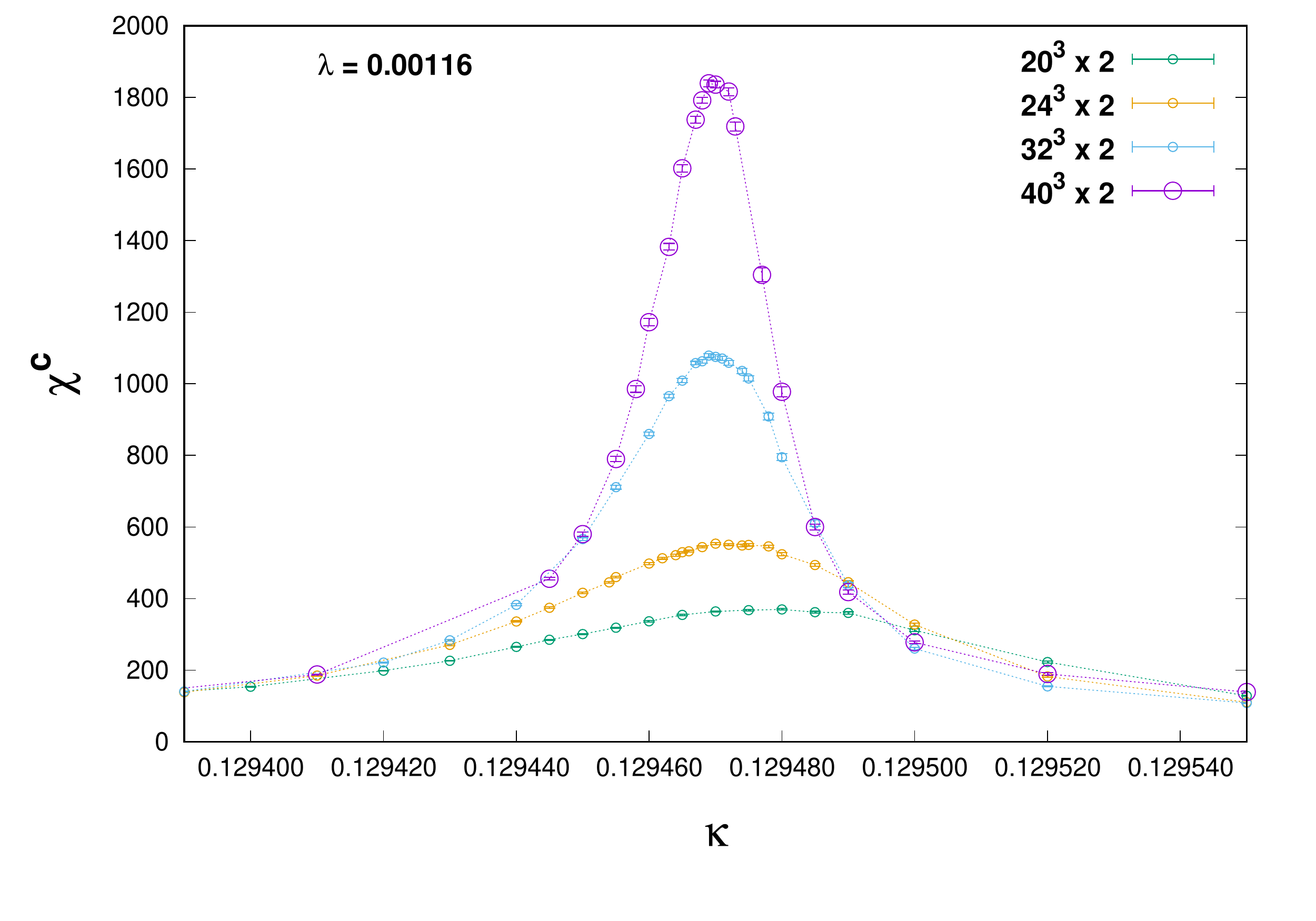}}
    \label{fig:orig_suscept_1160}
  }
  \subfigure[]{%
    %    \subfigtopskip 5pt
    \centering
    \rotatebox{270}{\includegraphics[width=0.26\hsize]
      {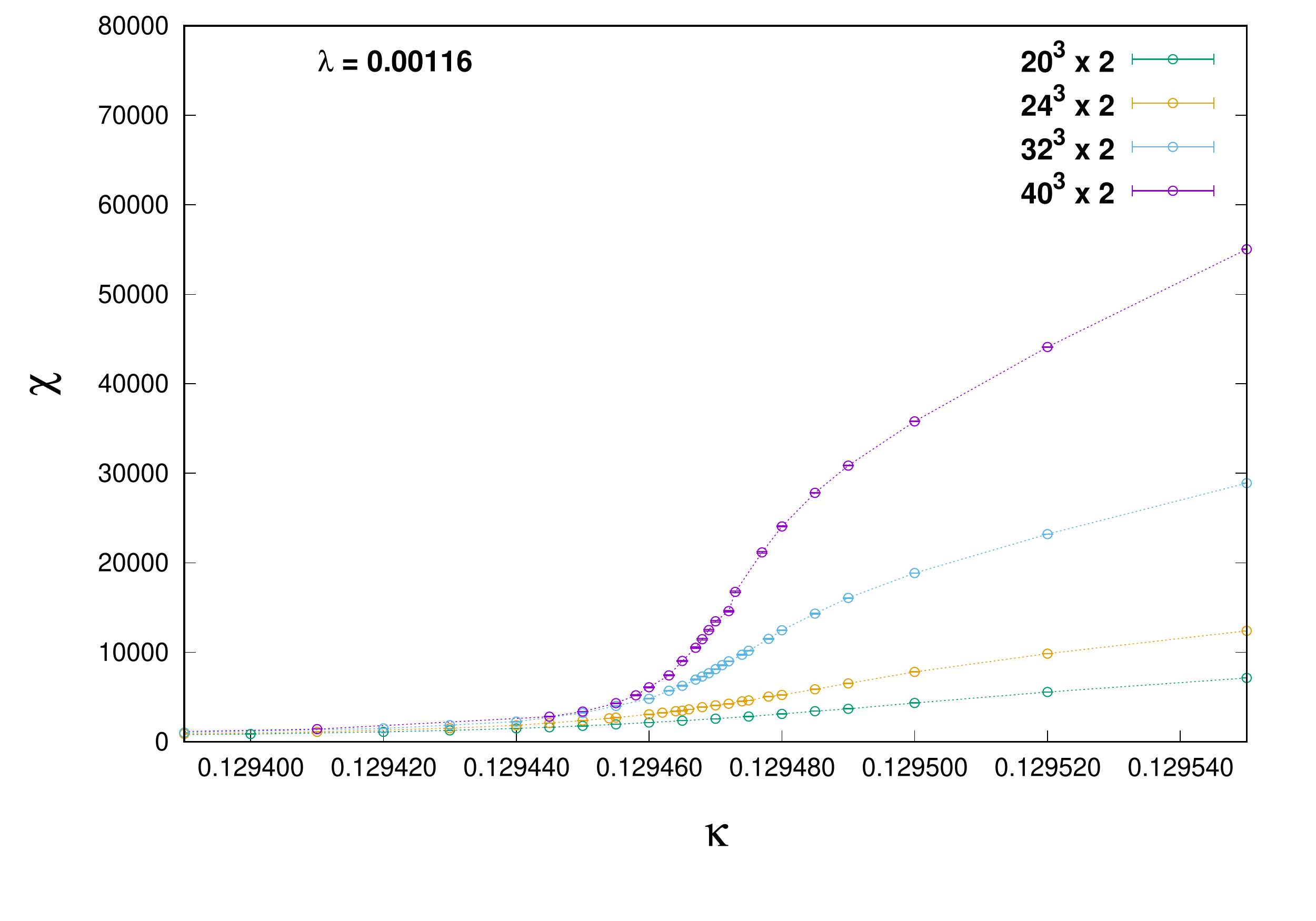}}
    \label{fig:orig_tot_suscept_1160}
  }
  \subfigure[]{%
    \centering
    \rotatebox{270}{\includegraphics[width=0.26\hsize]
      {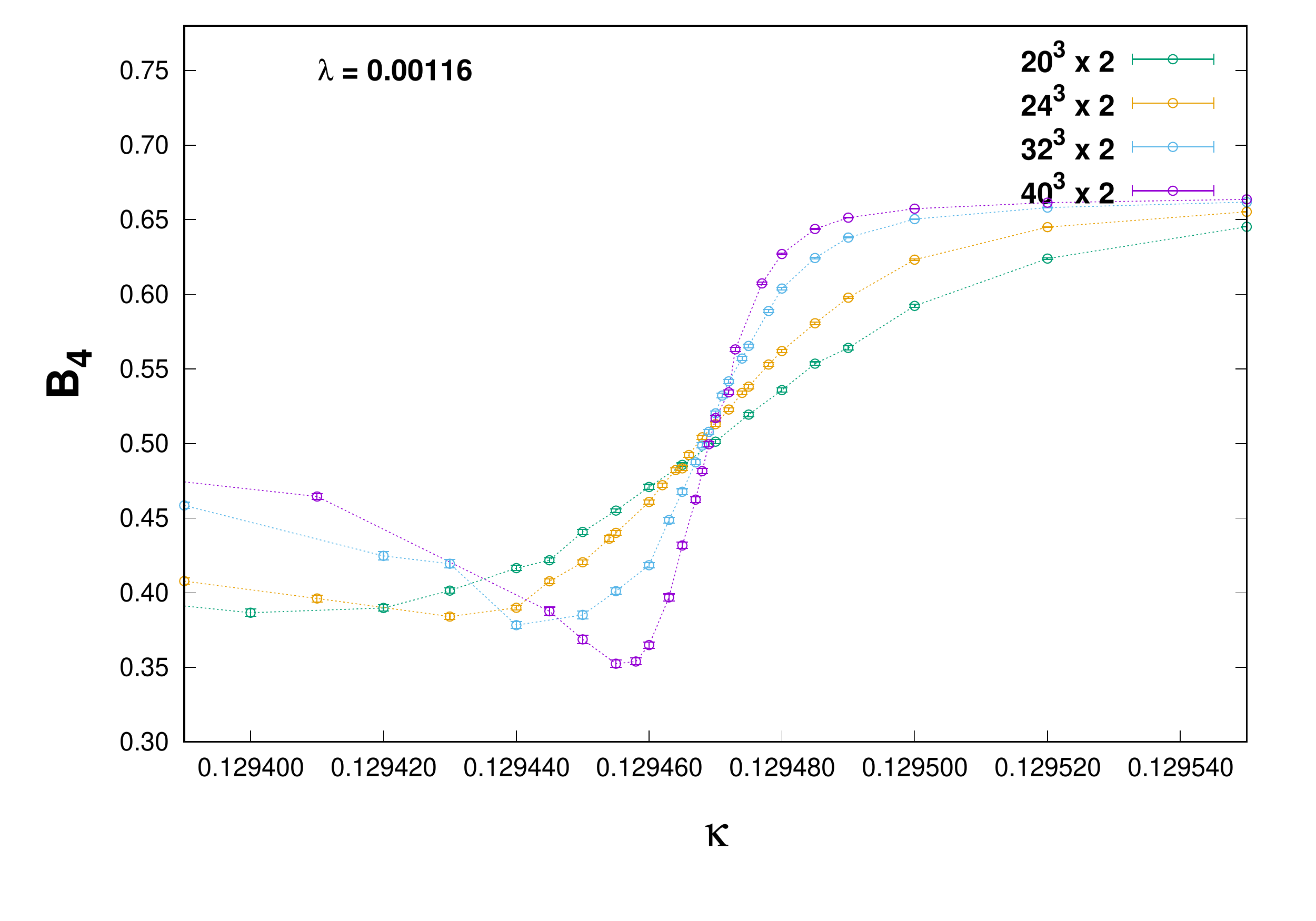}}
    \label{fig:orig_binder_1160}
  }
  \caption{
    (a) Magnetization
    (b) Connected susceptibility
    (c) Total susceptibility, and
    (d) Binder cumulant as functions of $\kappa$ at $\lambda = 0.00116$. The dotted
    lines are not fitted and are for eye guides only.}
  \label{fig:orig_1160}
\end{figure}
\begin{figure}[htbp]
    \centering
    \rotatebox{270}{\includegraphics[width=0.35\hsize]
      {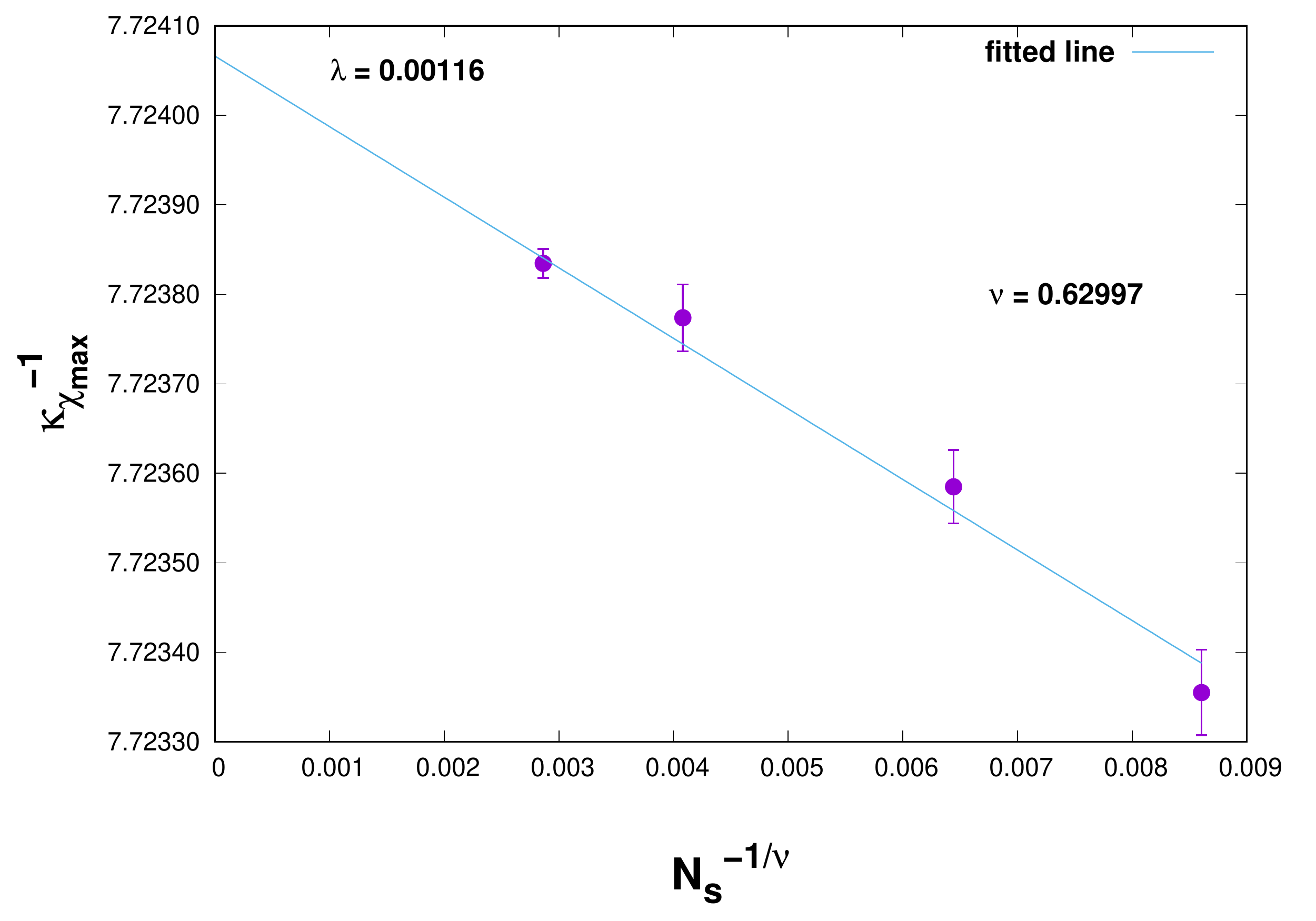}}
    \caption{The values of ${\kappa_\chi}_{\rmsmall{max}}$ as a function of $N_s^{-1/\nu}$
      at $\lambda = 0.00116$ for $N_s = 20, 24, 32$ and $40$. We set $\nu = 0.629971$. The
      intersection of the fitted straight line provides the value of $\kappa_c$.}
    \label{fig:fit_kceff_succept_max_ising_1160}
\end{figure}
\begin{figure}[htbp]
  \centering
  \subfigure[]{%
    %    \subfigtopskip 5pt
    \centering
    \rotatebox{270}{\includegraphics[width=0.26\hsize]
      {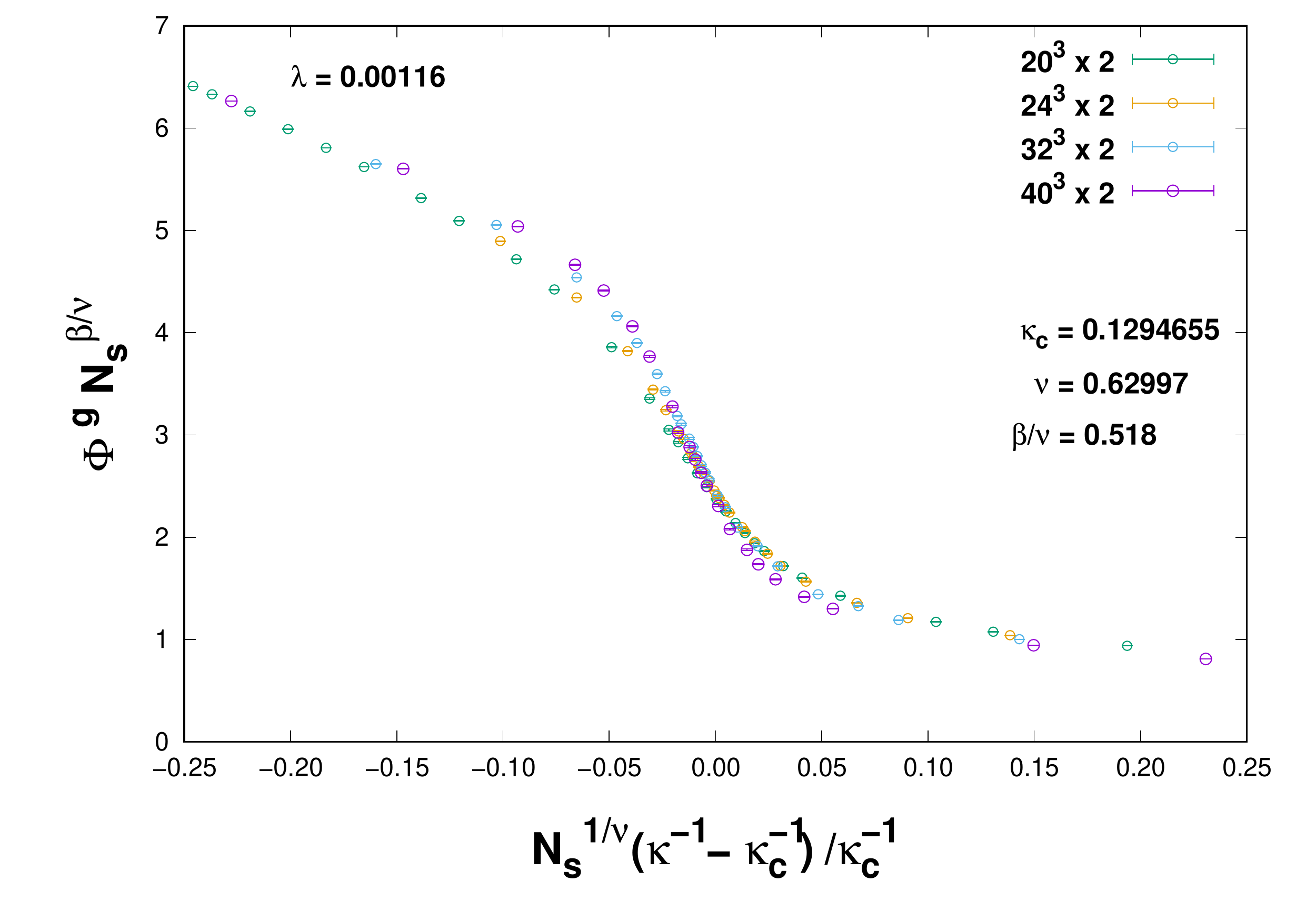}}
    \label{fig:rescaled_magtn_1160}
  }
  \subfigure[]{%
    \centering
    \rotatebox{270}{\includegraphics[width=0.26\hsize]
      {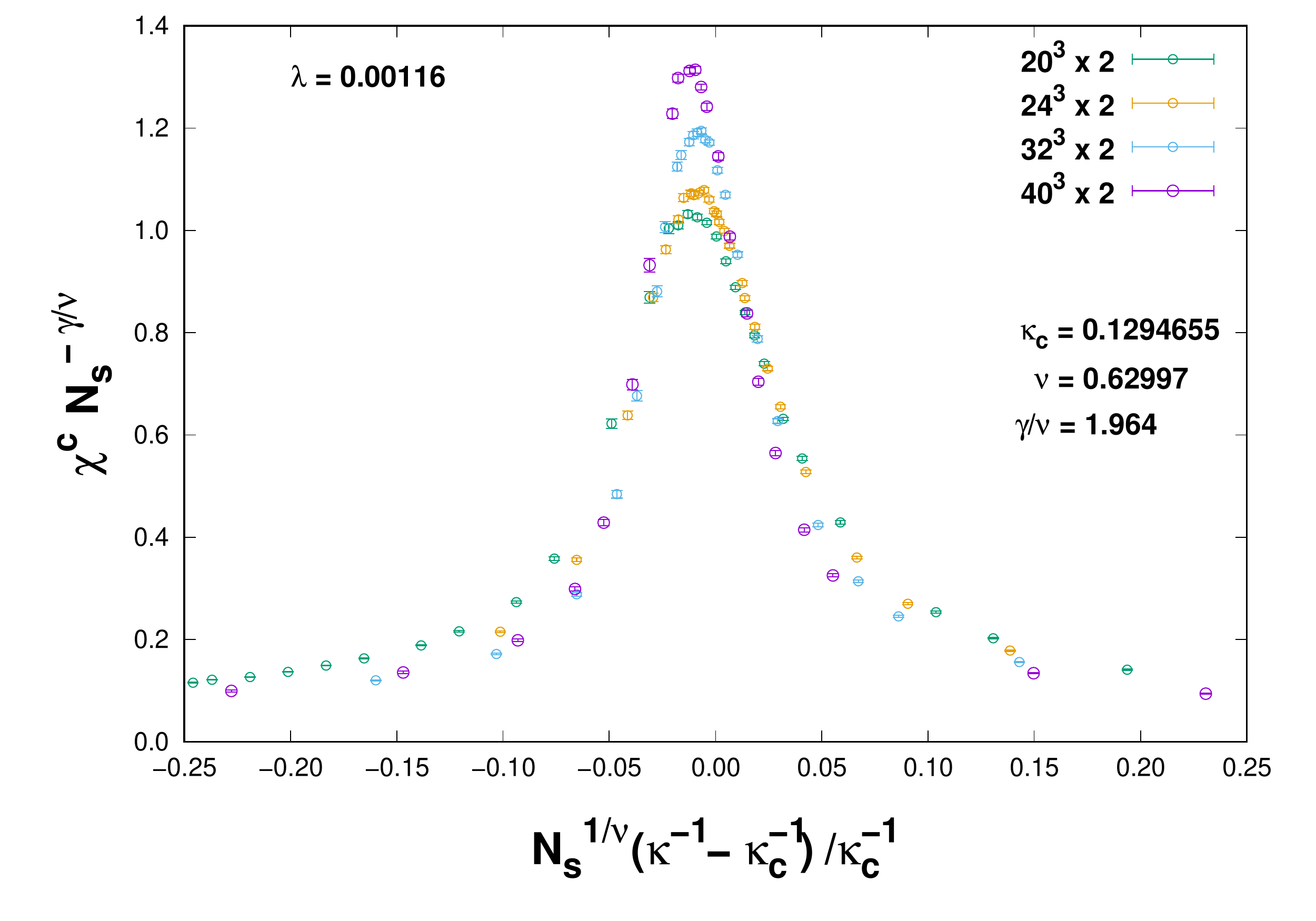}}
    \label{fig:rescaled_suscept_1160}
  }
  \subfigure[]{%
    %    \subfigtopskip 5pt
    \centering
    \rotatebox{270}{\includegraphics[width=0.26\hsize]
      {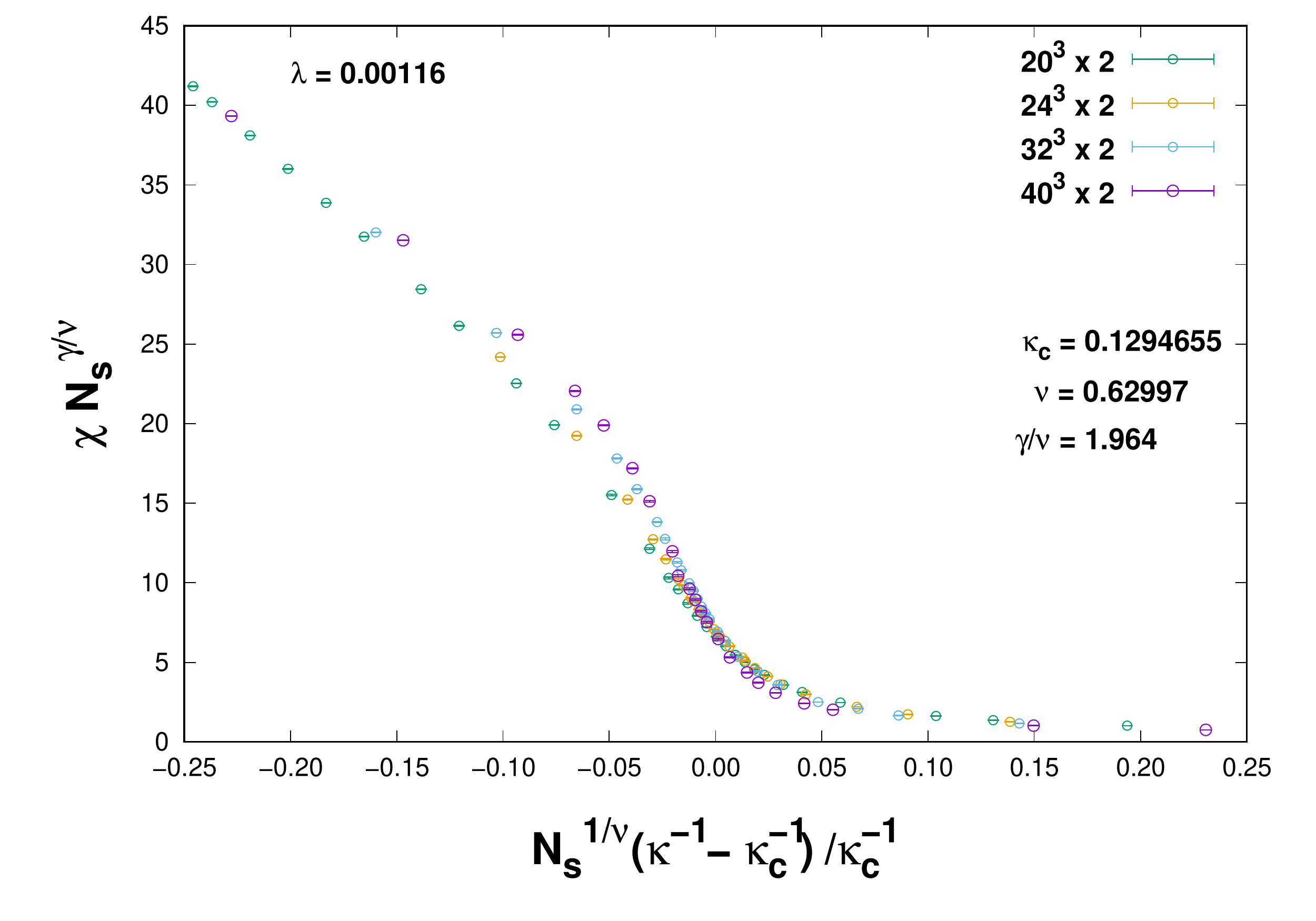}}
    \label{fig:rescaled_tot_suscept_1160}
  }
  \subfigure[]{%
    \centering
    \rotatebox{270}{\includegraphics[width=0.26\hsize]
      {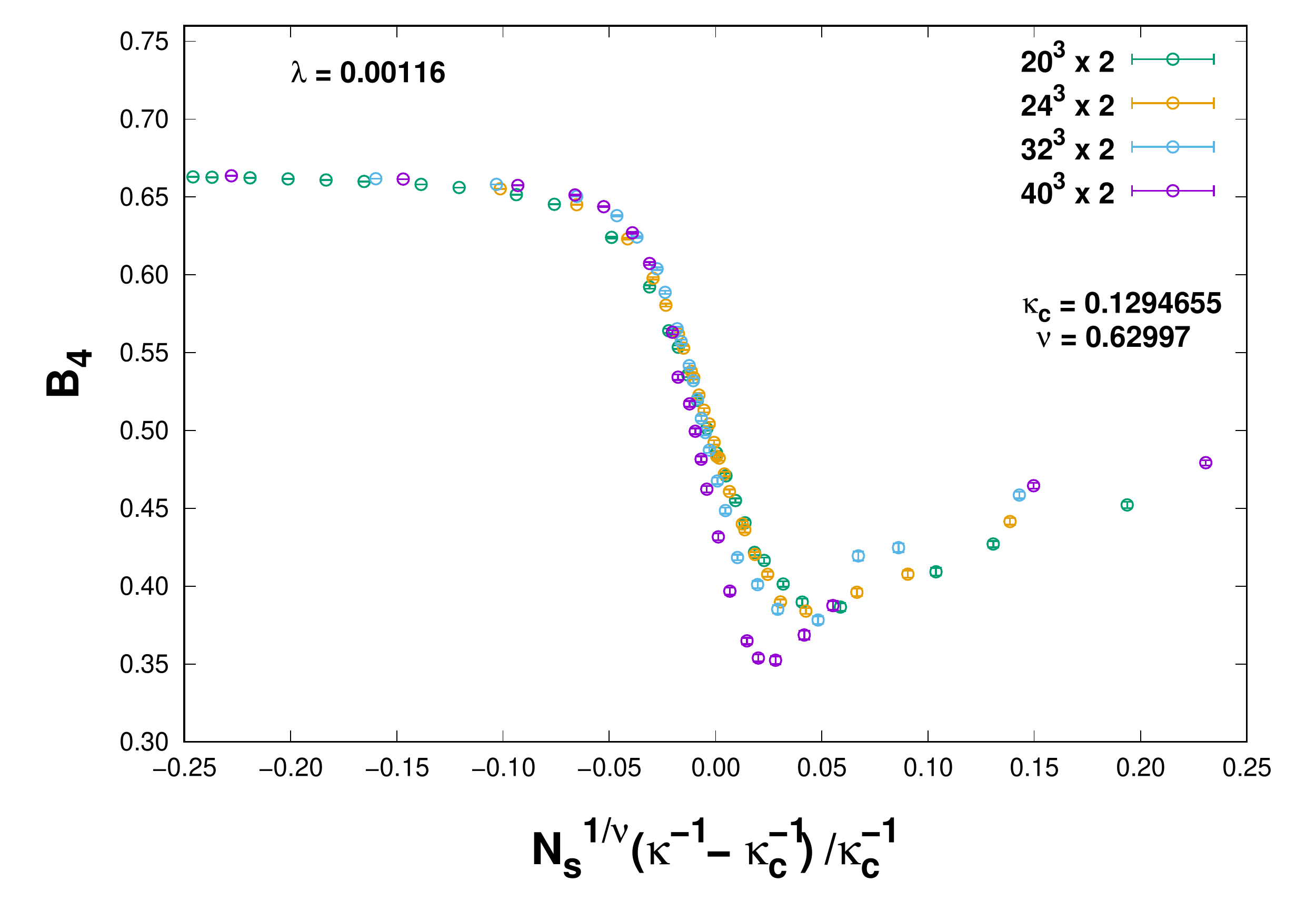}}
    \label{fig:rescaled_binder_1160}
  }
  \caption{Scaled
    (a) Magnetization
    (b) Connected susceptibility
    (c) Total susceptibility, and
    (d) Binder cumulant as functions of $\kappa$ at $\lambda = 0.00116$.}
  \label{fig:rescaled_1160}
\end{figure}
\subsection{{\boldmath$\lambda = 0.00116$}}
\label{subsec:lambda_1160}

In Fig.~\ref{fig:orig_1160}, we plot magnetization, susceptibility, total susceptibility
and Binder cumulant for $\lambda = 0.00116$. We see that $\kappa$ acts like an ``inverse
temperature''~\cite{Wurtz:2009gf} for all the four quantities, and there are indications
that  these four quantities behave similar to that of Ising Model.\ However,  it is necessary
to study further by scaling all the four quantities with the standard Ising exponents and see
whether the scaled quantities follow the FSS behavior. 

\smallskip

In order to carry out the scaling procedure, the corresponding critical value of $\kappa$
needs to be evaluated. To find $\kappa_c$, we first estimate the value of $\chi^c_{\rmsmall{max}}$
for each volume and obtain the corresponding ${\kappa_\chi}_{\rmsmall{max}}$. The standard procedure
to find $\chi^c_{\rmsmall{max}}$ is the Rewieghting method.\ It can be seen from
Fig.~\ref{fig:orig_suscept_1160} that we have a reasonable amount of data near the peak point
for every volume.\ Thus,\ we instead use the Cubic Spline Interpolation method to generate a
few hundred points close to ${\kappa_\chi}_{\rmsmall{max}}$ for every Jackknife sample.\ From these
sets of interpolated points,\ we find the values of $\chi^c_{\rmsmall{max}}$ and the corresponding
${\kappa_\chi}_{\rmsmall{max}}$ for each volume. The value of $\kappa_c$ is then obtained by using
the following FSS relation given by, 
\eqarray{
  {\kappa_\chi}_{\rmsmall{max}}^{-1} &=& \kappa_c^{-1} + a\, N_s^{-1/\nu}.
  \label{kmax_3d}
}
By using the standard value of $\nu = 0.629971$ for $\!3$d Ising Model, we get the value of
$\kappa_c$ from the linear fit of ${\kappa_\chi}_{\rmsmall{max}}$ as a function of $N_s^{-1/\nu}$
as (See Fig.~\ref{fig:fit_kceff_succept_max_ising_1160}),
\eqarray{
  \kappa_c &=& 0.1294655 (5).
  \label{eq:kappa_c.chimax_1160}
}
Using the value of $\kappa_c$ from Eq.~(\ref{eq:kappa_c.chimax_1160}), and the standard
$\!3$d Ising values of $\gamma = 1.237075$ and $\beta = 0.326419$, the quantities are
scaled. They are plotted in Fig.~\ref{fig:rescaled_1160}.

\smallskip

It can be clearly seen from Fig.~\ref{fig:rescaled_1160} that quantities do not properly
obey the scaling behavior. The differences are more visible in the cases of susceptibility
and Binder cumulant. Further, no improvement in the scaling is seen by changing the
universality class. We thus extend our study to other $\lambda$ values for which scaling
can be seen. We considered two other values of $\lambda$, namely $\lambda =  0.00132$ and
$0.00150$, which are around $\sim 1\sigma$ and $\sim 2\sigma$ from $\lambda=0.00116$,
respectively.
\begin{figure}[htbp]
  \centering
  \subfigure[]{%
    %    \subfigtopskip 5pt
    \centering
    \rotatebox{270}{\includegraphics[width=0.26\hsize]
      {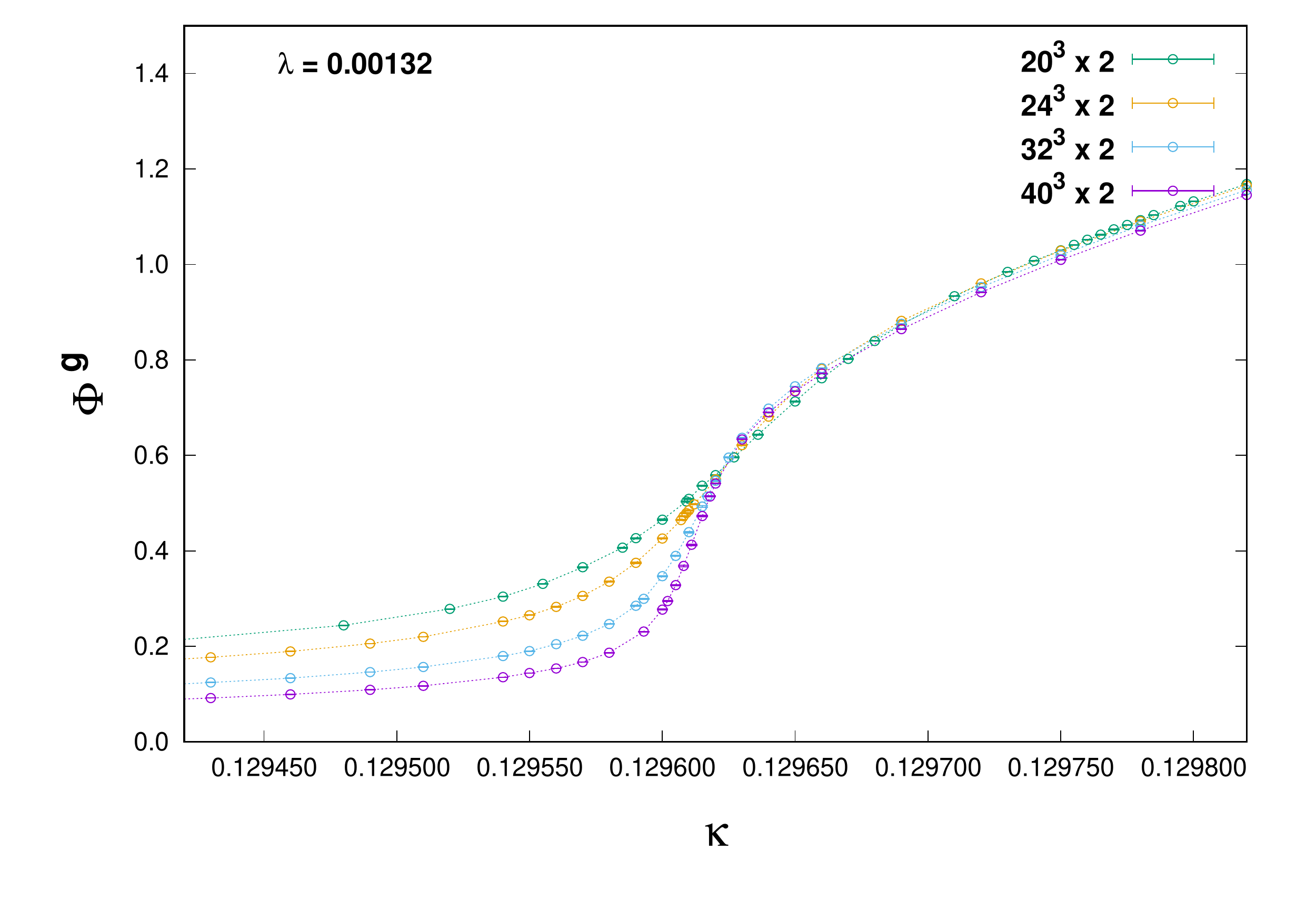}}
    \label{fig:orig_magtn_1320}
  }
  \subfigure[]{%
    \centering
    \rotatebox{270}{\includegraphics[width=0.26\hsize]
      {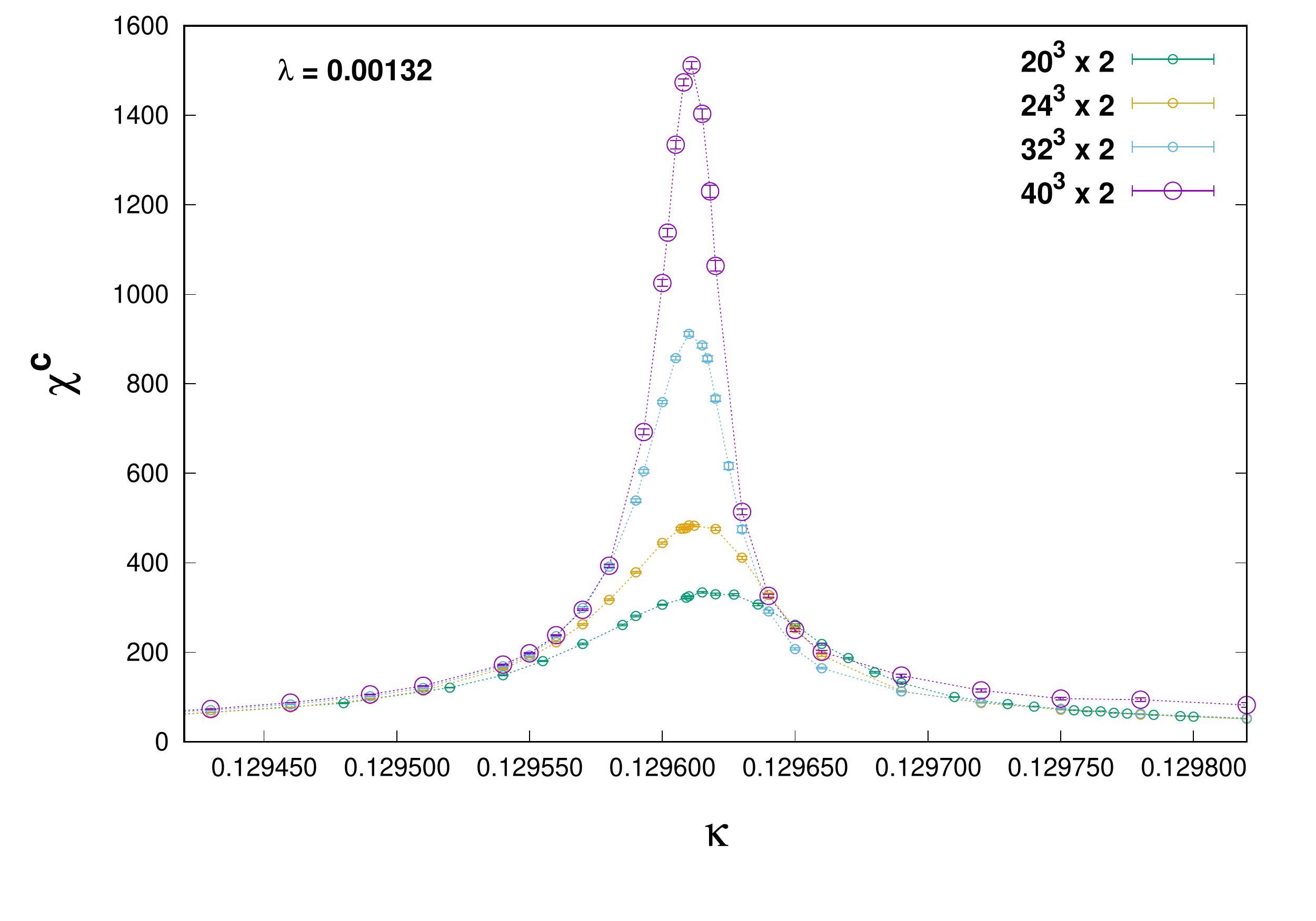}}
    \label{fig:orig_suscept_1320}
  }
  \subfigure[]{%
    %    \subfigtopskip 5pt
    \centering
    \rotatebox{270}{\includegraphics[width=0.26\hsize]
      {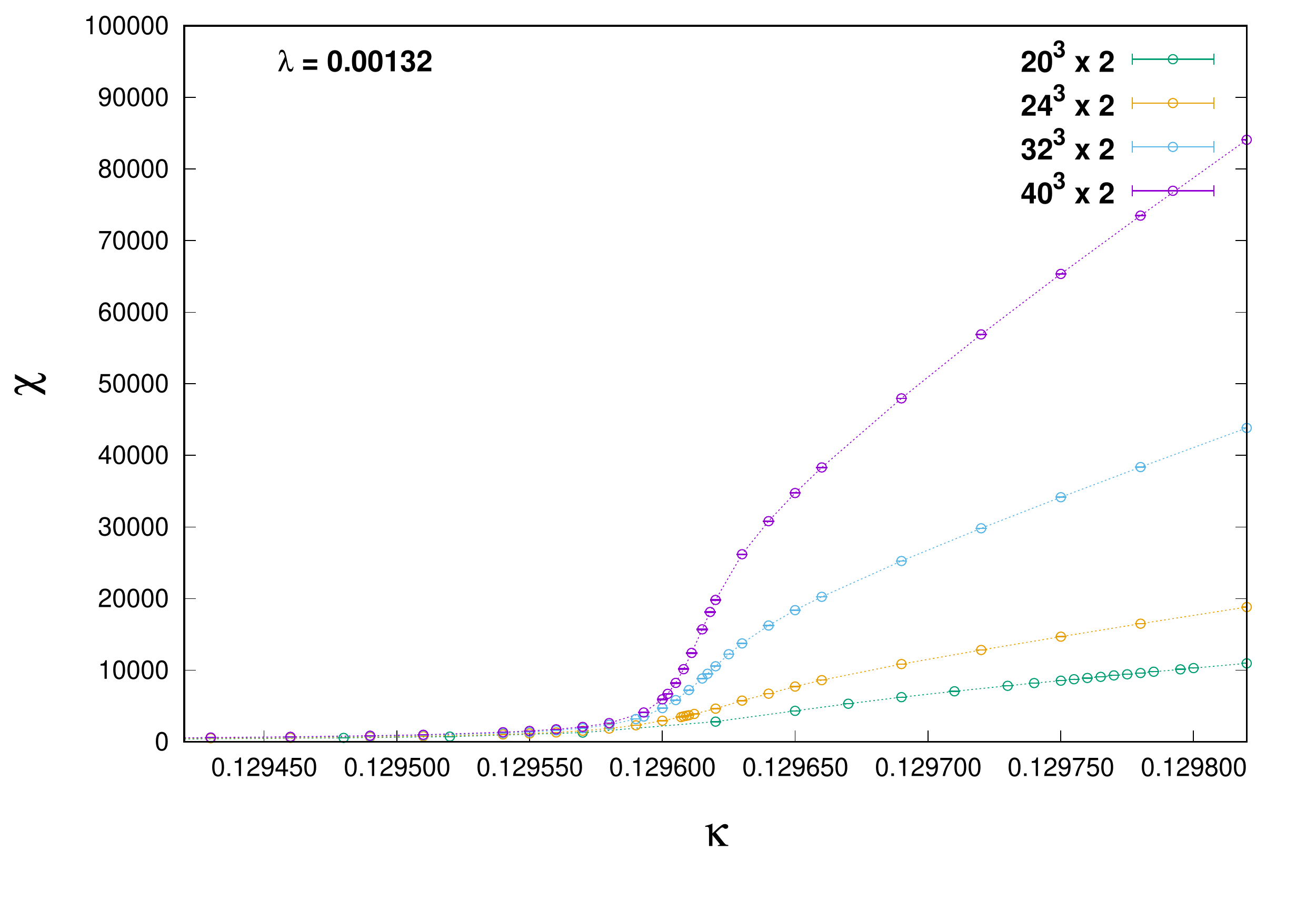}}
    \label{fig:orig_tot_suscept_1320}
  }
  \subfigure[]{%
    \centering
    \rotatebox{270}{\includegraphics[width=0.26\hsize]
      {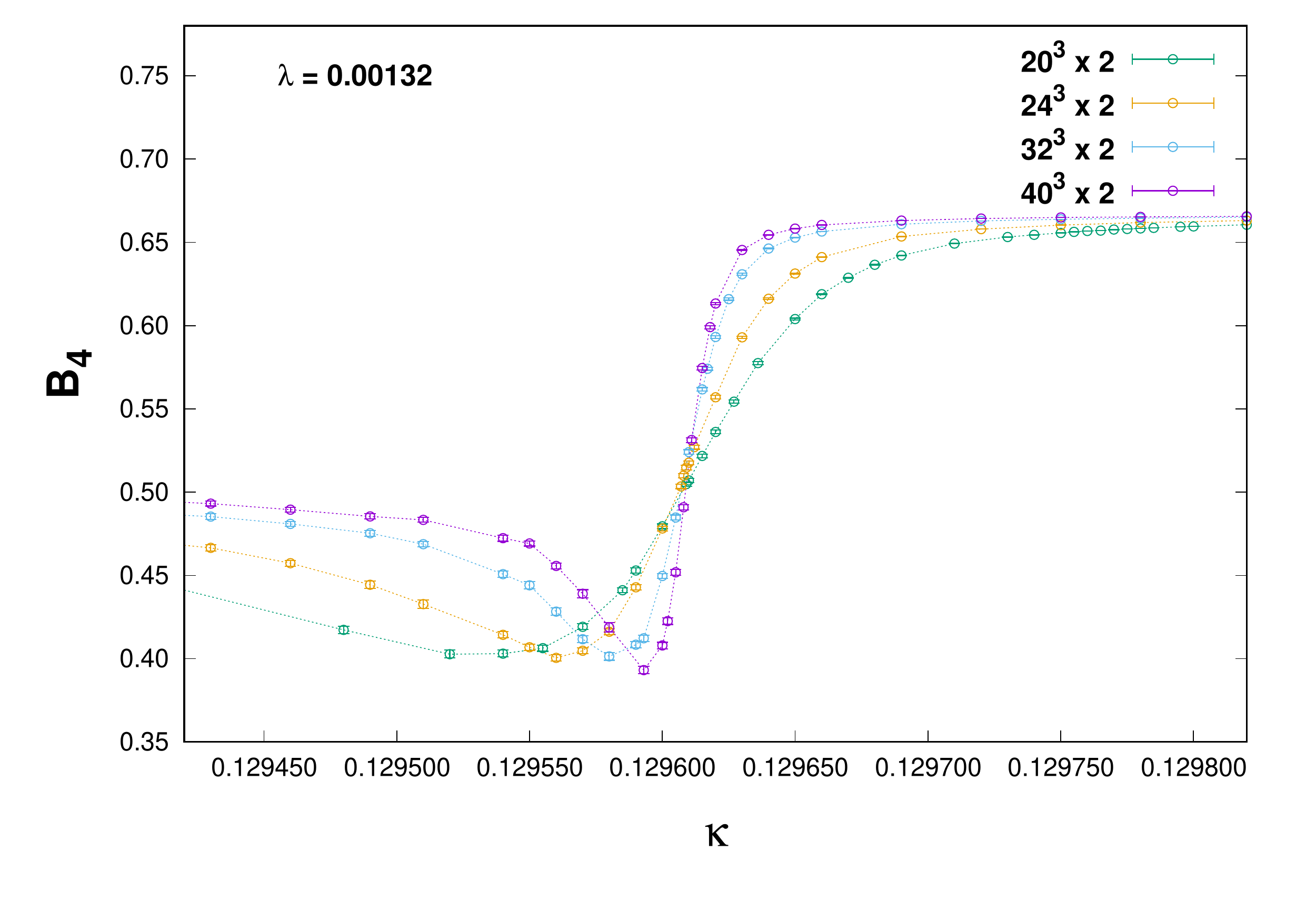}}
    \label{fig:orig_binder_1320}
  }
  \caption{
    (a) Magnetization
    (b) Connected susceptibility
    (c) Total susceptibility, and
    (d) Binder cumulant as functions of $\kappa$ at $\lambda = 0.00132$. The dotted
    lines are for eye guides only.}
  \label{fig:orig_1320}
\end{figure}
\begin{figure}[htbp]
  \centering
  \subfigure[]{%
    %    \subfigtopskip 5pt
    \centering
    \rotatebox{270}{\includegraphics[width=0.26\hsize]
      {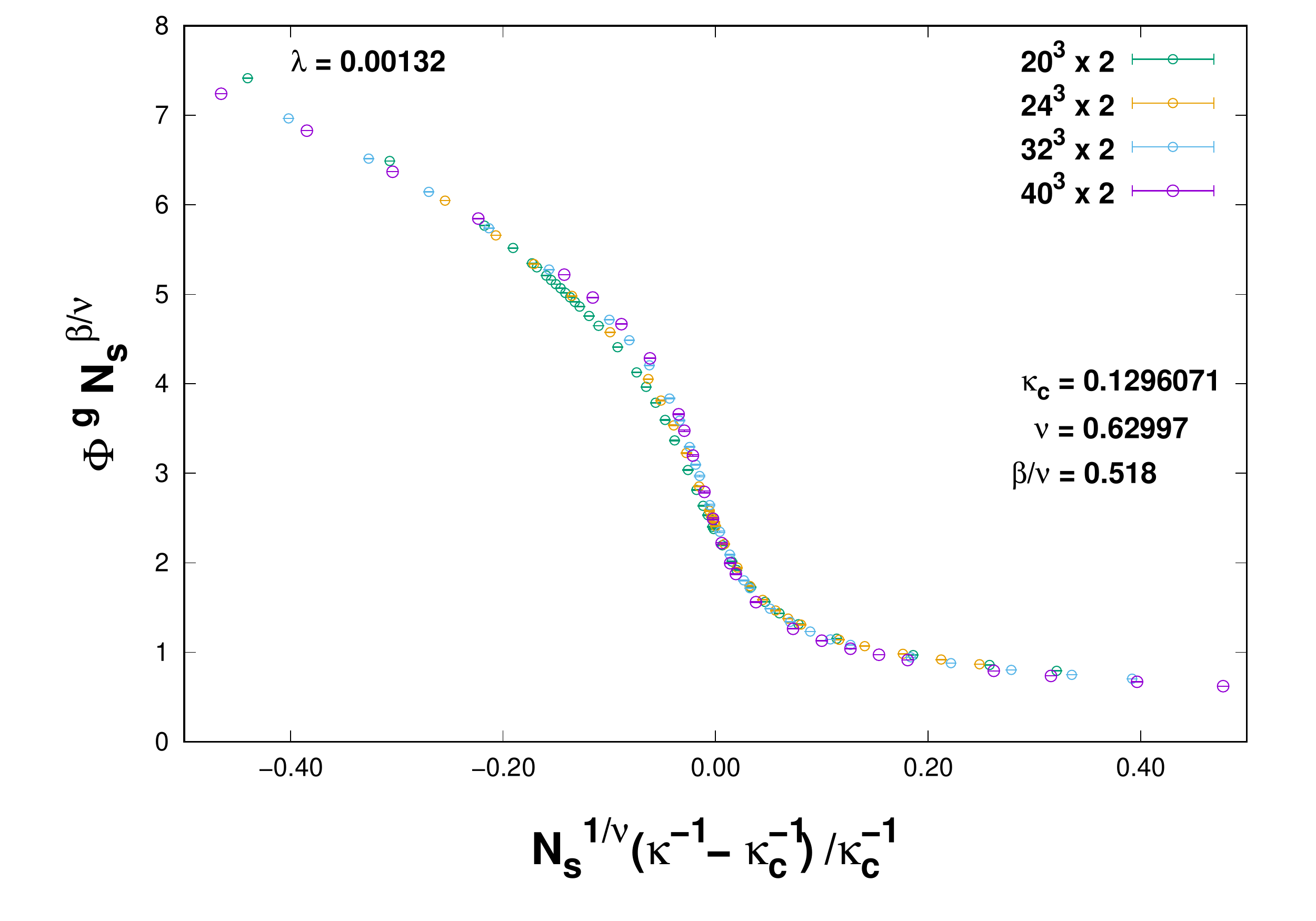}}
    \label{fig:rescaled_magtn_1320}
  }
  \subfigure[]{%
    \centering
    \rotatebox{270}{\includegraphics[width=0.26\hsize]
      {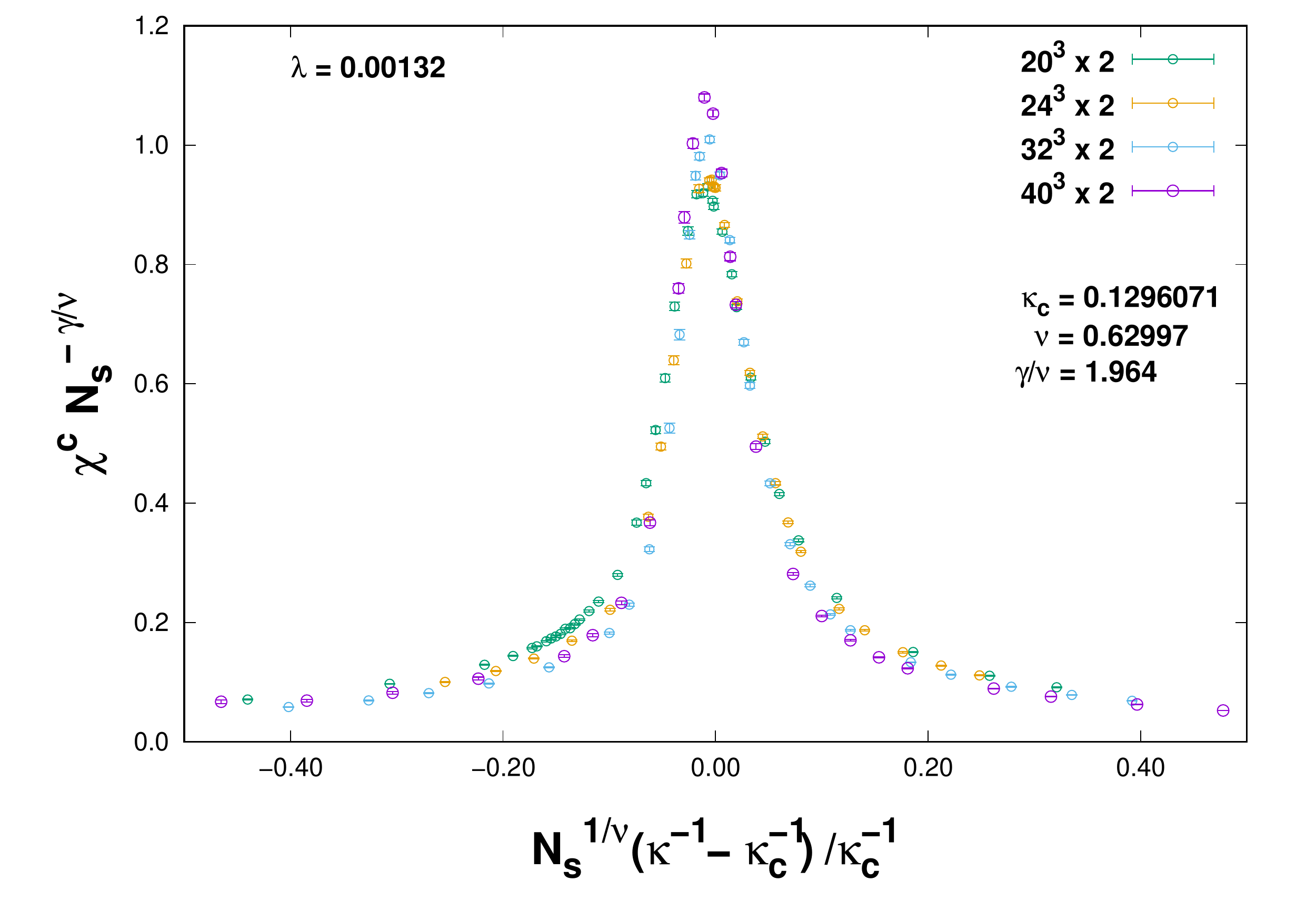}}
    \label{fig:rescaled_suscept_1320}
  }
  \subfigure[]{%
    %    \subfigtopskip 5pt
    \centering
    \rotatebox{270}{\includegraphics[width=0.26\hsize]
      {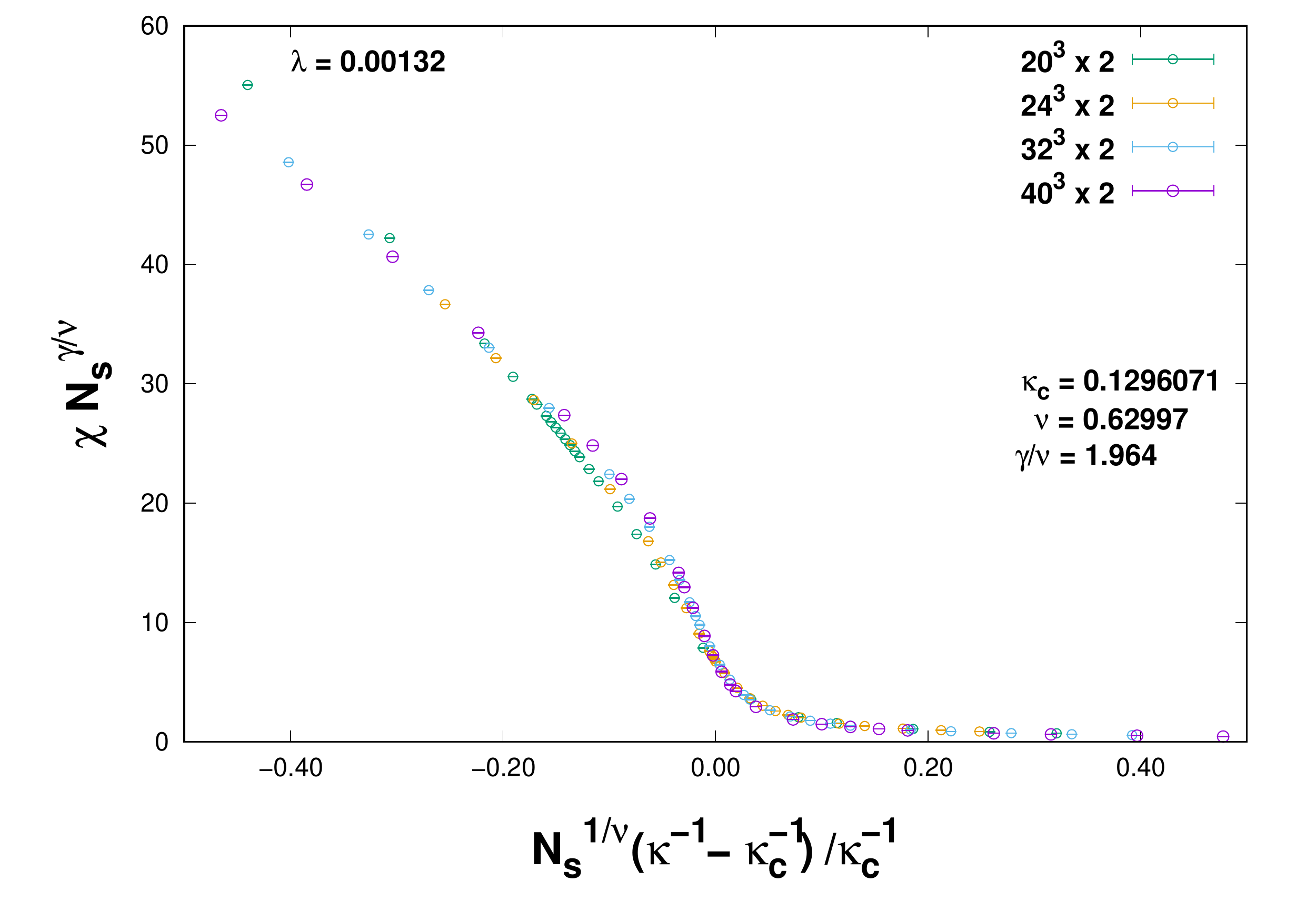}}
    \label{fig:rescaled_tot_suscept_1320}
  }
  \subfigure[]{%
    \centering
    \rotatebox{270}{\includegraphics[width=0.26\hsize]
      {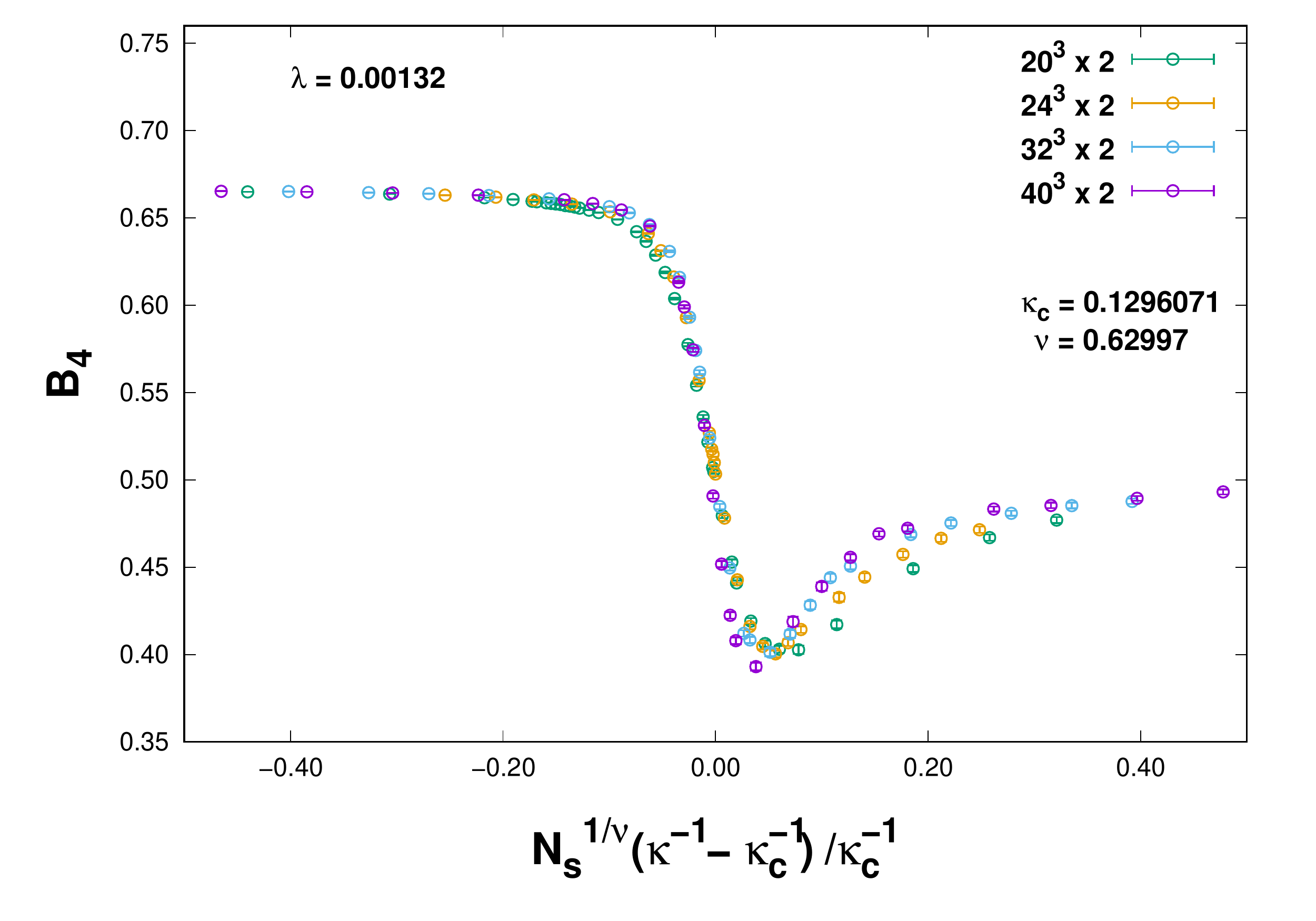}}
    \label{fig:rescaled_binder_1320}
  }
  \caption{Scaled
    (a) Magnetization
    (b) Connected susceptibility
    (c) Total susceptibility, and
    (d) Binder cumulant as functions of $\kappa$ at $\lambda = 0.00132$.}
  \label{fig:rescaled_1320}
\end{figure}
\begin{figure}[htbp]
  \centering
  \subfigure[]{%
    %    \subfigtopskip 5pt
    \centering
    \rotatebox{270}{\includegraphics[width=0.26\hsize]
      {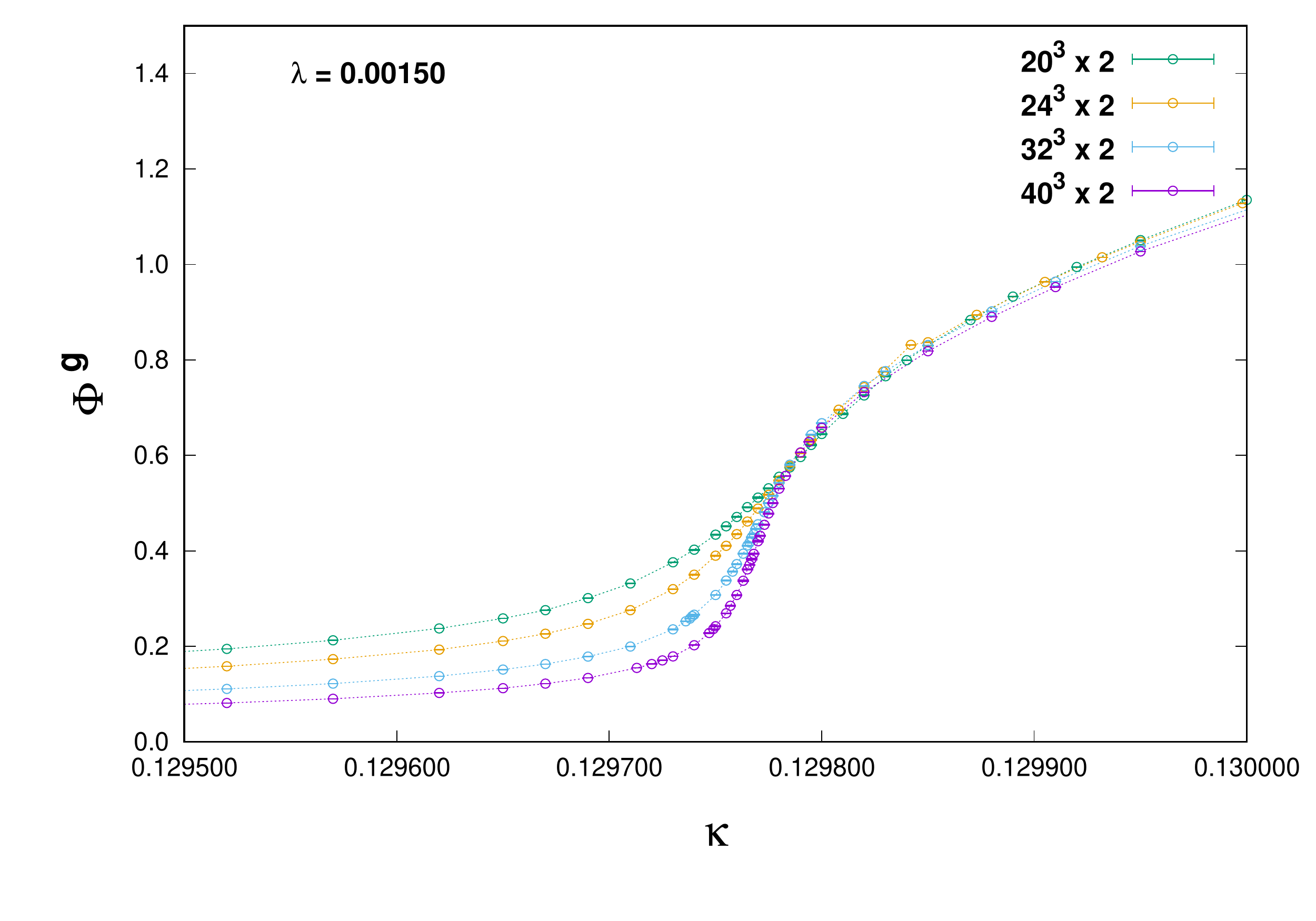}}
    \label{fig:orig_magtn_1500}
  }
  \subfigure[]{%
    \centering
    \rotatebox{270}{\includegraphics[width=0.26\hsize]
      {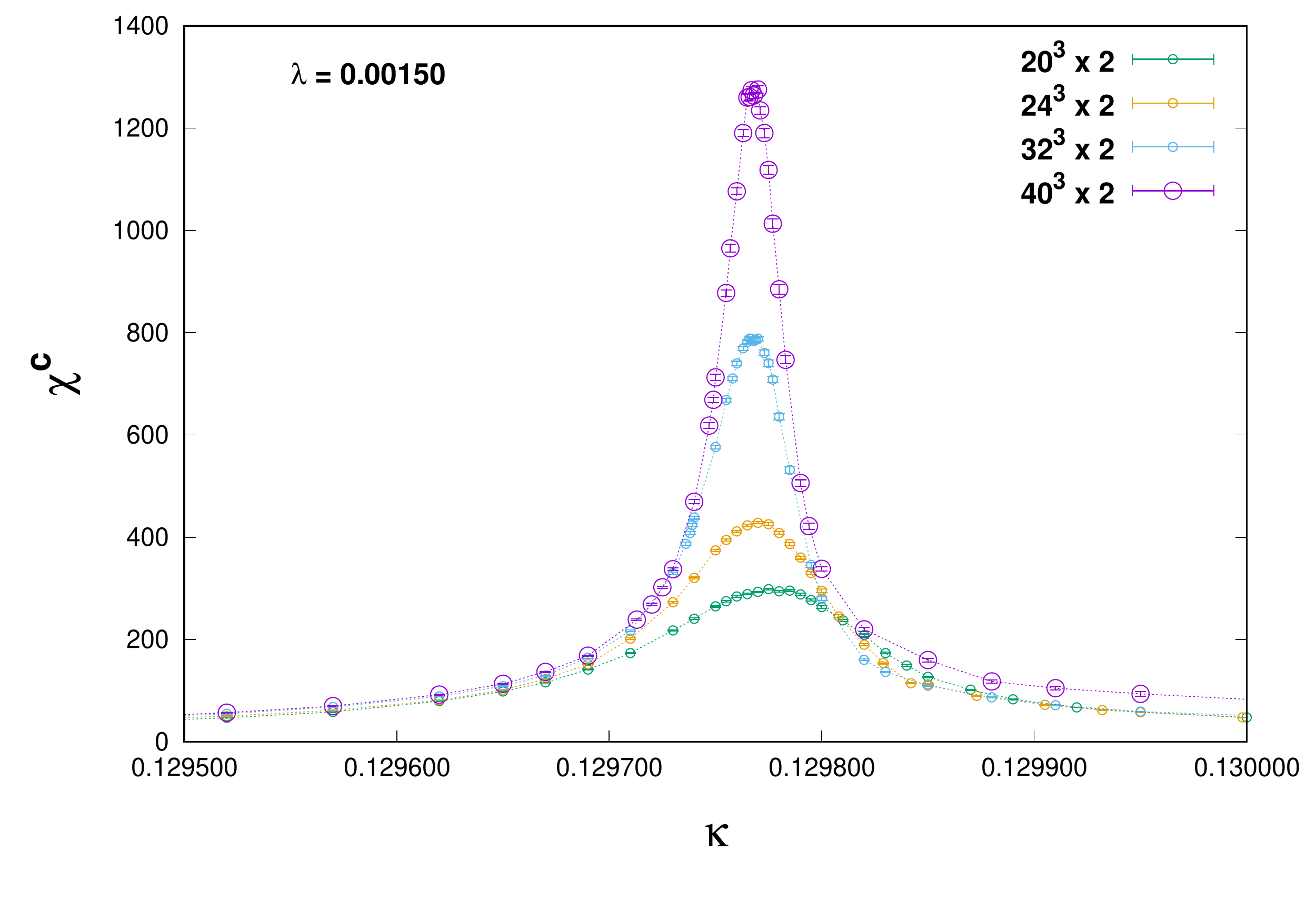}}
    \label{fig:orig_suscept_1500}
  }
  \subfigure[]{%
    %    \subfigtopskip 5pt
    \centering
    \rotatebox{270}{\includegraphics[width=0.26\hsize]
      {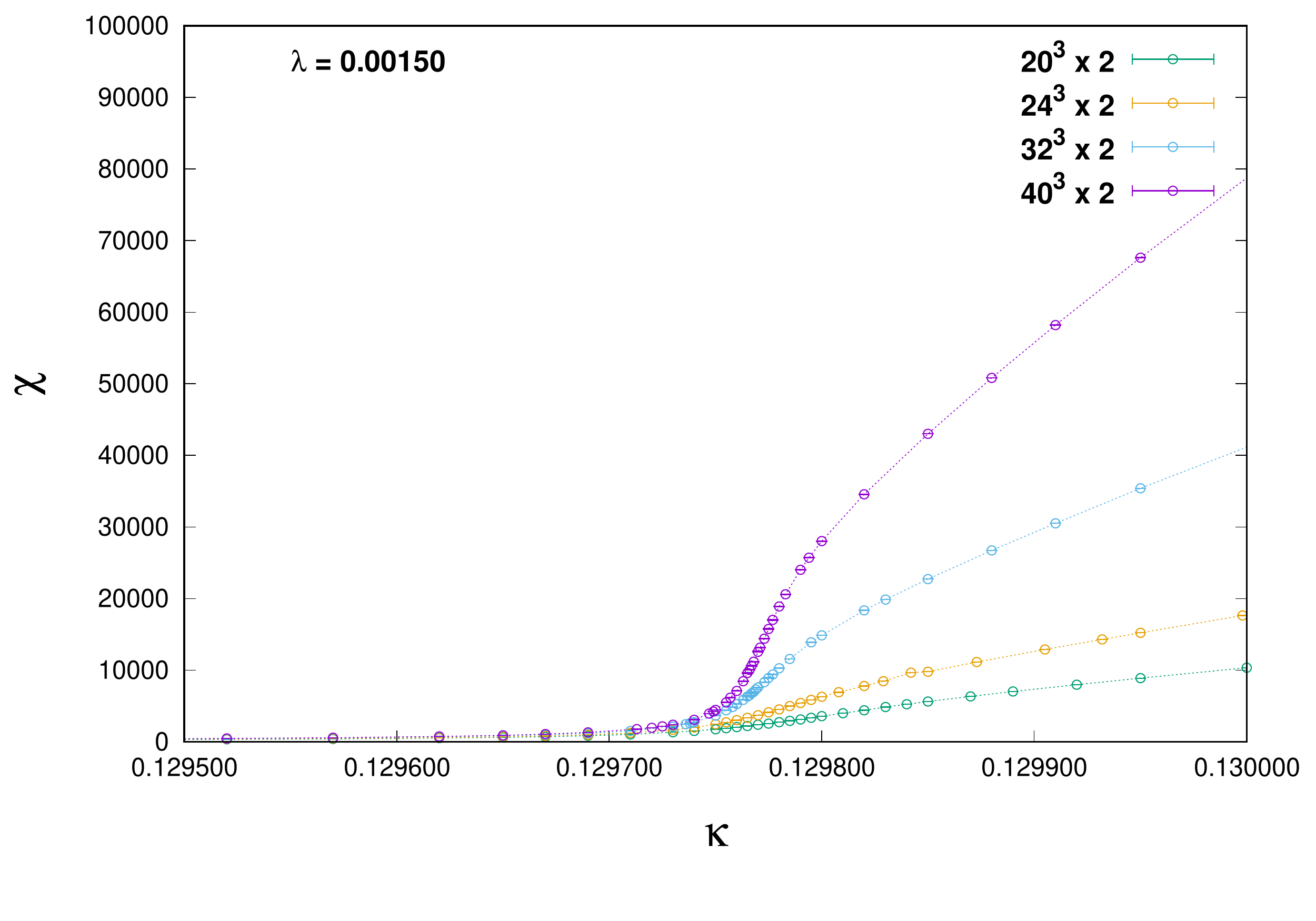}}
    \label{fig:orig_tot_suscept_1500}
  }
  \subfigure[]{%
    \centering
    \rotatebox{270}{\includegraphics[width=0.26\hsize]
      {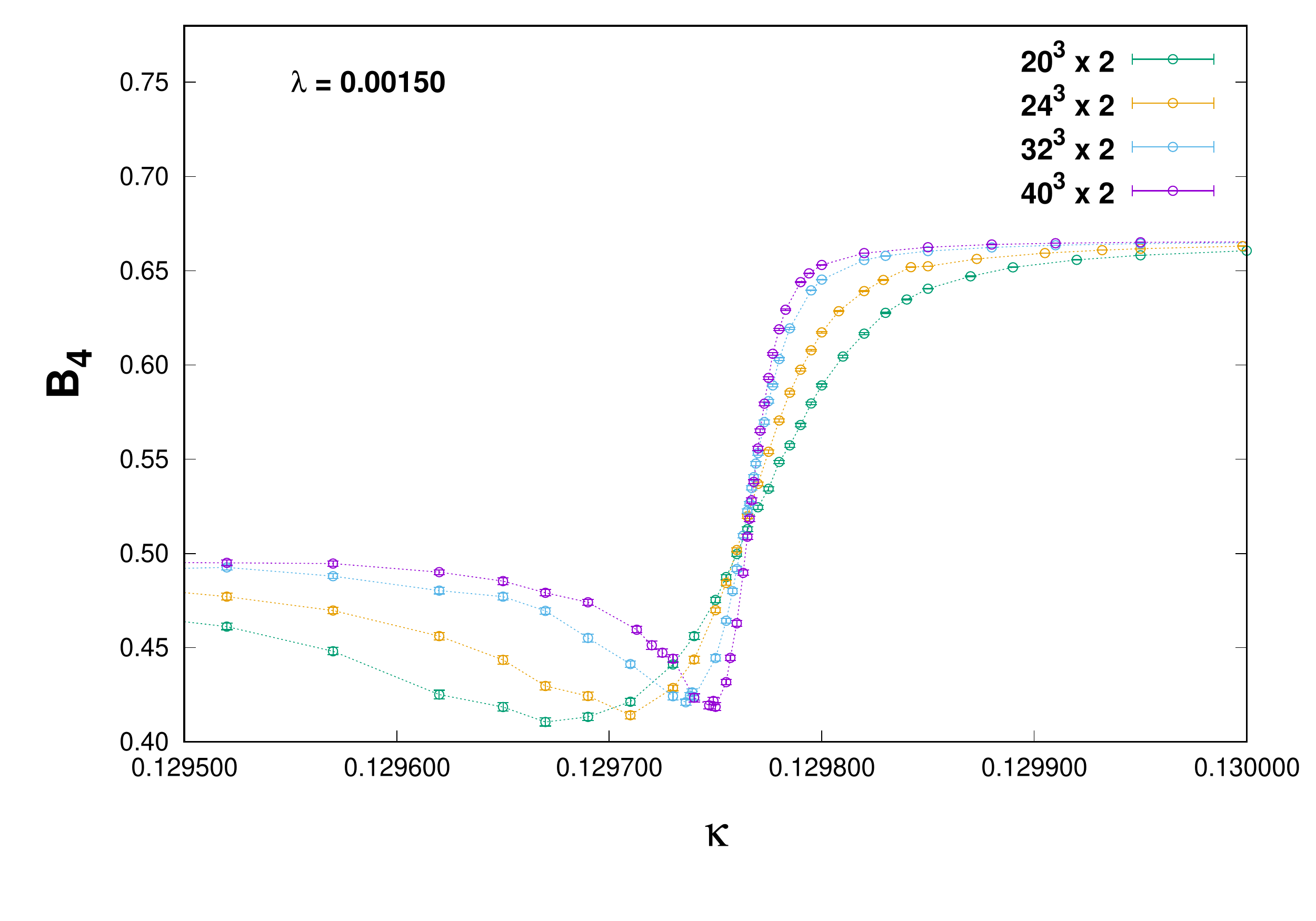}}
    \label{fig:orig_binder_1500}
  }
  \caption{
    (a) Magnetization
    (b) Connected susceptibility
    (c) Total susceptibility, and
    (d) Binder cumulant as functions of $\kappa$ at $\lambda = 0.00150$. The dotted
    lines are for eye guides only.}
  \label{fig:orig_1500}
\end{figure}
\begin{figure}[htbp]
  \centering
  \subfigure[]{%
    %    \subfigtopskip 5pt
    \centering
    \rotatebox{270}{\includegraphics[width=0.26\hsize]
      {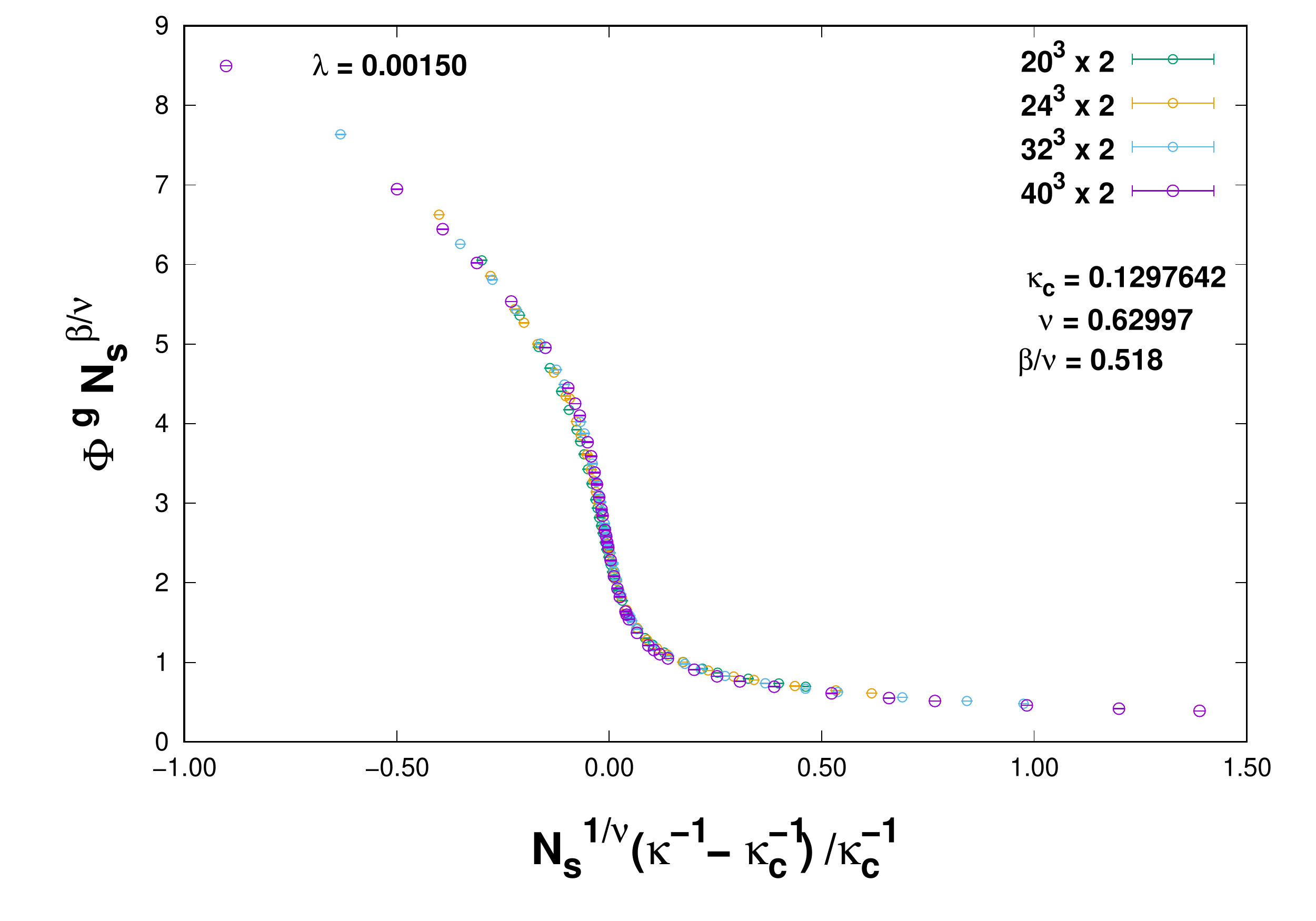}}
    \label{fig:rescaled_magtn_1500}
  }
  \subfigure[]{%
    \centering
    \rotatebox{270}{\includegraphics[width=0.26\hsize]
      {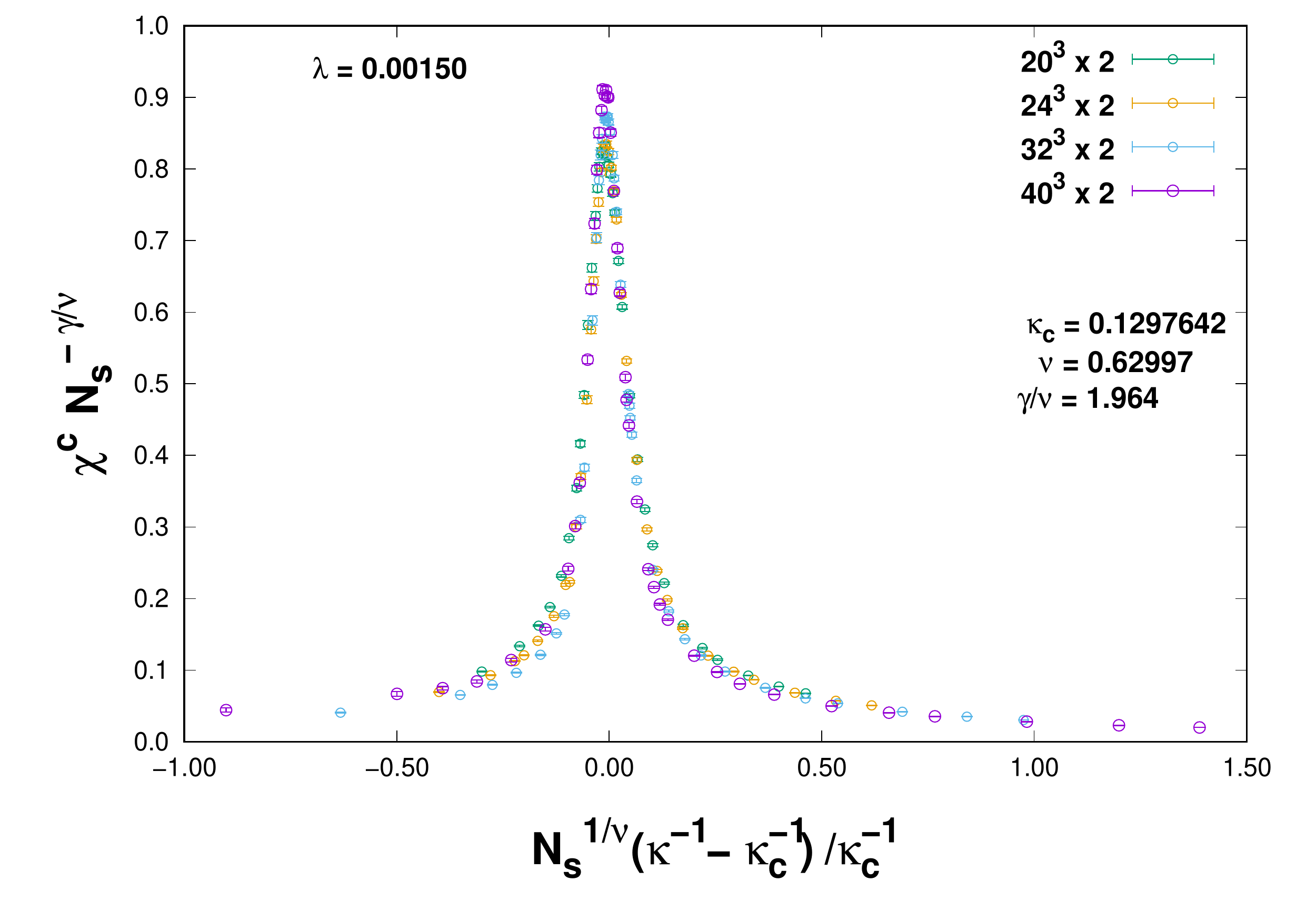}}
    \label{fig:rescaled_suscept_1500}
  }
  \subfigure[]{%
    %    \subfigtopskip 5pt
    \centering
    \rotatebox{270}{\includegraphics[width=0.26\hsize]
      {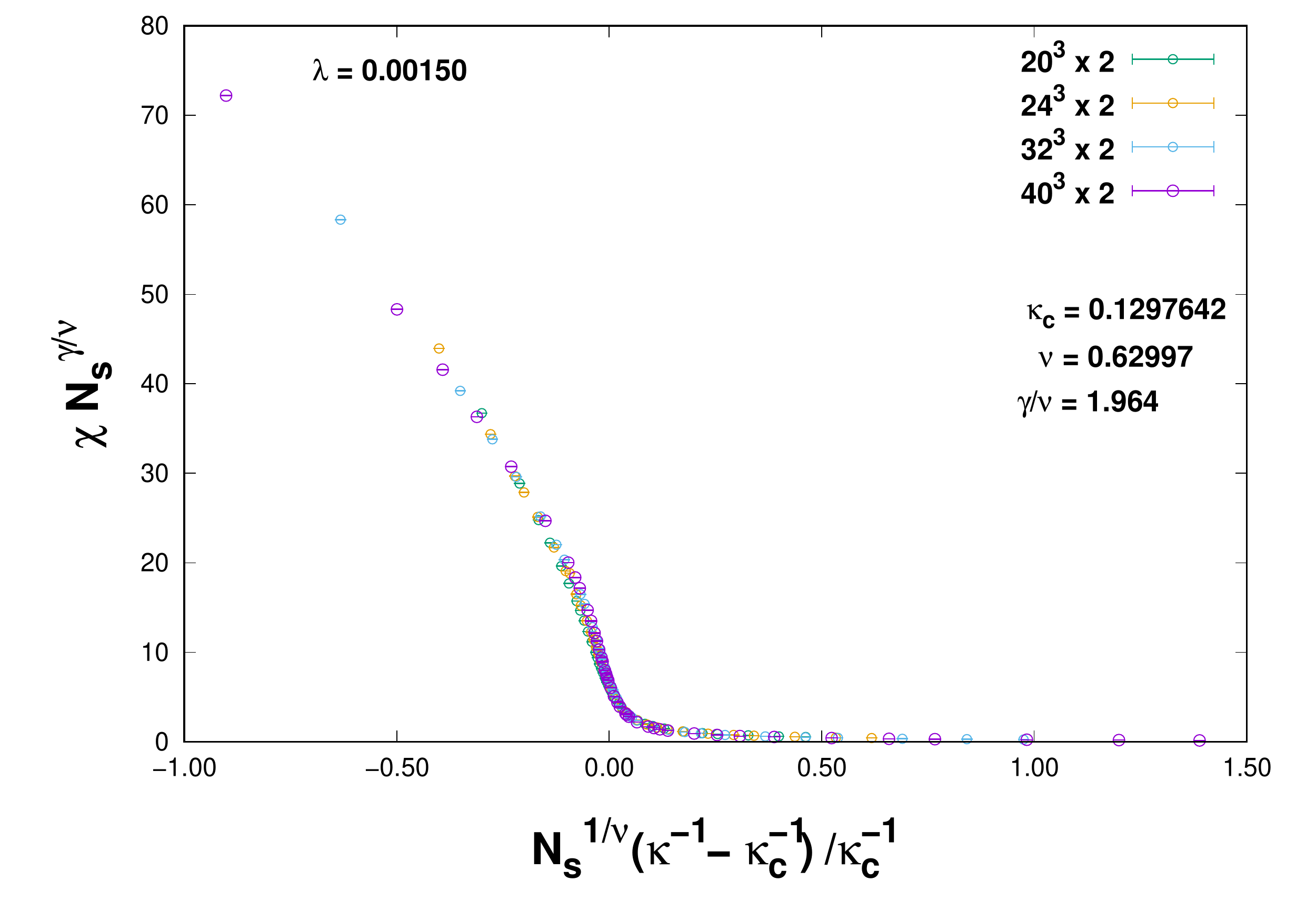}}
    \label{fig:rescaled_tot_suscept_1500}
  }
  \subfigure[]{%
    \centering
    \rotatebox{270}{\includegraphics[width=0.26\hsize]
      {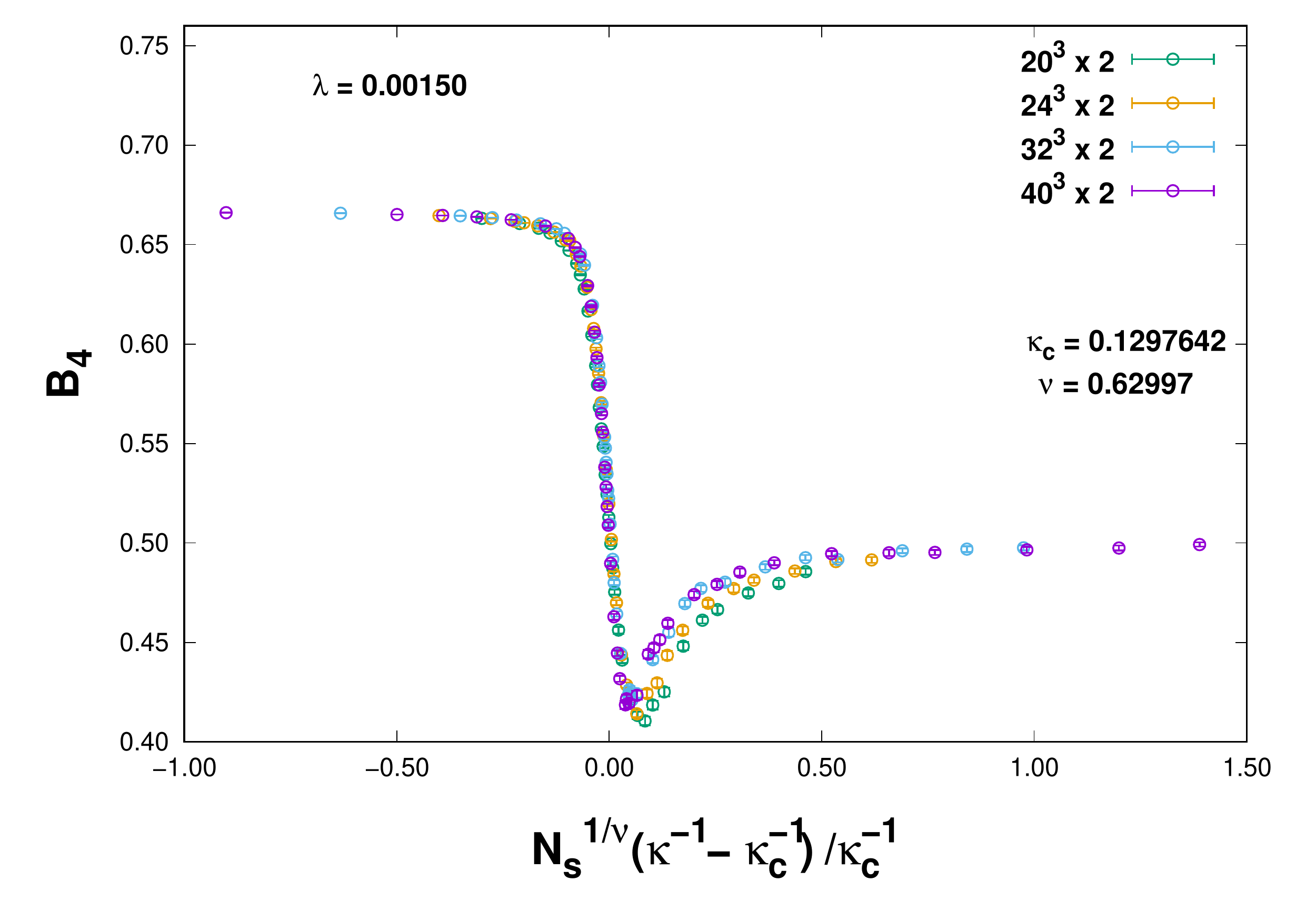}}
    \label{fig:rescaled_binder_1500}
  }
  \caption{Scaled
    (a) Magnetization
    (b) Connected susceptibility
    (c) Total susceptibility, and
    (d) Binder cumulant as functions of $\kappa$ at $\lambda = 0.00150$.}
  \label{fig:rescaled_1500}
\end{figure}

\subsection{{\boldmath$\lambda = 0.00132$} and {\boldmath$0.00150$}}
\label{subsec:lambda_1320_and_1500}

We compute magnetization, susceptibility, total susceptibility and Binder cumulant for
$\lambda = 0.00132$ and $0.00150$. They are plotted in Figs.~\ref{fig:orig_1320} and
\ref{fig:orig_1500}, respectively. We again see Ising like behavior of these quantities.
Note that the value of $\kappa_c$ changes with the value of $\lambda$.\ We follow
the same procedure as described in Sec.~\ref{subsec:lambda_1160} to determine $\kappa_c$
for $\lambda = 0.00132$ and $0.00150$. Thus, we get,
\eqarray{
  \kappa_c &=& 0.1296071 (8)\,\,\,\,\, \mbox{for $\lambda = 0.00132$}, \\
  \kappa_c &=& 0.1297642 (7)\,\,\,\,\, \mbox{for $\lambda = 0.00150$}.
}
Using these values of $k_c$ and the standard values of the exponents for $\!3$d Ising Model,
we scale the quantities as in Sec.~\ref{subsec:lambda_1160}. They are plotted in
Figs.~\ref{fig:rescaled_1320} and \ref{fig:rescaled_1500}, respectively. By comparing
Figs.~\ref{fig:rescaled_1160}, \ref{fig:rescaled_1320} and \ref{fig:rescaled_1500}, we
see scaling  for $\lambda = 0.00150$. This is an indication that
$\Phi^g$ is a good order parameter. Also this suggests that the critical end-point
is close to $\lambda = 0.00150$.

%%%%%%%%%%%%%%%%%%%%%%%%%%%%%%%%%%%%%%%%%%%%
%\lowercase{\boldmath{${\kappa_\chi}_{\rmsmall{max}}$}}}
\section{Magnetization at various other {\boldmath $\lambda$}'s}
\label{sec:mag_at_var_lambda}
%%%%%%%%%%%%%%%%%%%%%%%%%%%%%%%%%%%%%%%%%%%%

%
\begin{table}[htbp]
  \centering
  \caption{Run parameters for other $\lambda$'s.}
  \setlength{\tabcolsep}{9pt}
  \small
  \begin{tabular}{ccccccc}
    \hline\hline
    \multirow{4}{*}{\boldmath$\beta_g$}
    &\multirow{4}{*}{\boldmath$\lambda$}
    & \multirow{4}{*}{\boldmath$N_s$}
    & \multirow{4}{*}{\boldmath$N_\tau$}
    & \multirow{4}{*}{\boldmath$\kappa$}
    & \multirow{4}{*}{\bf Confs.}
    & {\bf Trajectories} \\
    &&&&&& {\bf between} \\
    &&&&&& {\bf consecutive} \\
    &&&&&& {\bf confs.} \\
    \hline
    \multirow{10}{*}{8.0}
    & 0.00010
    & 20, 24, 32, 40  &  2 & \{0.127800, 0.128890\} & 10 000 & 50 \\
    %\hline
    & 0.00050
    & 20, 24, 32, 40  &  2 & \{0.127800, 0.129200\} & 10 000 & 50 \\
    %\hline
    & 0.00080
    & 20, 24, 32, 40  &  2 & \{0.129000, 0.129400\} & 10 000 & 50 \\
     %\hline
    & 0.00095
    & 20, 24, 32, 40  &  2 & \{0.129000, 0.129400\} & 10 000 & 50 \\
    %\hline
    & 0.00100
    & 20, 24, 32, 40  &  2 & \{0.129120, 0.129500\} & 10 000 & 50 \\
    %\hline
    & 0.00170
    & 20, 24, 32, 40  &  2 & \{0.129690, 0.130300\} & 10 000 & 50 \\
    %\hline
    & 0.00220
    & 20, 24, 32, 40  &  2 & \{0.129450, 0.130600\} & 10 000 & 50 \\
    %\hline
    & 0.00270
    & 20, 24, 32, 40  &  2 & \{0.130000, 0.131340\} & 10 000 & 50 \\
    %\hline
    & 0.00310
    & 20, 24, 32, 40  &  2 & \{0.129500, 0.135000\} & 10 000 & 50 \\
    %\hline
    & 0.00350
    & 20, 24, 32, 40  &  2 & \{0.127500, 0.137000\} & 10 000 & 50 \\
    \hline\hline
  \end{tabular}
  \normalsize
  \label{tab:run_para_2}
\end{table}
\begin{figure}[htbp]
  \centering
  \subfigure[]{%
    %    \subfigtopskip 5pt
    \centering
    \rotatebox{270}{\includegraphics[width=0.26\hsize]
      {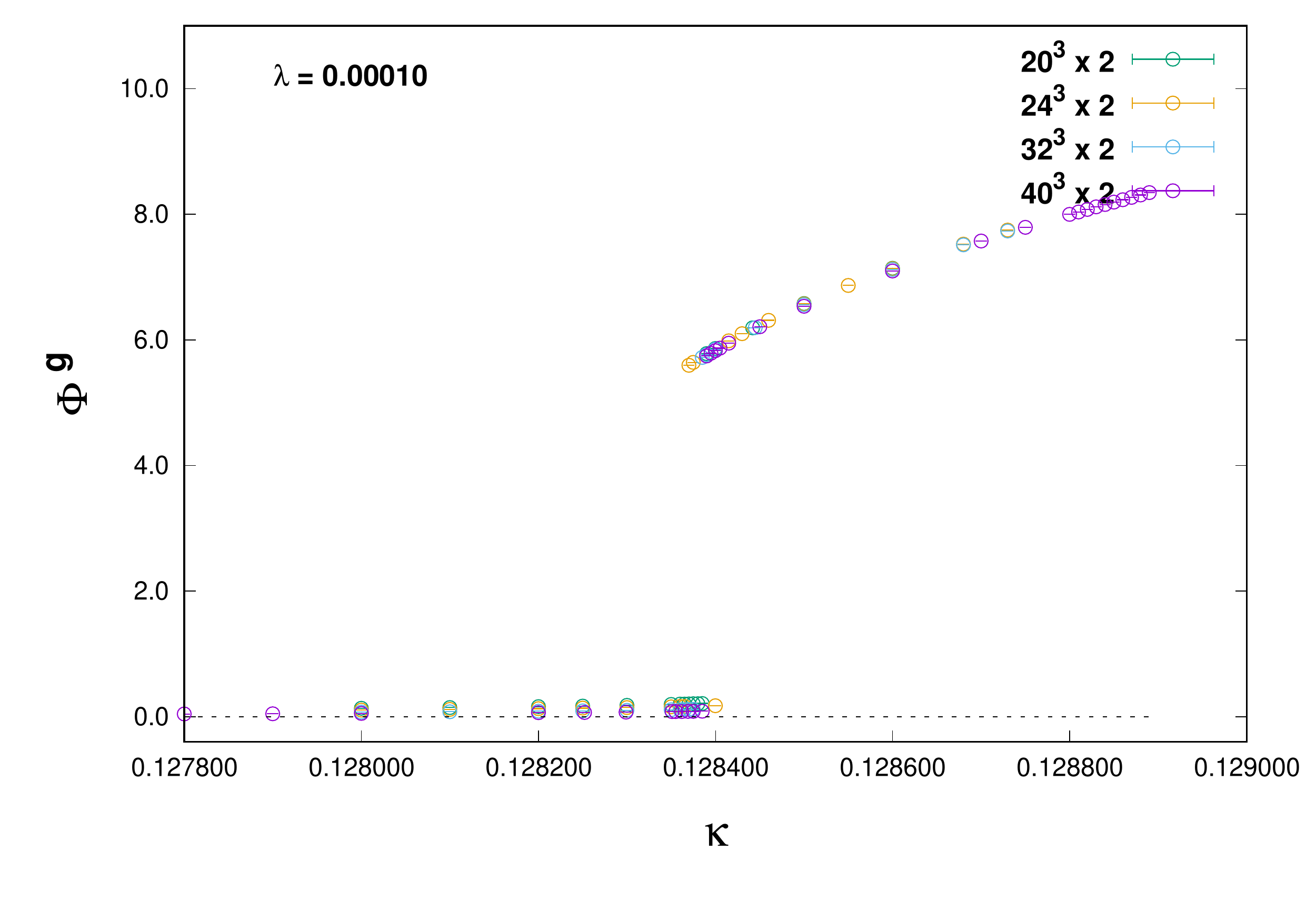}}
    \label{fig:orig_magtn_0100}
  }
  \subfigure[]{%
    \centering
    \rotatebox{270}{\includegraphics[width=0.26\hsize]
      {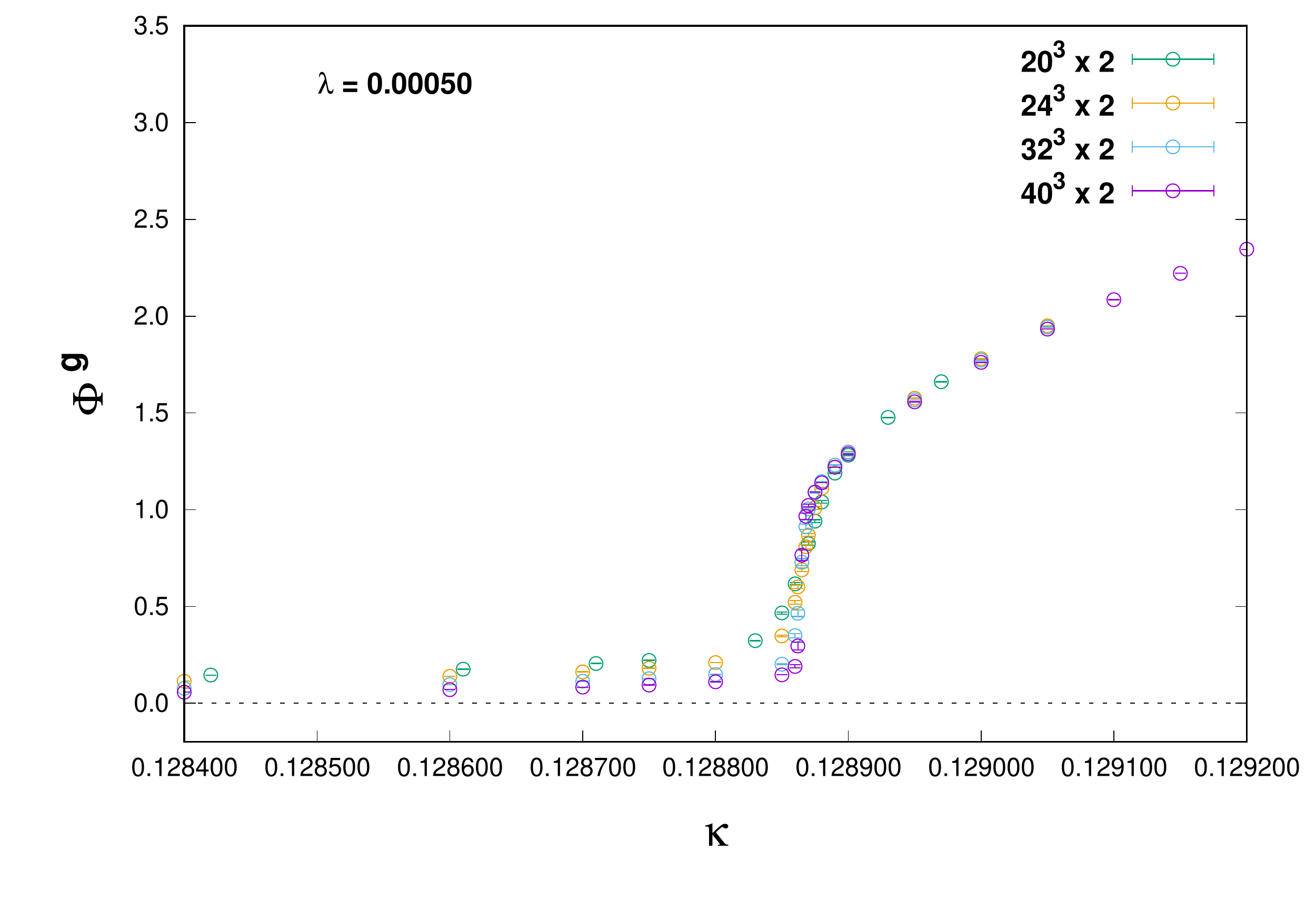}}
    \label{fig:orig_magtn_0500}
  }
  \subfigure[]{%
    %    \subfigtopskip 5pt
    \centering
    \rotatebox{270}{\includegraphics[width=0.26\hsize]
      {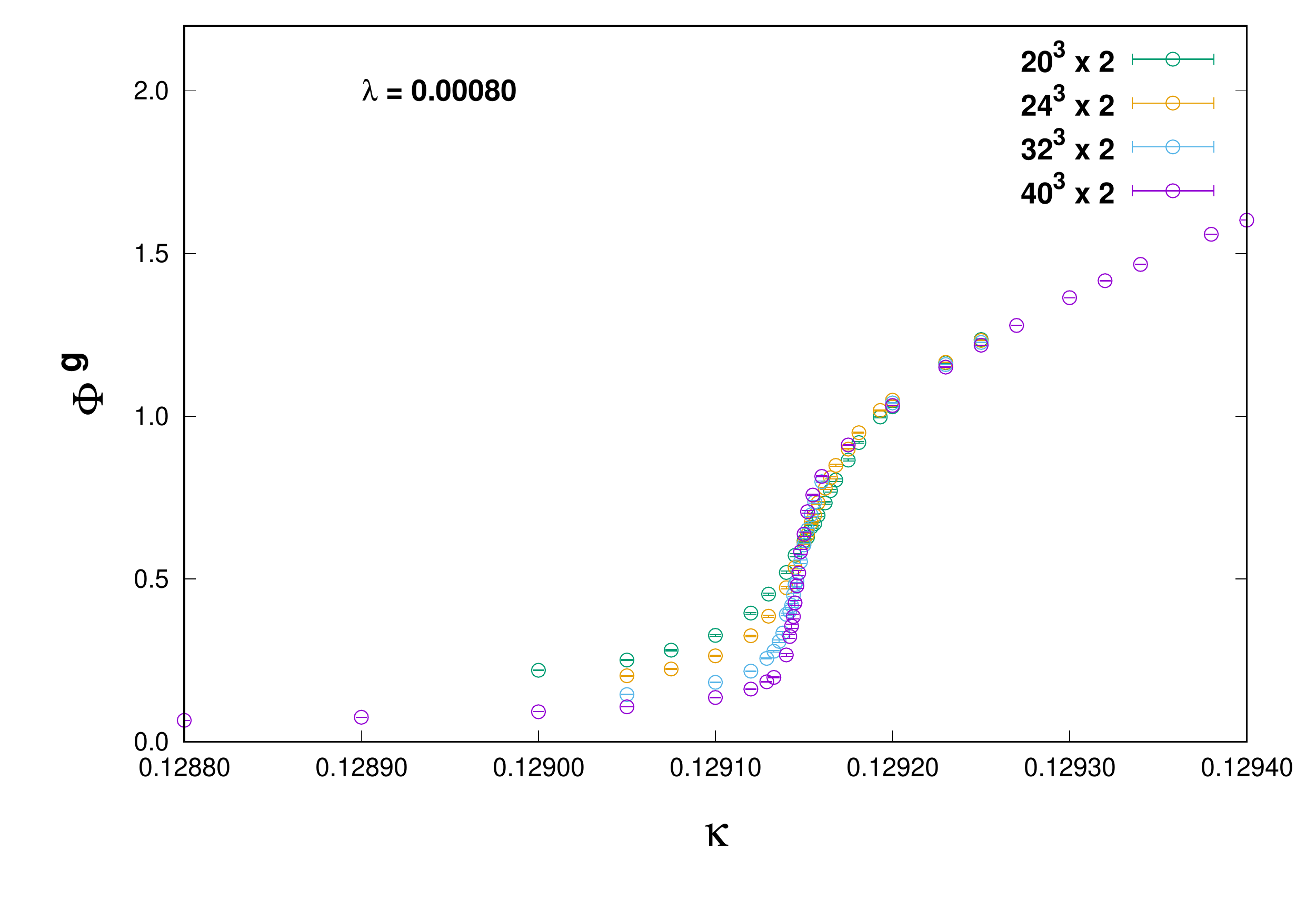}}
    \label{fig:orig_magtn_0800}
  }
  \subfigure[]{%
    \centering
    \rotatebox{270}{\includegraphics[width=0.26\hsize]
      {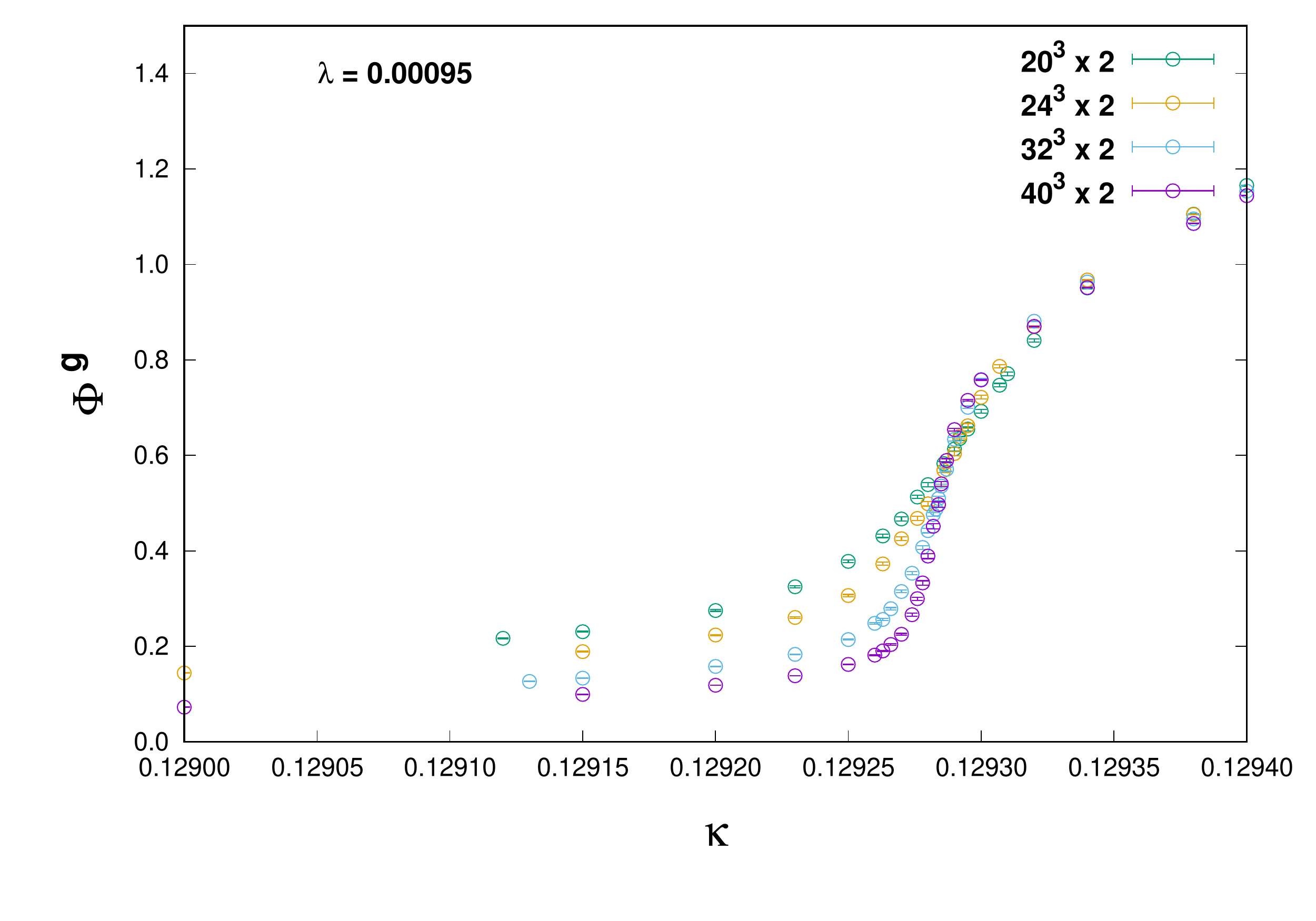}}
    \label{fig:orig_magtn_0950}
  }
  \subfigure[]{%
    \centering
    \rotatebox{270}{\includegraphics[width=0.26\hsize]
      {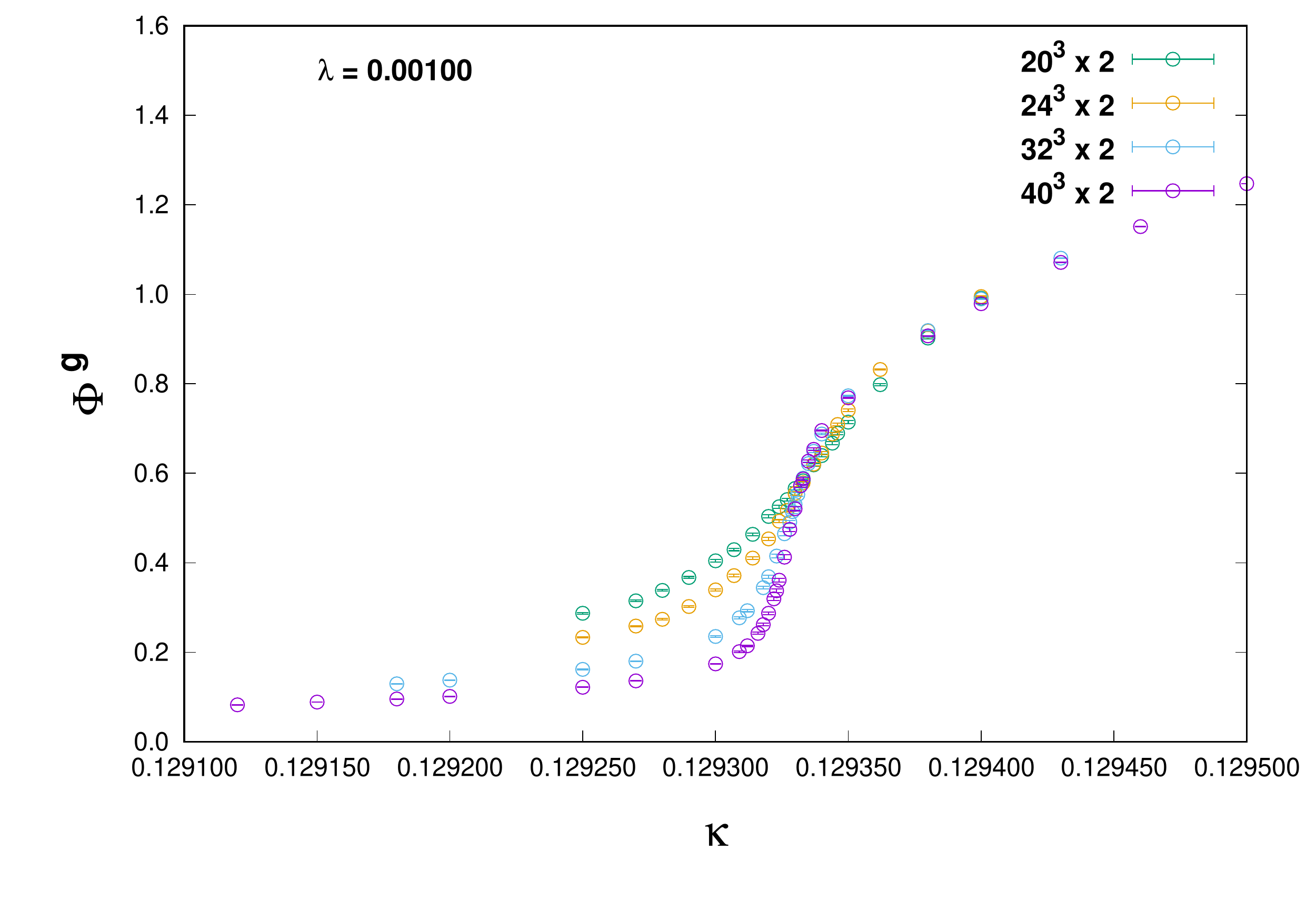}}
    \label{fig:orig_magtn_1000}
  }
  \subfigure[]{%
    %    \subfigtopskip 5pt
    \centering
    \rotatebox{270}{\includegraphics[width=0.26\hsize]
      {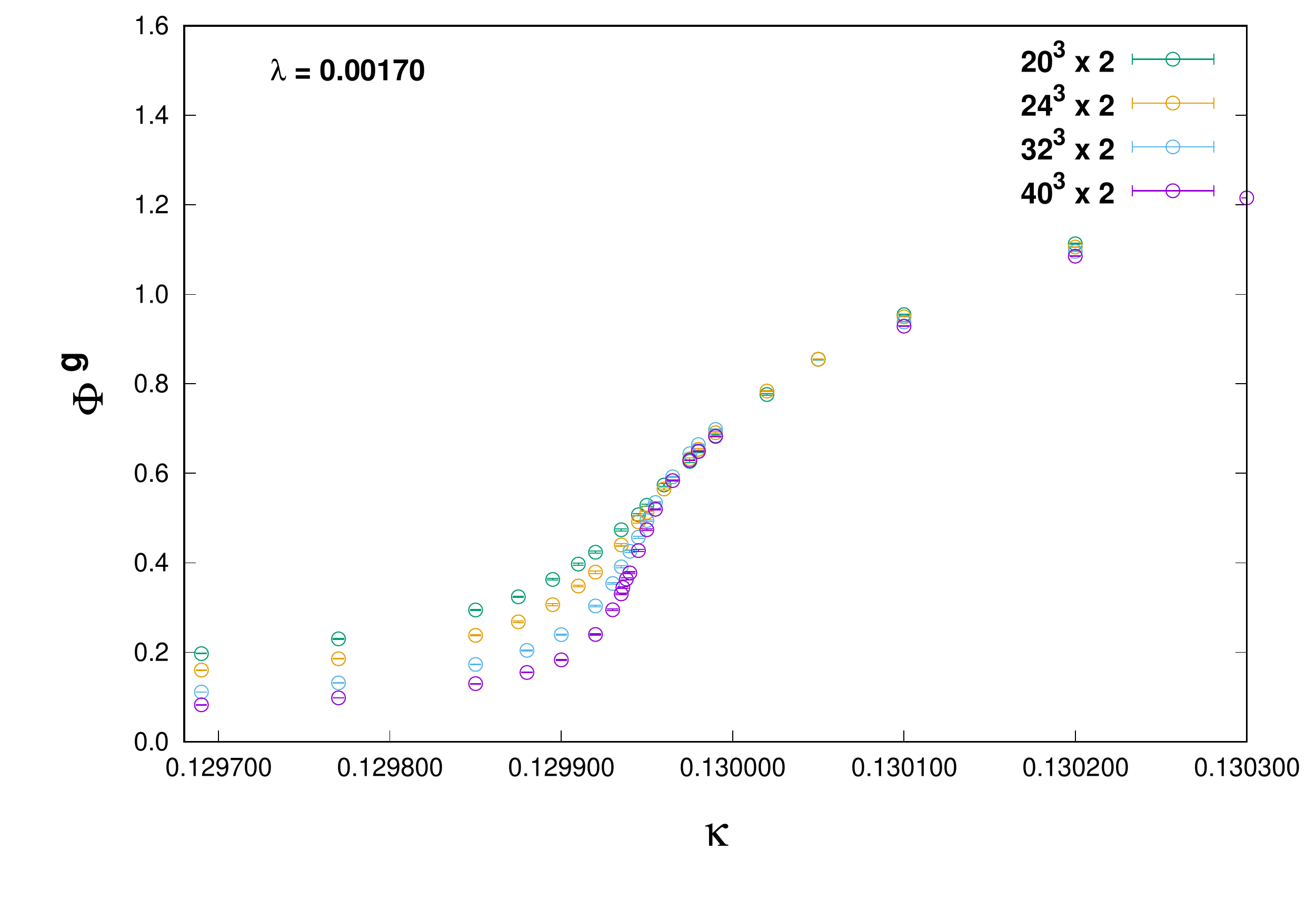}}
    \label{fig:orig_magtn_1700}
  }
  \subfigure[]{%
    \centering
    \rotatebox{270}{\includegraphics[width=0.26\hsize]
      {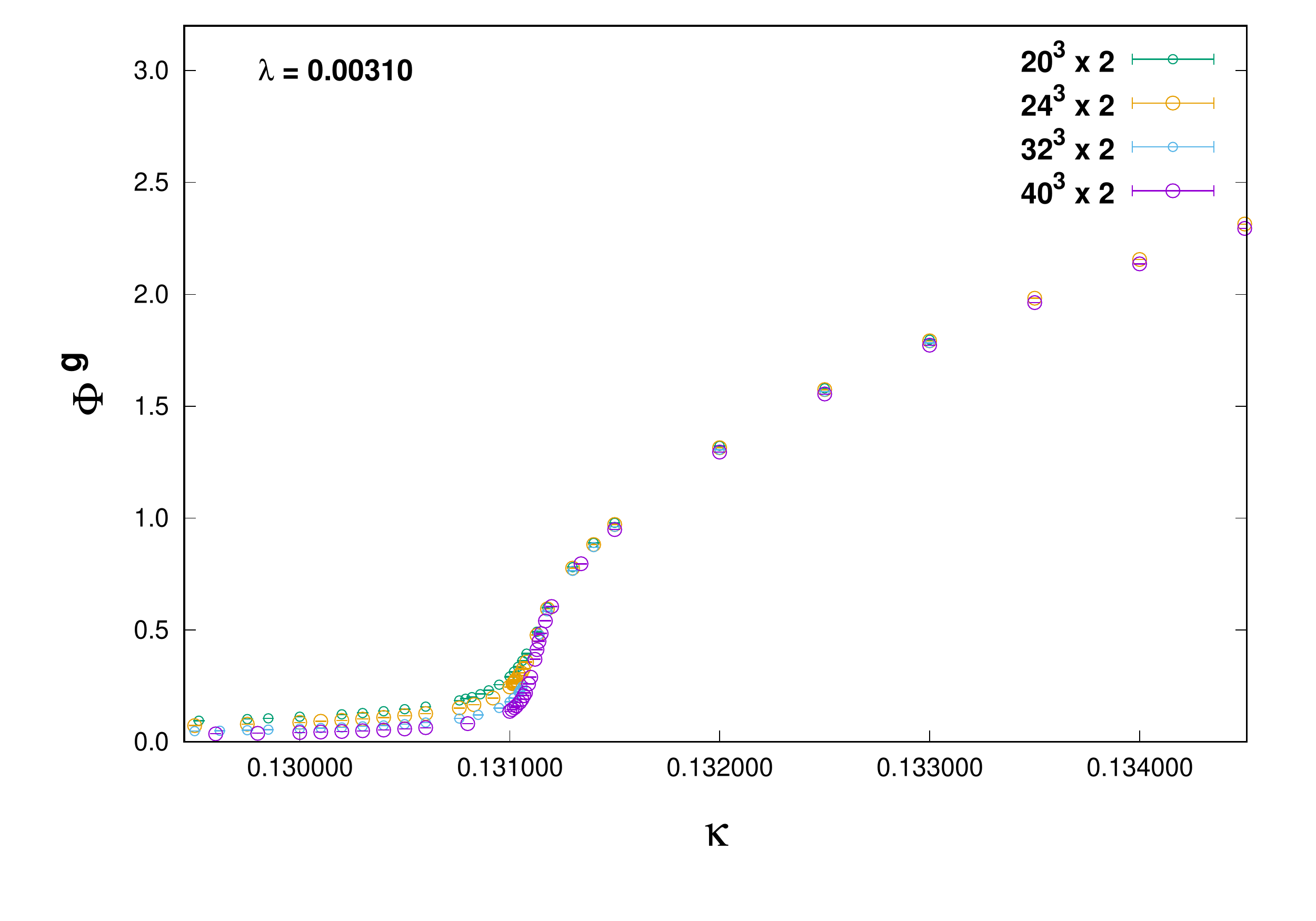}}
    \label{fig:orig_magtn_3100}
  }
  \subfigure[]{%
    \centering
    \rotatebox{270}{\includegraphics[width=0.26\hsize]
      {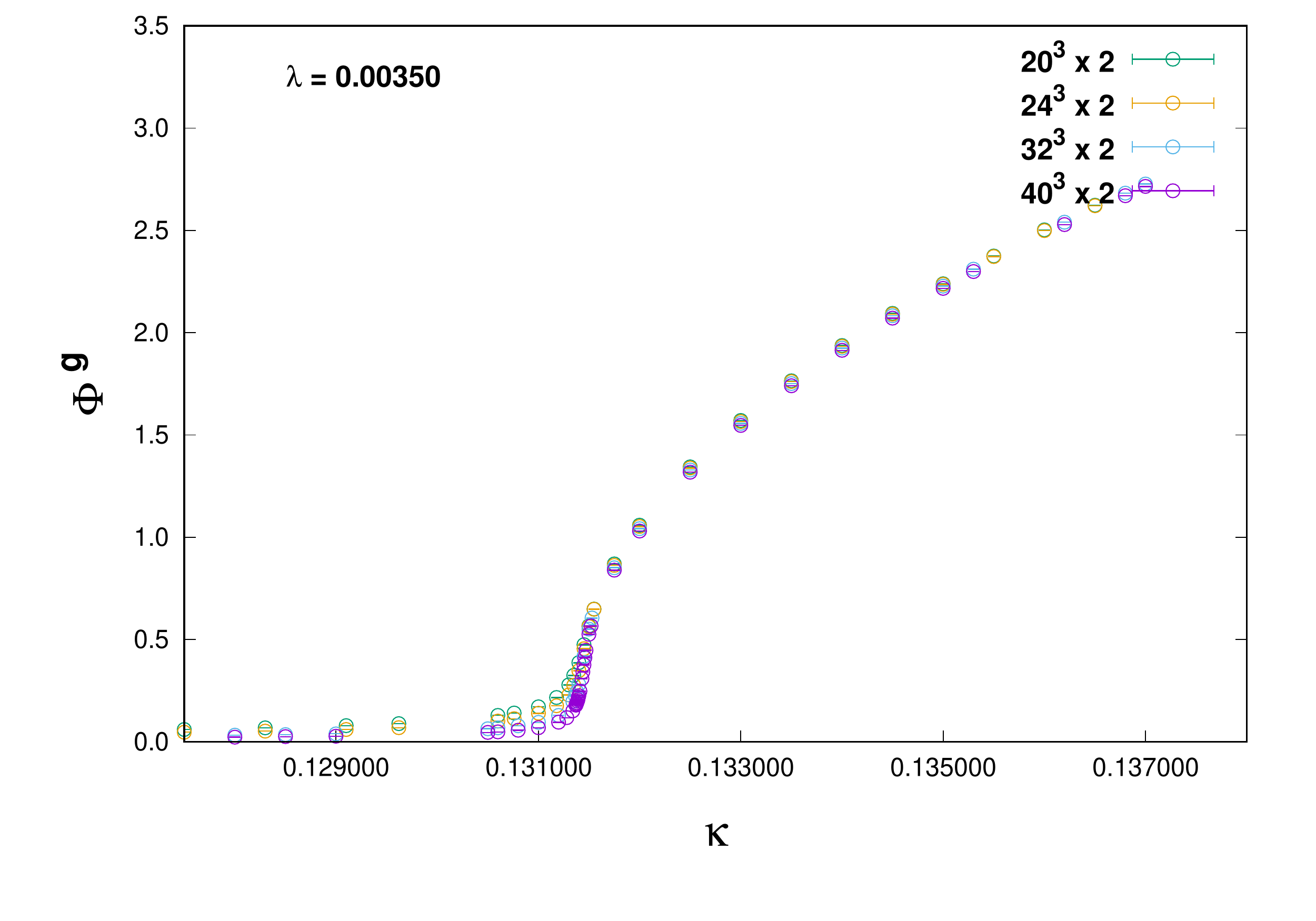}}
    \label{fig:orig_magtn_3500}
  } 
  \caption{
    Magnetization at various values of $\lambda$.
    (a) $\lambda = 0.00010$,
    (b) $\lambda = 0.00050$,
    (c) $\lambda = 0.00080$,
    (d) $\lambda = 0.00095$,
    (e) $\lambda = 0.00100$,
    (f) $\lambda = 0.00170$,
    (g) $\lambda = 0.00310$,
    and
    (h) $\lambda = 0.00350$.}
  \label{fig:orig_magtn_all_l}
\end{figure}

We further investigate whether $\Phi^g$ displays the behavior of a conventional
order parameter as a function of $\lambda$. We study the magnetization for various
$\lambda$ values between $\{0.0001, 0.0035\}$. The simulation parameters for this
study are listed in Table~\ref{tab:run_para_2}. As mentioned in
Sec.~\ref{sec:numerical_param}, the number of configurations is restricted to
$10\, 000$ only for each combination of $\lambda$ and $\kappa$.

\smallskip

From Figs.~\ref{fig:orig_magtn_0100} and \ref{fig:orig_magtn_0500} at $\lambda = 0.00010$
and $0.00050$, respectively, we clearly see that $\Phi^g$ displays a $1$st order
transition as a function of $\kappa$. Finally, we see from Figs.~\ref{fig:orig_magtn_3100}
and \ref{fig:orig_magtn_3500} that the volume dependence nearly disappears at the values
of $\lambda = 0.00310$ and $0.00350$, respectively, indicating a cross-over.

\smallskip

\begin{figure}[htbp]
    \centering
    \rotatebox{270}{\includegraphics[width=0.35\hsize]
      {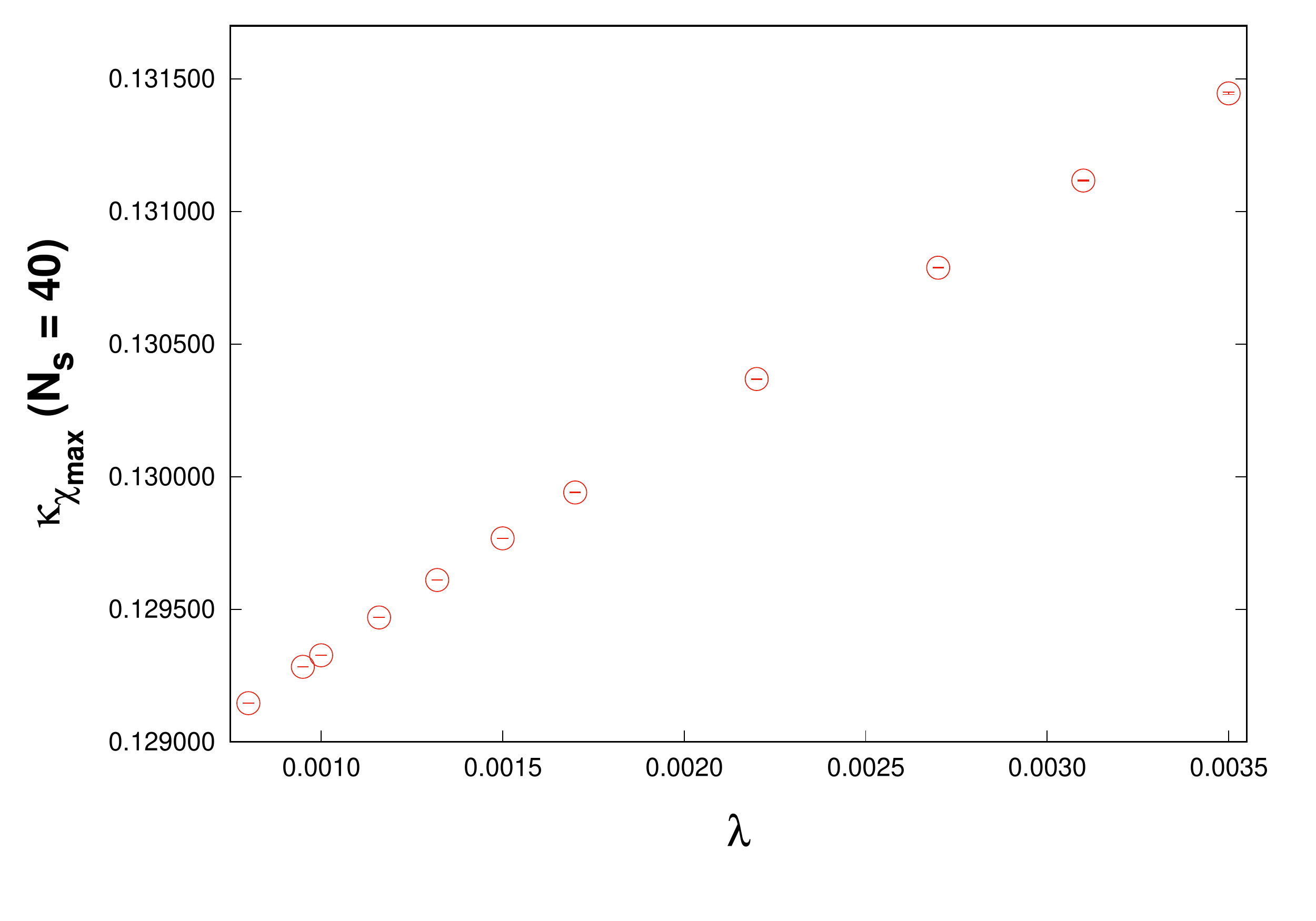}}
    \caption{The behavior of ${\kappa_\chi}_{\rmsmall{max}}$ for $N_s = 40$ as a function
      of $\lambda$.}
    \label{fig:kmx_vs_lambda}
\end{figure}

We also evaluate ${\kappa_\chi}_{\rmsmall{max}}$'s from the maximum value of susceptibility
for $N_s = 40$ for $\lambda =\{0.0008, 0.0035\}$. They are plotted in
Fig.~\ref{fig:kmx_vs_lambda}.

%\newpage

%%%%%%%%%%%%%%%%%%%%%%%%%%%%%%%%%%%%%%%%%%%%
%\lowercase{\boldmath{${\kappa_\chi}_{\rmsmall{max}}$}}}
\section{Discussion and Conclusion}
\label{sec:discussion_conclusion}
%%%%%%%%%%%%%%%%%%%%%%%%%%%%%%%%%%%%%%%%%%%%

In this work, we study the behavior of the volume average of the Landau gauge-fixed
$\Phi^g$ and related quantities in $SU(2)-$Higgs theory. The aim of our study
is to see  if $\Phi^g$ behaves like an order parameter for the Higgs transition. We
consider $\beta_g=8$ as in~\cite{Aoki:1999fi}, and study $\Phi^g$ at various values of
$\lambda$, from $\lambda = 0.00010$ to $\lambda = 0.0035$. Our results show that, $\Phi^g$
for different $\lambda$ accurately identifies the transitions points ($\kappa_c$'s).
Across first order transition, $\Phi^g$ varies discontinuously, as expected from an
order parameter.  This is clearly seen at $\lambda = 0.00010$ and$\lambda = 0.00050$
in agreement with previous study~\cite{Aoki:1999fi}. From $\lambda = 0.0008$ onward, we
see  weakening of transition, with the jump in $\Phi^g$ decreasing with $\lambda$.
At much higher values of $\lambda$, such as $\lambda = 0.00310$ and $\lambda = 0.00350$,
we see a gradual disappearance of volume dependence signifying a possible crossover. It is
interesting to note that a different lattice study~\cite{Wurtz:2013ova} at zero temperature
performed at $\beta_g = 8, \lambda = 0.0033, \kappa =0.131$ with a lattice size of
$24^3\times 48$ has found the physical Higgs mass to be $123 (1)$~GeV which is closer to
the experimental value.

\smallskip

A significant part our numerical simulations are devoted to study behavior of $\Phi^g$
near the end-point. The magnetization, susceptibility,
total susceptibility and binder cumulant etc. of $\Phi^g$ as functions of $\kappa$ give
an impression of an order parameter like behavior at $\lambda_c$. The
scaling of these quantities for
$\nu=0.62997$, $\beta/\nu=0.518$ and $\gamma/\nu=1.964$, shows that the end-point
belongs to the $3$d Ising universality class. From the scaling of the Binder cumulant
from Fig.~\ref{fig:rescaled_binder_1500}, we find its value at
$(\lambda_c,\kappa_c)$ to be consistent with the standard $3$d Ising value of
$\sim 0.47$~\cite{ Ferrenberg:1991,Hasenbusch:1998gh}.

\smallskip

We mention here that the FSS of $\Phi^g$ and its various cumulants
suggests that the $\!3$d Ising like behavior seen close to $\lambda = 0.00150$ which
is within $2\sigma$ of the end-point found in~\cite{Aoki:1999fi}. The FSS of $\Phi^g$
seen in Fig.~\ref{fig:rescaled_1500}
suggests that exactly at the critical point $(\lambda_c,\kappa_c)$ $\Phi^g$ vanishes. Also
the numerical results suggest that for $\lambda < 0$ in the Higgs symmetric phase
and at $\kappa_c$, $\Phi^g$ vanishes in the infinite volume limit. This is in contrast
to behavior of magnetization in $3$-state Potts model, where the magnetization vanishes
in the symmetric phase only in the absence of external field. In the presence of external
field, in the symmetric phase at transition point, magnetization increases with external
field. This behavior of magnetization $vs$ external field can be be obtained by adding a
simple explicit breaking term to Ginzburg-Landau type of mean field free energy. Such an
effective description of $\Phi^g$ require non-trivial term(s) such that $\Phi^g$ is zero in
the Higgs symmetric phase at least for $\lambda < \lambda_c$. 

%\clearpage

%\newpage

%%%%%%%%%%%%%%%%%%%%%%%%%%%%%%%%%%%%%%%%%%%
%%%%%%%%%%%%%%%%%%%%%%%%%%%%%%%%%%%%%%%%%%%

%%%%%%%%%%%%%%%%%%%%%%%%%%%%%%%%%%%%%%%%%%%
\acknowledgments
%%%%%%%%%%%%%%%%%%%%%%%%%%%%%%%%%%%%%%%%%%%

We thank Saumen Datta, Mikko Laine and Kiyoshi Sasaki for helpful discussions. M. Deka
thanks for the hospitality of the Institute of Mathematical Sciences, India  where
a part of this work has been completed. All our numerical computations have
been performed at Govorun super-cluster at Joint Institute for Nuclear Research, Dubna
and Annapurna super-cluster based at the Institute of Mathematical Sciences, India. We
have used the {\tt MILC} collaboration's public lattice gauge theory code
(version 6)~\cite{milc} as our base code.

%%%%%%%%%%%%%%%%%%%%%%%%%%%%%%%%%%%%%%%%%%%
%%%%%%%%%%%%%%%%%%%%%%%%%%%%%%%%%%%%%%%%%%%

\vspace{15mm}

%\newpage

\phantomsection


\addcontentsline{toc}{chapter}{References}

\centerline{\bf  REFERENCES}\vskip -20pt

\begin{thebibliography}{99}     
\vspace{-4mm}   
%       
%
\bibitem{Kuzmin:1985mm}
V.~A.~Kuzmin, V.~A.~Rubakov and M.~E.~Shaposhnikov,
%``On the Anomalous Electroweak Baryon Number Nonconservation in the Early Universe,''
Phys. Lett. B \textbf{155}, 36 (1985)
doi:10.1016/0370-2693(85)91028-7
%
\bibitem{Matveev:1988pj}
V.~A.~Matveev, V.~A.~Rubakov, A.~N.~Tavkhelidze and M.~E.~Shaposhnikov,
%``Nonconservation of Baryon Number Under Extremal Conditions,''
Sov. Phys. Usp. \textbf{31}, 916-939 (1988)
doi:10.1070/PU1988v031n10ABEH005633
%
\bibitem{Rubakov:1996vz}
V.~A.~Rubakov and M.~E.~Shaposhnikov,
%``Electroweak baryon number nonconservation in the early universe and in high-energy collisions,''
Usp. Fiz. Nauk \textbf{166}, 493-537 (1996)
doi:10.1070/PU1996v039n05ABEH000145
[arXiv:hep-ph/9603208 [hep-ph]].
%
\bibitem{Fodor:1998if}
Z.~Fodor,
%``Status of the electroweak phase transitions,''
KEK-TH-558.
%
\bibitem{Damgaard:1986qe}
P.~H.~Damgaard and U.~M.~Heller,
%``The Fundamental SU(2) Higgs Model at Finite Temperature,''
Nucl. Phys. B \textbf{294}, 253 (1987)
doi:10.1016/0550-3213(87)90582-7
%
\bibitem{Evertz:1986af}
H.~G.~Evertz, J.~Jersak and K.~Kanaya,
%``FINITE TEMPERATURE SU(2) HIGGS MODEL ON A LATTICE,''
Nucl. Phys. B \textbf{285}, 229-252 (1987)
doi:10.1016/0550-3213(87)90336-1.
%
\bibitem{Farakos:1994xh}
K.~Farakos, K.~Kajantie, K.~Rummukainen and M.~E.~Shaposhnikov,
%``3-d physics and the electroweak phase transition: A Framework for lattice Monte Carlo analysis,''
Nucl. Phys. B \textbf{442}, 317-363 (1995)
doi:10.1016/0550-3213(95)80129-4
[arXiv:hep-lat/9412091 [hep-lat]].
%
\bibitem{Kajantie:1995kf}
K.~Kajantie, M.~Laine, K.~Rummukainen and M.~E.~Shaposhnikov,
%``The Electroweak phase transition: A Nonperturbative analysis,''
Nucl. Phys. B \textbf{466}, 189-258 (1996)
doi:10.1016/0550-3213(96)00052-1
[arXiv:hep-lat/9510020 [hep-lat]].
%
\bibitem{Rummukainen:1998as}
K.~Rummukainen, M.~Tsypin, K.~Kajantie, M.~Laine and M.~E.~Shaposhnikov,
%``The Universality class of the electroweak theory,''
Nucl. Phys. B \textbf{532}, 283-314 (1998)
doi:10.1016/S0550-3213(98)00494-5
[arXiv:hep-lat/9805013 [hep-lat]].
%
\bibitem{Gurtler:1997hr}
M.~Gurtler, E.~M.~Ilgenfritz and A.~Schiller,
%``Where the electroweak phase transition ends,''
Phys. Rev. D \textbf{56}, 3888-3895 (1997)
doi:10.1103/PhysRevD.56.3888
[arXiv:hep-lat/9704013 [hep-lat]].
%
\bibitem{Bunk:1992xt}
B.~Bunk, E.~M.~Ilgenfritz, J.~Kripfganz and A.~Schiller,
%``Lattice studies at zero and finite temperature in the SU(2) Higgs model at small couplings,''
Phys. Lett. B \textbf{284}, 371-376 (1992)
doi:10.1016/0370-2693(92)90447-C
%
\bibitem{Bunk:1992kf}
B.~Bunk, E.~M.~Ilgenfritz, J.~Kripfganz and A.~Schiller,
%``The Finite temperature phase transition in lattice SU(2) Higgs theory at weak couplings,''
Nucl. Phys. B \textbf{403}, 453-474 (1993)
doi:10.1016/0550-3213(93)90043-O
%
\bibitem{Bunk:1994xs} 
  B.~Bunk,
  %``Monte Carlo methods and results for the electro-weak phase transition,''
  Nucl.\ Phys.\ Proc.\ Suppl.\  {\bf 42}, 566 (1995).\ 
  doi:10.1016/0920-5632(95)00313-X.
%
\bibitem{Gurtler:1997ki}
M.~Gurtler, E.~M.~Ilgenfritz, A.~Schiller and C.~Strecha,
%``Hot electroweak matter near to the endpoint of the phase transition,''
Nucl. Phys. B Proc. Suppl. \textbf{63}, 563-565 (1998)
doi:10.1016/S0920-5632(97)00834-7
[arXiv:hep-lat/9709020 [hep-lat]].
%
\bibitem{Buchmuller:1994qy}
W.~Buchmuller and O.~Philipsen,
%``Phase structure and phase transition of the SU(2) Higgs model in three-dimensions,''
Nucl. Phys. B \textbf{443}, 47-69 (1995)
doi:10.1016/0550-3213(95)00124-B
[arXiv:hep-ph/9411334 [hep-ph]].
%
\bibitem{Kajantie:1996mn}
K.~Kajantie, M.~Laine, K.~Rummukainen and M.~E.~Shaposhnikov,
%``Is there a~ hot electroweak phase transition at $m_H \gtrsim m_W$?,''
Phys. Rev. Lett. \textbf{77}, 2887-2890 (1996)
doi:10.1103/PhysRevLett.77.2887
[arXiv:hep-ph/9605288 [hep-ph]].
%
\bibitem{Fodor:1994dm}
Z.~Fodor, J.~Hein, K.~Jansen, A.~Jaster, I.~Montvay and F.~Csikor,
%``Numerical simulations and the strength of the electroweak phase transition,''
Phys. Lett. B \textbf{334}, 405-411 (1994)
doi:10.1016/0370-2693(94)90706-4
[arXiv:hep-lat/9405021 [hep-lat]].
%
\bibitem{Fodor:1994sj}
Z.~Fodor, J.~Hein, K.~Jansen, A.~Jaster and I.~Montvay,
%``Simulating the electroweak phase transition in the SU(2) Higgs model,''
Nucl. Phys. B \textbf{439}, 147-186 (1995)
doi:10.1016/0550-3213(95)00038-T
[arXiv:hep-lat/9409017 [hep-lat]].
%
\bibitem{Csikor:1995jj}
F.~Csikor, Z.~Fodor, J.~Hein and J.~Heitger,
%``Interface tension of the electroweak phase transition,''
Phys. Lett. B \textbf{357}, 156-162 (1995)
doi:10.1016/0370-2693(95)00886-P
[arXiv:hep-lat/9506029 [hep-lat]].
%
\bibitem{Karsch:1996aw}
F.~Karsch, T.~Neuhaus, A.~Patkos and J.~Rank,
%``Gauge boson masses in the 3-D, SU(2) gauge Higgs model,''
Nucl. Phys. B \textbf{474}, 217-234 (1996)
doi:10.1016/0550-3213(96)00224-6
[arXiv:hep-lat/9603004 [hep-lat]].
%
\bibitem{Karsch:1996yh}
F.~Karsch, T.~Neuhaus, A.~Patkos and J.~Rank,
%``Critical Higgs mass and temperature dependence of gauge boson masses in the SU(2) gauge Higgs model,''
Nucl. Phys. B Proc. Suppl. \textbf{53}, 623-625 (1997)
doi:10.1016/S0920-5632(96)00736-0
[arXiv:hep-lat/9608087 [hep-lat]].
%
\bibitem{Aoki:1996cu} 
  Y.~Aoki,
  %``Four-dimensional simulation of the hot electroweak phase transition with the SU(2) gauge Higgs model,''
  Phys.\ Rev.\ D {\bf 56}, 3860 (1997).\ 
  doi:10.1103/PhysRevD.56.3860
  [hep-lat/9612023].
%
\bibitem{Aoki:1999fi} 
  Y.~Aoki, F.~Csikor, Z.~Fodor and A.~Ukawa,
  %``The Endpoint of the first order phase transition of the SU(2) gauge Higgs model on a four-dimensional isotropic lattice,''
  Phys.\ Rev.\ D {\bf 60}, 013001 (1999).\ 
  doi:10.1103/PhysRevD.60.013001
  [hep-lat/9901021].
%
\bibitem{Bonati:2009pf}
C.~Bonati, G.~Cossu, M.~D'Elia and A.~Di Giacomo,
%``Phase diagram of the lattice SU(2) Higgs model,''
Nucl. Phys. B \textbf{828}, 390-403 (2010)
doi:10.1016/j.nuclphysb.2009.12.003
[arXiv:0911.1721 [hep-lat]].
%
\bibitem{DOnofrio:2014rug}
M.~D'Onofrio, K.~Rummukainen and A.~Tranberg,
%``Sphaleron Rate in the Minimal Standard Model,''
Phys. Rev. Lett. \textbf{113}, no.14, 141602 (2014)
doi:10.1103/PhysRevLett.113.141602
[arXiv:1404.3565 [hep-ph]].
%
\bibitem{DOnofrio:2015gop}
M.~D'Onofrio and K.~Rummukainen,
%``Standard model cross-over on the lattice,''
Phys. Rev. D \textbf{93}, no.2, 025003 (2016)
doi:10.1103/PhysRevD.93.025003
[arXiv:1508.07161 [hep-ph]].
%
\bibitem{DOnofrio:2012yxq}
M.~D'Onofrio, K.~Rummukainen and A.~Tranberg,
%``The sphaleron rate at the electroweak crossover with 125 GeV Higgs mass,''
PoS \textbf{LATTICE2012}, 055 (2012)
doi:10.22323/1.164.0055
[arXiv:1212.3206 [hep-ph]].
%
\bibitem{Laine:2012jy}
M.~Laine, G.~Nardini and K.~Rummukainen,
%``Lattice study of an electroweak phase transition at $m_h \backsimeq$ 126 GeV,''
JCAP \textbf{01}, 011 (2013)
doi:10.1088/1475-7516/2013/01/011
[arXiv:1211.7344 [hep-ph]].
%
\bibitem{Gould:2022ran}
O.~Gould, S.~G\"uyer and K.~Rummukainen,
%``First-order electroweak phase transitions: a nonperturbative update,''
[arXiv:2205.07238 [hep-lat]].
%
\bibitem{Kajantie:1994mt}
K.~Kajantie,
%``Hot electroweak matter,''
Nucl. Phys. B Proc. Suppl. \textbf{42}, 103-112 (1995)
doi:10.1016/0920-5632(95)00192-C
[arXiv:hep-lat/9412072 [hep-lat]].
%
\bibitem{Jansen:1995yg}
K.~Jansen,
%``Status of the finite temperature electroweak phase transition on the lattice,''
Nucl. Phys. B Proc. Suppl. \textbf{47}, 196-211 (1996)
doi:10.1016/0920-5632(96)00045-X
[arXiv:hep-lat/9509018 [hep-lat]].
%
\bibitem{Rummukainen:1996sx}
K.~Rummukainen,
%``Finite T electroweak phase transition on the lattice,''
Nucl. Phys. B Proc. Suppl. \textbf{53}, 30-42 (1997)
doi:10.1016/S0920-5632(96)00597-X
[arXiv:hep-lat/9608079 [hep-lat]].
%
\bibitem{Pisarski:1983ms}
R.~D.~Pisarski and F.~Wilczek,
%``Remarks on the Chiral Phase Transition in Chromodynamics,''
Phys. Rev. D \textbf{29}, 338-341 (1984)
doi:10.1103/PhysRevD.29.338.
%
\bibitem{Karsch:2000xv}
F.~Karsch and S.~Stickan,
%``The Three-dimensional, three state Potts model in an external field,''
Phys. Lett. B \textbf{488}, 319-325 (2000)
doi:10.1016/S0370-2693(00)00902-3
[arXiv:hep-lat/0007019 [hep-lat]].
%
\bibitem{Alonso:1993tv}
J.~L.~Alonso, V.~Azcoiti, I.~Campos, J.~C.~Ciria, A.~Cruz, D.~Iniguez, F.~Lesmes, C.~Piedrafita, A.~Rivero and A.~Tarancon, \textit{et al.}
%``The U(1) Higgs model: Critical behavior in the confining Higgs region,''
Nucl. Phys. B \textbf{405}, 574-592 (1993)
doi:10.1016/0550-3213(93)90560-C
[arXiv:hep-lat/9210014 [hep-lat]].
%
\bibitem{Janke:1996qb}
W.~Janke and R.~Villanova,
%``Three-dimensional three state Potts model revisited with new techniques,''
Nucl. Phys. B \textbf{489}, 679-696 (1997)
doi:10.1016/S0550-3213(96)00710-9
[arXiv:hep-lat/9612008 [hep-lat]].
%
\bibitem{Wilding:1994zkq}
N.~B.~Wilding,
%``Critical point and coexistence curve properties of the Lennard-Jones fluid: A Finite-size scaling study,''
Phys. Rev. E \textbf{52}, 602-611 (1995)
doi:10.1103/PhysRevE.52.602
[arXiv:cond-mat/9503145 [cond-mat]].
%
\bibitem{Wilding:1994ud}
N.~B.~Wilding and M.~Mueller,
%``Liquid-vapour asymmetry in pure fluids: A Monte Carlo simulation study,''
J. Chem. Phys. \textbf{102}, 2562 (1995)
doi:10.1063/1.468686
[arXiv:cond-mat/9410077 [cond-mat]].
%
\bibitem{Wilding:1996}
  Nigel B Wilding 1997 J. Phys.: Condens. Matter 9 585.
%
\bibitem{Rehr:1973zz}
  J.~J.~Rehr and N.~D.~Mermin,
%``Revised Scaling Equation of State at the Liquid-Vapor Critical Point,''
Phys. Rev. A \textbf{8}, 472-480 (1973)
doi:10.1103/PhysRevA.8.472.

%
\bibitem{Kanaya:1994qe}
K.~Kanaya and S.~Kaya,
%``Critical exponents of a three dimensional O(4) spin model,''
Phys. Rev. D \textbf{51}, 2404-2410 (1995)
doi:10.1103/PhysRevD.51.2404
[arXiv:hep-lat/9409001 [hep-lat]].
%
\bibitem{Toussaint:1996qr}
D.~Toussaint,
%``Scaling functions for O(4) in three-dimensions,''
Phys. Rev. D \textbf{55}, 362-366 (1997)
doi:10.1103/PhysRevD.55.362
[arXiv:hep-lat/9607084 [hep-lat]].
%
\bibitem{milc}
  \milcurl.
%
\bibitem{Creutz:1980zw} 
  M.~Creutz,
  %``Monte Carlo Study of Quantized SU(2) Gauge Theory,''
  Phys.\ Rev.\ D {\bf 21}, 2308 (1980).\ 
  doi:10.1103/PhysRevD.21.2308.
%
\bibitem{Cabibbo:1982zn} 
  N.~Cabibbo and E.~Marinari,
  %``A New Method for Updating SU(N) Matrices in Computer Simulations of Gauge Theories,''
  Phys.\ Lett.\ B {\bf 119}, 387 (1982).\ 
  doi:10.1016/0370-2693(82)90696-7.
%
\bibitem{Whitmer:1984he} 
  C.~Whitmer,
  %``Overrelaxation Methods For Monte Carlo Simulations Of Quadratic And Multiquadratic Actions,''
  Phys.\ Rev.\ D {\bf 29}, 306 (1984).\ 
  doi:10.1103/PhysRevD.29.306.
%
\bibitem{Aoki:2003ip} 
  S.~Aoki {\it et\ al.} [CP-PACS Collaboration],\ 
  %``Nonperturbative calculation of Z(V) and Z(A) in domain wall QCD on a finite box,''
  Phys.\ Rev.\ D {\bf 70}, 034503 (2004).\ 
  doi:10.1103/PhysRevD.70.034503
  [hep-lat/0312011].
%
\bibitem{Wurtz:2009gf}
  M.~Wurtz, R.~Lewis and T.~G.~Steele,
  %``Effect of Multiple Higgs Fields on the Phase Structure of the SU(2)-Higgs Model,''
  Phys. Rev. D \textbf{79}, 074501 (2009)
  doi:10.1103/PhysRevD.79.074501
  [arXiv:0902.1167 [hep-lat]].
  %
\bibitem{Wurtz:2013ova}
  M.~Wurtz and R.~Lewis,
  %``Higgs and W boson spectrum from lattice simulations,''
  Phys. Rev. D \textbf{88}, 054510 (2013)
  doi:10.1103/PhysRevD.88.054510
  [arXiv:1307.1492 [hep-lat]].
  %
\bibitem{Ferrenberg:1991}
  Alan M. Ferrenberg and D. P. Landau,
  % "Critical behavior of the three-dimensional Ising model: A high-resolution Monte Carlo study".
  Phys. Rev. B \textbf{44}, 5081, 1991.
  %
\bibitem{Hasenbusch:1998gh}
  M.~Hasenbusch, K.~Pinn and S.~Vinti,
  %``Critical exponents of the three-dimensional Ising universality class from finite-size scaling with standard and improved actions,''
  Phys. Rev. B \textbf{59}, 11471-11483 (1999)
  doi:10.1103/PhysRevB.59.11471
  [arXiv:hep-lat/9806012 [hep-lat]].
  %
\end{thebibliography}
\end{document}